\documentclass[acmlarge, natbib=false, nonacm]{acmart}
\usepackage[datamodel=acmdatamodel, style=acmnumeric, sorting=none, backend=biber]{biblatex}
\usepackage{amsmath, amsfonts}

\usepackage{graphicx}
\usepackage{subcaption}

\usepackage[nameinlink]{cleveref}

\usepackage[inline]{enumitem}
\setlist*[itemize]{labelindent=10pt, itemindent=0pt, leftmargin=*}

\usepackage{booktabs}
\usepackage{multirow}
\usepackage{tabularx}

\usepackage{pgfplots}
\pgfplotsset{compat=1.18}
\usepgfplotslibrary{statistics}

\usepackage{csquotes}
\usepackage{textcomp}

\usepackage{balance}

\setcopyright{rightsretained}
\copyrightyear{2024}

\begin{document}

\title[Transformer-Based Particle Tracking for High-Luminosity LHC]{TrackFormers: In Search of Transformer-Based Particle Tracking for the High-Luminosity LHC Era}

\author{Sascha {Caron}}
\orcid{0000-0003-2941-2829}
\affiliation{%
	\institution{High-Energy Physics, Radboud University}
	\city{Nijmegen}
	\country{The Netherlands}}
\affiliation{%
	\institution{National Institute for Subatomic Physics (Nikhef)}
	\city{Amsterdam}
	\country{The Netherlands}}
\email{scaron@nikhef.nl}

\author{Nadezhda {Dobreva}}
\orcid{0009-0006-0923-6054}
\affiliation{%
	\institution{Institute for Computing and Information Sciences, Radboud University}
	\city{Nijmegen}
	\country{The Netherlands}}
\email{nadezhda.dobreva@ru.nl}

\author{Antonio {Ferrer Sánchez}}
\orcid{0000-0002-2305-5261}
\affiliation{%
	\institution{Intelligent Data Analysis Laboratory (IDAL), Department of Electronic Engineering, ETSE-UV, University of Valencia}
	\city{Valencia}
	\country{Spain}}
\affiliation{%
	\institution{Valencian Graduate School and Research Network of Artificial Intelligence (ValgrAI)}
	\city{Valencia}
	\country{Spain}}
\email{Antonio.Ferrer-Sanchez@uv.es}

\author{José D. {Martín-Guerrero}}
\orcid{0000-0001-9378-0285}
\affiliation{%
	\institution{Intelligent Data Analysis Laboratory (IDAL), Department of Electronic Engineering, ETSE-UV, University of Valencia}
	\city{Valencia}
	\country{Spain}}
\affiliation{%
	\institution{Valencian Graduate School and Research Network of Artificial Intelligence (ValgrAI)}
	\city{Valencia}
	\country{Spain}}
\email{jose.d.martin@uv.es}

\author{Uraz {Odyurt}}
\orcid{0000-0003-1094-0234}
\affiliation{%
	\institution{Faculty of Engineering Technology, University of Twente}
	\city{Enschede}
	\country{The Netherlands}}
\affiliation{%
	\institution{National Institute for Subatomic Physics (Nikhef)}
	\city{Amsterdam}
	\country{The Netherlands}}
\email{uodyurt@nikhef.nl}

\author{Roberto {Ruiz de Austri Bazan}}
\orcid{0000-0003-3688-9609}
\affiliation{%
	\institution{Instituto de Física Corpuscular, University of Valencia}
	\city{Valencia}
	\country{Spain}}
\email{rruiz@ific.uv.es}

\author{Zef {Wolffs}}
\orcid{0000-0001-5100-2522}
\affiliation{%
	\institution{Institute of Physics, University of Amsterdam}
	\city{Amsterdam}
	\country{The Netherlands}}
\affiliation{%
	\institution{National Institute for Subatomic Physics (Nikhef)}
	\city{Amsterdam}
	\country{The Netherlands}}
\email{zwolffs@uva.nl}

\author{Yue {Zhao}}
\orcid{0009-0005-4074-5116}
\affiliation{%
	\institution{SURF}
	\city{Amsterdam}
	\country{The Netherlands}}
\email{yue.zhao@surf.nl}

\renewcommand{\shortauthors}{S. Caron et al.}

\begin{abstract}
High-Energy Physics experiments are facing a multi-fold data increase with every new iteration. This is certainly the case for the upcoming High-Luminosity LHC upgrade. Such increased data processing requirements force revisions to almost every step of the data processing pipeline. One such step in need of an overhaul is the task of particle track reconstruction, a.k.a., \emph{tracking}. A Machine Learning-assisted solution is expected to provide significant improvements, since the most time-consuming step in tracking is the assignment of hits to particles or track candidates. This is the topic of this paper.

We take inspiration from large language models. As such, we consider two approaches: the prediction of the next word in a sentence (next hit point in a track), as well as the one-shot prediction of all hits within an event. In an extensive design effort, we have experimented with three models based on the Transformer architecture and one model based on the U-Net architecture, performing track association predictions for collision event hit points. In our evaluation, we consider a spectrum of simple to complex representations of the problem, eliminating designs with lower metrics early on. We report extensive results, covering both prediction accuracy (score) and computational performance. We have made use of the REDVID simulation framework, as well as reductions applied to the TrackML data set, to compose five data sets from simple to complex, for our experiments. The results highlight distinct advantages among different designs in terms of prediction accuracy and computational performance, demonstrating the efficiency of our methodology. Most importantly, the results show the viability of a one-shot encoder-classifier based Transformer solution as a practical approach for the task of tracking.

\end{abstract}

%
%


\maketitle


\section{Introduction}
\label{sec:introduction}
Today's scientific research, especially in experimental physics, always involves computational components. In fields such as experimental High-Energy Physics (HEP), data-intensive analysis is essential. Consequently, data science and Machine Learning (ML) techniques have become an integral part of these analyses. The application of deep learning in HEP offers several opportunities for the incorporation of new network architectures that may significantly improve prediction accuracy, computational efficiency, or both. Our focus is on the use of the \emph{Transformer} ML model architecture~\cite{Vaswani:2017:Attention} and the \emph{U-Net} architecture~\cite{Ronneberger:2015:UNet} to overcome the challenge of tracking in HEP.

\emph{Tracking} refers to the reconstruction of the trajectory (track) of subatomic particles present in particle physics experiments. The task of tracking involves reconstructing the trajectory of a particle based on detector data. A sensor may or may not be \emph{hit} by a particle, and sequences of these hits comprise tracks. In HEP, the sensory data can come from detectors such as, ALICE~\cite{Collaboration:2008:ALICE}, ATLAS~\cite{Collaboration:2008:ATLAS}, CMS~\cite{Collaboration:2008:CMS}, and LHCb~\cite{Collaboration:2008:LHCb}, installed at the Large Hadron Collider (LHC). At the same time, tracking is important for experiments in neutrino physics, astroparticle physics and other fields, e.g.,~\cite{Woodruff:2018:APTI, Aguilar:2011:FAMT, Patterson:2009:ETER, González-Díaz:2015:AGMT}.

Since the tracking algorithms are time-consuming for most experiments, the reconstruction of events is mainly done offline, i.e., after data acquisition. The reconstruction of the trajectories of charged particles often takes place in two steps.  In the first step, the hits of the sensors belonging to this particle are identified, and in the second step, the kinematic properties of the particle trajectory are determined using a mathematical model of the possible trajectories. This paper deals with novel methods for the first of these two steps, i.e., the fast and accurate assignment of hits to particles or track candidates.

Traditional tracking methods, such as Kalman filters~\cite{Kalman:1960:NALF, Abreu:1996:DELPHI} are currently fundamental in particle tracking for the LHC. These approaches rely on statistical modelling to predict the future state of a particle based on its past states using an iterative process and assuming Gaussian uncertainties. Interestingly, they can be considered as autoregressive models and therefore have properties similar to large language models that predict the next state based on the previous states. With the increase of the integrated luminosity expected during the High-Luminosity phase of the LHC, the number of tracks and the density of detected hits are correspondingly expected to increase significantly. This makes the discrimination between overlapping tracks more difficult, so that more sophisticated and above all faster algorithms are needed to maintain the accuracy of tracking. Track reconstruction, in its current form, is also one of the most computationally intensive components of event reconstruction, as it exhibits quadratic growth with the number of particles in the detector. There is therefore a real need for faster solutions that can be deployed for the upcoming High-Luminosity upgrade of the LHC.

\subsection{Our contribution}
This paper focuses on the development of first solutions for the assignment of hits to track/particle candidates using the \emph{Transformer} and the \emph{U-Net} ML model architectures. Note that for our current solutions, we have opted to address the formation of hit clusters matching tracks and do not intend to cover the actual parameterisation of a track function, which would be a separate non-trivial task. The challenge is to achieve high solution accuracy, which corresponds to the accuracy of assigning hits to correct tracks, while achieving better computational efficiency. Here, computational efficiency is measured by the mean execution time per event.

Inspired by \emph{large language models} predicting the next word from a large set of words (a certain vocabulary), we have developed an autoregressive Transformer encoder-decoder model that predicts the next hit from a set of hits (EncDec). We argue however that the inference time of such a model scales with the total number of hits, or worse. Therefore, we have also investigated methods that can assign particle/track labels to all hits within an event, \emph{in a single step}. Such models can be considered as \enquote{translation} or \enquote{mapping} models, translating sensor hits into track candidates, in one step. Accordingly, we have developed a Transformer encoder-based track classifier that maps or translates hits into track candidates (EncCla). Additionally, we have developed a U-Net based architecture, inspired by models assigning labels in images, in a single step.

As an alternative approach, we have developed a model that translates hits into track parameters using a Transformer encoder-based track regressor (EncReg). This approach is conceptually similar to the Hough Transform~\cite{Duda:1972:HTLC}, a classic method used in image processing to detect shapes like lines and circles by mapping points to a parameter space. 
Similarly, the EncReg model predicts track parameters directly from the hit data. To assign labels to the hits, our model is followed by a clustering algorithm that clusters hit candidates in the track parameter space. This combined approach allows for effective track reconstruction by first mapping hits to parameters and then grouping them into track candidates.

Our goal is to offer fast, i.e., computationally-efficient and high-throughput, solutions for tracking. Therefore, we strive for models that assign hits to trajectories in a single inference step. We determine the accuracy and speed of these models for the tracking problem and use simulated data of increasing complexity to compare our approaches. Finally, we discuss how our initial approaches can improve their performance through post-processing steps and argue that our approaches are fast and accurate enough to be considered as candidates for future tracking pipelines. Our data sets~\cite{Odyurt:2024:DATASET} and code~\cite{Dobreva:2024:CODE}\footnote{Also available at: \url{https://github.com/TrackFormers/TrackFormers-Models}} are publicly available.

\subsection{Advances and applications in particle tracking for high-energy physics}
Particle tracking is a crucial task in HEP experiments. Over the years, various methods have been developed and refined to enhance the accuracy and efficiency of particle tracking.

\subsubsection{HEP experiments and tracking}
In this context, by HEP experiments, we refer to accelerator experiments in which high-energy subatomic particle collisions occur. The Large Hadron Collider (LHC) is perhaps the most well-known particle accelerator, and provides the highest-energy collisions to date. In the LHC, either protons or ions are made to smash into one another in so-called \emph{events}. In this paper, we focus our attention on proton-proton (\emph{pp}) collision events.

These events in turn release a plethora of subatomic particles for which the behaviour is studied through \emph{tracking} and \emph{calorimetry}. Sophisticated detectors, such as ALICE~\cite{Collaboration:2008:ALICE}, ATLAS~\cite{Collaboration:2008:ATLAS}, CMS~\cite{Collaboration:2008:CMS}, and LHCb~\cite{Collaboration:2008:LHCb}, allow us to measure the footprint of individual particles as they travel through space. They are each equipped with dedicated tracking detectors designed to measure the trajectories of charged particles. These consist of layers of sensitive material, such as silicon detectors, generating electrical signals---hits---when charged particles pass through them. These hits are not continuous, but discrete recordings and limited by detector density.

\subsubsection{Simulations for HEP}
Any research/design work with the aim of developing algorithmic solutions or improving existing algorithms requires large amounts of (labelled) data. These data sets are to be used for extensive testing and validation of the algorithms' expected characteristics, such as correctness, data processing capacity and performance, computational efficiency and power consumption. The same applies, or rather is strictly required, when it comes to solutions involving ML models.

As HEP experiments are not of the kind to be performed on demand, simulations are suitable and often necessary alternative. Simulations for HEP can be used to study the effects of physics phenomena through the generation of data sets for analyses and algorithm design efforts. There are numerous simulations available, predominantly focusing on physics-accuracy and detector specificity. Examples relevant to the ATLAS detector are Geant4~\cite{Agostinelli:2003:GST}, FATRAS~\cite{Edmonds:2008:FATS} and ATLFAST~\cite{Richter-Was:1998:AFSP}.

\subsubsection{Traditional methods and performance}
Traditional methods to assign hits to track candidates include the Kalman Filter (KF) and Hough Transforms. The KF assumes Gaussian uncertainties and is considered an optimal solution under certain conditions, although it can be slow. The KF is an algorithm that determines the internal state of a linear dynamic system by recursively processing discrete measurements with random perturbations present in the measurements and the system itself~\cite{Lantz:2020:SPTR}. Both stages of tracking rely on the KF~\cite{Braun:2019:CKFH}, with the track finding phase using a specific version called the combinatorial KF (CKF). This starts with a track seed (a recorded hit) and updates it with the information from subsequent hits. The next hit is selected from a pool of candidates, evaluated, and scored to find the most fitting one. In the track fitting stage, a standard KF algorithm is used, but the overall procedure is similar.

Variations of Kalman filtering have been used at LHC for decades~\cite{Kalman:1960:NALF, Frühwirth:1987:AKFT, Braun:2019:CKFH} (and similar GPU-based algorithms~\cite{Aaij:2020:ALLEN}) due to its robustness and excellent performance. However, its combinatorial nature in the track finding stage and inherent sequential execution make KF unsuitable for the upcoming High-Luminosity stage of the LHC, as it scales poorly with the number of recorded hits. The currently utilised algorithm is tested on simulated data of events with 200 points of origin of tracks (pile-up)~\cite{Atlas:2019:FTRH}, which is similar in complexity to the largest data evaluated in this paper. The KF pipeline takes 214.3 HS06 × seconds for a single event, translating to around 12 seconds CPU-time\footnote{The CPU-time is multiplied by the HS06 factor of 17.8 for single-threaded execution, since the considered CPU is an Intel Xeon E5-2620v2. HS06 is a benchmark for measuring CPU performance in HEP. More information can be found at: \url{https://w3.hepix.org/benchmarking/HS06.html}}. An optimised algorithm with tighter track selection in the track finding stage achieves a 7x speed-up, requiring 1.8 seconds of CPU-time per event~\cite{Atlas:2019:FTRH}.

\subsubsection{Machine learning competitions and community engagement}
To engage data scientists and machine learning experts, a tracking competition (TrackML)~\cite{Kiehn:2019:TrackML} was launched on the Kaggle platform. A training data set was created based on the simulation of a generic High-Luminosity LHC (HL-LHC) experiment tracker, listing the measured 3D hits/points for each event and the list of 3D hits belonging to a real track. The events are top-pair events with a "pile-up" of $\mu=200$ using Poisson statistics for the superposition of Quantum ChromoDynamics (QCD) events, in which, there are concentrations of events in close proximity.

\subsubsection{Advances in deep learning and GNNs}
Recent advances in deep learning have led to the adoption of Deep Neural Networks (DNNs) for particle tracking. These models can learn complex, non-linear relationships from data, potentially improving track reconstruction accuracy. Various architectures, including Convolutional Neural Networks (CNNs) and Recurrent Neural Networks (RNNs), have also been explored for this purpose.

Given that particle interactions can be conveniently represented as graphs, Graph Neural Networks (GNNs) have emerged as a promising approach. These networks operate on graph-structured data, capturing the relational information between hits in a detector. The state-of-the-art in recent years has focused on GNN-based solutions~\cite{Bakina:2022:DLTR, Biscarat:2021:TRTR, Choma:2020:TSLE, DeZoort:2021:CPTE, Duarte:2022:GNNP, Elabd:2022:GNNC, Farrell:2018:NDLM, Heintz:2020:ACPT, Ju:2020:GNNP, Ju:2021:PGDL, Mieskolainen:2023:HNCH, Murnane:2023:EGNN, Thais:2021:ISGO, Caillou:2024:PPAG}, or see this reference for a review on GNNs in HEP~\cite{Shlomi:2021:GNNP}.

In GNN-based methods, edges between vertices (hits) are predicted to determine actual physical trajectories. A graph is generated based on an event, with all hits as nodes connected by edges based on some constraint\footnote{For example, geometric constraints or a preprocessing algorithm such as Hough Transform~\cite{Duda:1972:HTLC}.}. A GNN is trained to assign weights to edges or prune unlikely ones, ensuring a fitting connectivity between vertices to represent particle trajectories~\cite{DeZoort:2021:CPTE, Ju:2021:PGDL, Lazar:2023:AIEP}.

A recent successful approach~\cite{Lieret:2024:OCPC, Lieret:2023:HPPT} involves constructing a graph by connecting hits from different detector layers that satisfy certain geometric constraints. A fully connected neural network estimates the weights of all edges, pruning those below a threshold. Next, object condensation clusters hits of the same track in a learned space, regressing the properties of the reconstructed objects~\cite{Kieseler:2020:OCOG}. Clustering is done using Density-Based Spatial Clustering of Applications with Noise (DBSCAN). Performance is assessed on the inner detector hits from the TrackML data. Lieret et al.~\cite{Lieret:2023:HPPT} defined multiple custom metrics to assess physics performance. No time performance is reported, but the authors suggest that using Transformer models would significantly reduce inference time. An interesting benchmark for GNNs is~\cite{Ju:2021:PGDL}, which reports an inference time of 2.2 seconds wall-clock time (including data transfer to GPUs) on the full TrackML events, using an NVIDIA A100 GPU. This approach involved significant pre and post-processing of the data and reports a TrackML score of about $0.87$. In our approach, we consider an iteration of the TrackML score, i.e., FitAccuracy, as a metric for scoring the success rates of our solutions. FitAccuracy score and its relation with TrackML score is described in detail in \Cref{sec:results}.

Finally, we would like to point out that Transformers have already been used in many applications in HEP and have proven their versatility and efficiency~\cite{Qu:2022:PTJT, Mikuni:2021:PCTC, Butter:2023:JDMN, Leigh:2024:DPCG, Builtjes:2024:ASPI, Caron:2024:UADL}. Though there are more visionary efforts towards the use of Transformers for foundational models, or to further generalise their application in collider physics, these efforts are not as mature yet~\cite{Huang:2024:LMPT}.

\subsubsection{Software frameworks and other ML-assisted solutions}
Other approaches to improve tracking include software frameworks like \enquote{A Common Tracking Software (ACTS)}~\cite{Ai:2022:ACTS}, tested on the Open Data Detector (ODD). ACTS is an experiment-independent toolkit for particle track reconstruction in HEP.

ML-assisted solutions have led to specific improvements, such as ambiguity resolution at the end of the tracking chain, determining which track candidates to keep or discard~\cite{Allaire:2023:RNNA}. This insight generation can be considered a side benefit of researching ML-assisted solutions. Incorporating data-driven, iterative improvements to traditional algorithms or partial inclusion of ML models is a conservative approach. We seek a solution predominantly relying on ML models as its core building block, operating as a single-pass algorithm. As in many computer science topics, there is also the question of monolithic versus hybrid/modular deployment of ML models. A promising detector-specific hybrid solution focuses on finding Primary Vertices (PVs)~\cite{Akar:2021:PDHD, Akar:2023:CIHD}. The approach of replacing parts of traditional tracking solutions will eventually lead to hybrid/modular solutions.

\subsubsection{Data reduction and standardisation}
The most notable trend is the de facto use of TrackML~\cite{Amrouche:2020:TMLC} data for experimentation. However, ML model training is computationally expensive and requires large amounts of hardware resources, such as GPU memory. Therefore, authors often reduce the data. The most common reduction is to only consider data associated with the inner detector, also known as the pixel detector~\cite{Choma:2020:TSLE, DeZoort:2021:CPTE, Elabd:2022:GNNC, Heintz:2020:ACPT, Ju:2020:GNNP, Murnane:2023:EGNN, Thais:2021:ISGO}. A few authors apply further reductions, such as noise hit reduction~\cite{Choma:2020:TSLE, Ju:2020:GNNP} or filtering for limited $P_T$ values~\cite{Heintz:2020:ACPT}. Some examples consider the full detector but apply lower pile-up alongside noise hit reduction~\cite{Mieskolainen:2023:HNCH}.

The application of reduction to a de facto standard data set (TrackML) and the lack of consensus for a common data reduction protocol, render the data used by different authors non-standard. This makes it challenging to perform direct comparisons between different results. Our data reduction protocol and the considered performance metrics are elaborated in \Cref{sec:data_sets,sec:results}, respectively.

\section{Data sets}
\label{sec:data_sets}
In this section, we describe the data sets used to train and evaluate our particle tracking methods. These data sets cover a spectrum of scale and track representation complexity, ranging from simple linear tracks to more complex helical and closer to real-world tracks. By increasing the problem complexity in two dimensions (scale and track representation) we can thoroughly assess the performance and robustness of the tracking algorithms. As such, we consider five data sets, covering a spectrum of scale and track representation complexity from low to high. Data set titles are,
\begin{itemize}
    \item 10-50 (variable count) linear tracks per event, generated with REDVID,
    \item 10-50 (variable count) helical tracks per event, generated with REDVID,
    \item 50-100 (variable count) helical tracks per event, generated with REDVID,
    \item 10-50 (variable count) tracks per event, extracted from the TrackML data set,
    \item 200-500 (variable count) tracks per event, extracted from the TrackML data set.
\end{itemize}

\subsection{REDVID data set}
The first three data sets are the result of simulations using REDVID simulation framework~\cite{Odyurt:2024:RSHE}. REDVID simulations are fully configurable with an extensive set of options. We have considered events with random track counts. Simulations with minimum and maximum track count boundaries as $[10, 50]$ and $[50, 100]$ have been executed. Track function complexity varies between linear and expanding helical, with the latter representing a simple emulation of charged particles in a magnetic field. Note that the 3D geometric space and elements contained within are defined in cylindrical coordinate system, with $r$, $\theta$ and $z$ coordinates as radius, angle with the X-axis and location on the Z-axis, respectively. The above REDVID data sets contain 100\,000 events each, with noise (smearing) applied to calculated hit coordinates. Data set headers include 15 fields, covering
\begin{itemize}
    \item event, sub-detector, track and hit ids,
    \item track function parameters, whether linear or helical, same number of parameters are present, with different interpretations per track type,
    \item hit coordinates, and
    \item a few type descriptors for sub-detector and track types.
\end{itemize}
For more detailed descriptions covering field data types, we refer readers to the README file in the shared Zenodo record~\cite{Odyurt:2024:DATASET}.

\subsubsection{REDVID detector geometry}
\begin{figure}[htbp]
	\centering
	\includegraphics[width=0.4\linewidth]{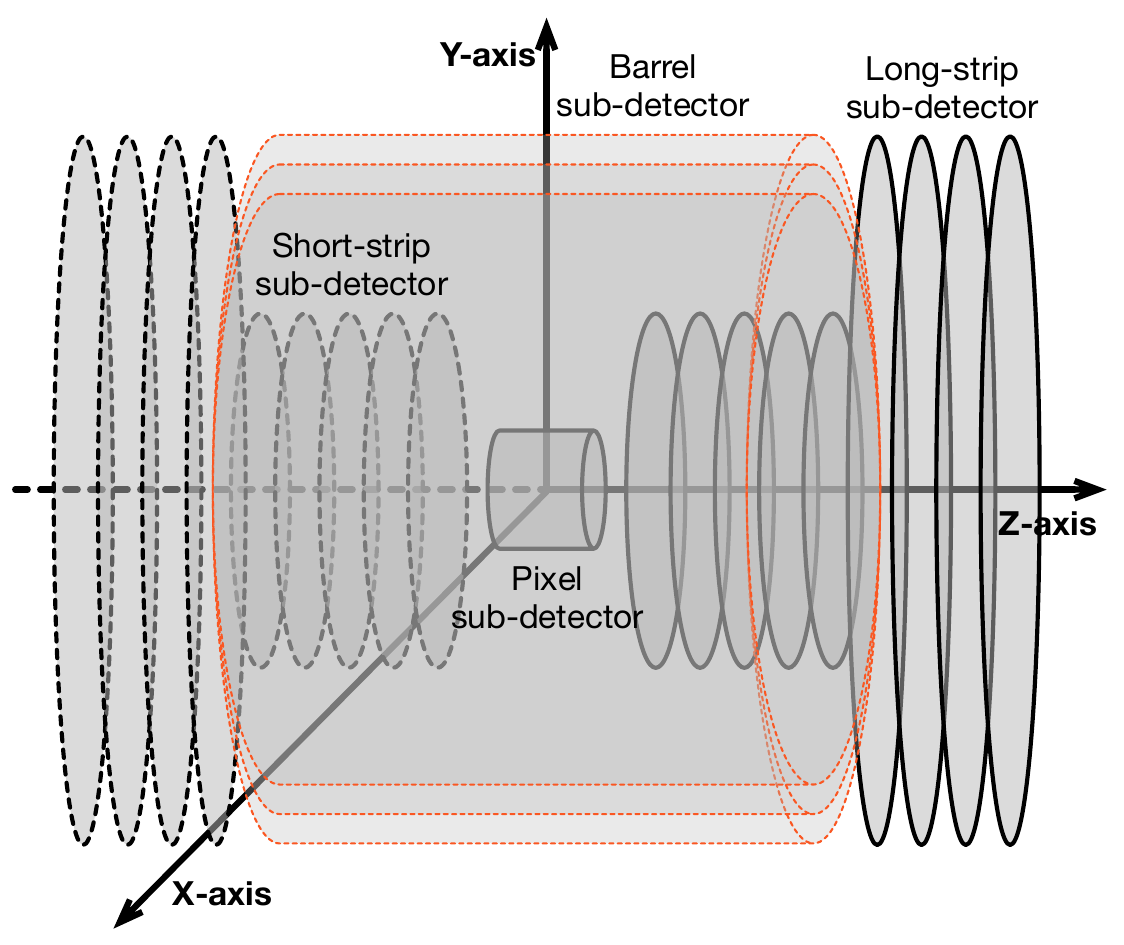}
	\caption{The fully parametric detector geometry, allowing for inclusion/exclusion of different sub-detector types, with full control over sub-layer counts, sizes and placements.}
	\label{fig:detector_geometry}
\end{figure}

At its core, a detector model is comprised of the geometric definitions of the included elements, shapes, sizes, and placements in space. Although we can support a variety of detector geometries, the overall structure, especially for our experimental results, resembles the ATLAS detector. Accordingly, there are four sub-detector types, \emph{Pixel}, \emph{Short-strip}, \emph{Long-strip} and \emph{Barrel}. The pixel and the barrel types have cylindrical shapes with the pixel being a filled cylinder, while the barrel being a cylinder shell with open caps. These are not hard requirements, as the geometry is fully parametric, and differing definitions can be opted for, e.g., a pixel as a cylinder shell. The long-strip and the short-strip types are primarily intended as flat disks, but can be defined as having a thickness, rendering them as cylinders. Sub-detector types can be selectively present or absent. \Cref{fig:detector_geometry} depicts a representative variation of the detector geometry involving the aforementioned elements.

Structurally speaking, in a real-world detector, like ATLAS, the internals of short-strip and long-strip sub-detector types are different. We on the other hand, reduce such complexities to placement location and size, i.e., distance from the origin and sub-detector disk radius. Note that our geometric model does support disk thickness, which basically would turn disks into shallow cylinders. However, we have considered flat disks for our experiments.

\subsubsection{REDVID event simulation}
One of the simplifications for our complexity reduction approach is to consider a single collision per event (no pile-up), either aligning exactly to the origin point of the detector geometry, or at the presence of smearing for this alignment. This can be configured. However, the list of complexities, even without the polluting effects of multiple collisions, is extensive. Particles travelling through the detector matter could lead to secondary collisions, resulting in drastic changes in their trajectory. Such secondary collisions could also lead to the release of particles not originating from the collision event itself. These will show up as tracks with unusual starting points within the detector space, rather distant from the collision point. Some particles could also come to a halt, which would be seen as abruptly terminating tracks. Such physics-accurate behaviour of particles interacting with the present matter in detectors is not considered for our simulator. It must be noted that the generation of tracks originating far away from the origin and prematurely terminating tracks, can be emulated in our simulator in a randomised fashion.

\subsection{TrackML data set}
To move forward in the complexity spectrum, beyond the \enquote{50-100 helical tracks REDVID} data set, we have opted to switch to the data associated with the TrackML Kaggle challenge~\cite{Kiehn:2019:TrackML}. Reduced versions of this data set are frequently used to assess the state-of-the-art models for this task. Released in 2018, the goal of this challenge was to identify machine learning solutions to the track reconstruction problem. 

The TrackML data set is a simulated collection of proton-proton collision events. It simulates specifically the production of top quark-antiquark pairs in the high-pileup conditions expected at the HL-LHC, with each event incorporating an average of 200 pileup interactions. The simulated detector is designed as a generalised representation of an LHC tracker, featuring discrete layers of cylinder and disk-shaped sensor arrays operating within a magnetic field as layers. The data set provides 3D hit positions along with truth-level information about the particles responsible for generating these hits. Some simplifications with respect to more realistic data were made. For example, it does not account for reconstructed hit merging, as such occurrences are rare (less than 0.5\%). Moreover, it excludes complex smaller detector elements such as electronic systems, cooling tubes, or cables.

The TrackML data set consists of 3D hit coordinates in a Cartesian coordinate system, with the global Z-axis defined along the beam direction~\cite{Amrouche:2020:TMLC}. The data set contains ten values associated with every particle: particle identifier $pid$, the vertex of origin ($v_x, x_y, x_z$), the initial momentum (in GeV/c) along each global axis ($p_x, p_y, p_z$), the particle charge $q$ and the number of hits this particle generated $nhits$. We make use of only four of those as track defining parameters: $q, p_x, p_y, p_z$. However, the momenta are transformed into the spherical coordinate system according to 
\begin{align*}
    p &= \sqrt{p_x^2 + p_y^2 + p_z^2}\texttt{,} \\
    \theta &= \mathrm{arccos}\left(\frac{p_z}{p}\right)\texttt{, and} \\
    \phi &= \mathrm{arctan2}\left(\frac{p_y}{p_x}\right)\texttt{.}
\end{align*}
As a preprocessing step, in addition to converting the track parameters to another coordinate system, we normalise the data.

Though resulting from a simulation, the TrackML data has considerable scale. It contains 8\,850 individual events with an average of 10\,780 particles, between 6\,898 and 16\,197, per event, which could be interpreted as the average track count. It includes particles represented with as many as 28 hit points. At the same time, there are particles with 0 associated hit points. The average particle hit count throughout the events is 8.52. The number of individual particle vertices composing a bunch crossing varies from 1300 to 3048. Substantial amount of \emph{noise} hits are present (between 18\,590 and 20\,518 per event), which is one of the main factors making TrackML a challenging data set. Almost every publication considering this data set has performed some sort of noise reduction. For our highest complexity data set, \enquote{200-500 tracks per event from the TrackML data set}, we have selected between 200 to 500 tracks at random per event. This process is repeated five times, resulting in 43\,725 reduced events. The tracks and hits are selected completely at random without any special attention to, e.g., number of noisy hits, i.e., we do not apply any noise reduction.

The TrackML data set is described extensively in literature. For further information please refer to~\cite{Kiehn:2019:TrackML} or~\cite{Amrouche:2020:TMLC}.

\section{ML model designs}
\label{sec:ml_model_designs}
We cover four different designs, three based on the Transformer model architecture and one based on the U-Net model architecture. Note that the ease of evaluation is enabled by the reduction of complexity and simplified simulations.

\subsection{Transformer designs}
The Transformer is a deep learning architecture that allows us to model pair-wise relationships among elements in sequential data by leveraging the attention mechanism~\cite{Vaswani:2017:Attention}. It can be used to process sequences with permutation equivariance and work with variable input lengths, which makes it suitable for the task of trajectory reconstruction. Furthermore, due to the wide success of the Transformer architecture in various machine learning fields, many techniques have been invented and implemented that reduce its complexity to a sub-quadratic level, e.g.,~\cite{Dao:2023:FABP, Child:2019:GLSS, Beltagy:2020:LLDT, Kitaev:2020:RTET, Choromanski:2022:RAWP}.

The model's input sequence consists of \emph{tokens} $\vec{x_1}, \vec{x_2}, ..., \vec{x_n}$ and the initial embedding layer projects each token to a higher dimensional representation. Optionally, positional encoding is applied, to include the position of each token in the input into its representation. Next, the continuous values are fed into the encoder of the Transformer, which comprises a number of identical encoder layers. Each encoder layer consists of the multi-head attention operation, normalisation and a feed forward network, i.e., composed of fully connected layers.

The attention mechanism allows the model to embed the context of the entire sequence into the representation of each token. It operates using key ($K$), value ($V$), and query ($Q$) vectors, which are linear transformations of the input, derived from the input to the encoder block in the first layer and from the output of the previous layer in subsequent layers.

$Q$ represents the elements for which we want to compute attention scores with respect to other elements in the sequence. The Key, $K$, provides information about the elements in the sequence, helping to determine their importance relative to the current element when computing attention scores. In self-attention, $K$ represents the same elements as $Q$, but each is multiplied with different learned weights. The value, $V$, contains learned representations of the content of each element in the sequence.

In the scaled dot-product attention mechanism, the $Q$ and $K$ vectors are multiplied to compute attention scores, capturing dependencies within the sequence. These scores are then used to weight the $V$ vectors, allowing the model to focus on the relevant information from the sequence. The attention function is given by,
\[
\text{Attention}(Q, K, V ) = \text{softmax}(\frac{QK^T}{\sqrt{d_k}})V\texttt{,}
\]
where $d_k$ is the dimensionality of the query and key vectors.

Typically, attention is computed on a set of queries at the same time, making $K$, $V$ and $Q$ matrices. Multi-Head Attention (MHA) allows for jointly attending to information from different representation sub-spaces at different positions in the data. MHA is summarised by
\[
\text{MultiHead}(Q, K, V ) = \text{Concat}(\text{head}_1, ..., \text{head}_h)W\texttt{,}
\]
for which $h$ is the number of heads, $W$ is a parameter matrix, with $W \in \mathbb{R}^{hd_k\times d_{\text{model}}}$ and $d_{\text{model}}$ as the dimensionality of the embedding. The attention heads are used in parallel with dimensionality $d_k = \frac{d_{\text{model}}}{h}$ -- this optimisation reduces the computational cost, making it comparable to that of single-head attention. Overall, the shape of the $K, V, Q$ matrices is then $(B \times h \times N \times \frac{d_{\text{model}}}{h})$, equivalent to $(B \times h \times N \times d_k)$, with batch size $B$, number of attention heads $h$, and sequence length $N$. 

The decoder has a similar architecture to the encoder, with the additional operation of multi-head attention over the output of the encoder stack. Its purpose is to generate sequences, and it is auto-regressive, meaning that it makes use of its previously generated symbols as additional input. The decoder takes a start token, its previous outputs and the encoder output as input, and generates the sequence's next token as output. The multi-head attention layers employ masking to prevent it from conditioning on future tokens. Masking can also be utilised in the encoder, e.g., for ensuring the attention mechanism does not attend to padding values.

Note that the attention mechanism creates a $N\times N$ matrix for each attention head, for each encoder and decoder layer. This leads to a quadratic memory and time complexity of the Transformer and can restrict application for very long sequences, which has motivated research into optimisation techniques that reduce the cost of attention computations. Flash attention is one such method~\cite{Dao:2022:FAFM}: It splits $Q$, $K$ and $V$ into smaller blocks, loads them into fast static RAM, and only then computes the attention matrices with respect to these blocks. Each block's output is scaled by an appropriate factor then added up, which leads to the same correct result as the normal attention mechanism ends up with. This approach boosts performance, with the authors reporting Flash attention as 3x faster and 20x more memory efficient than exact attention. Another technique that can optimise computational complexity involves using the Induced Set Attention Block from Set Transformers~\cite{Lee:2019:SetTransformer}. In that case, there is a set of inducing points, learnable parameters, which the high-dimensional input gets projected onto, which reduces the size of the attention matrix.

\subsection{U-Net design}
The U-Net~\cite{Ronneberger:2015:UNet} is a CNN architecture that primarily segments images. It aims to classify each pixel in an image and allows it to distinguish and separate objects in complicated visual data. The \enquote{U} in U-Net represents its U-shaped structure: it has a contracting path (encoder) and a wide-ranging path (decoder) with a bottleneck in between. On the one hand, contextual information is caught by the contracting path through the progressive downsampling of the input image using convolutional layers and pooling operations that reduce the spatial dimensions and increase feature abstraction. This process allows the network to grasp the global structure of objects in the image while improving its analytical capabilities. The most abstracted features are extracted and the smallest resolution is reached at the bottleneck. On the other hand, the decoder part aims to reconstruct the image at the original resolution by employing transposed convolutions or upsampling layers. It gradually refines the output by combining contextual information from the encoder with finer details provided by the skip connections thereby ensuring that the spatial details lost during downsampling are satisfactorily recovered.

Although U-Net architectures are being widely used for image segmentation tasks it is important to note that traditional convolutional networks usually work on densely populated spaces. For most of the ordinary tasks this does not pose a real problem in terms of memory usage and computational effort. However, in problems where the nature of the data involves relatively sparse information spanning along very large physical spaces, modifications to traditional CNN may be considered. These adaptations can help mitigate physical memory limitations in hardware and accelerate the mathematical operations performed. More recently, a novel convolution operator named as \emph{Sparse Convolution} (SC) has been introduced in order to work with sparse data which may represent a suitable option for processing event detection data in three-dimensional geometries, such as those proposed in this manuscript. Interested readers about how sparse convolutions work and how they are implemented are referred to~\cite{Graham:2017:SSCN} and the references therein.

\subsection{Design choices}
\begin{figure}[htbp]
    \centering
    \begin{subfigure}{0.9\textwidth}
      \centering
      \includegraphics[width=0.7\linewidth]{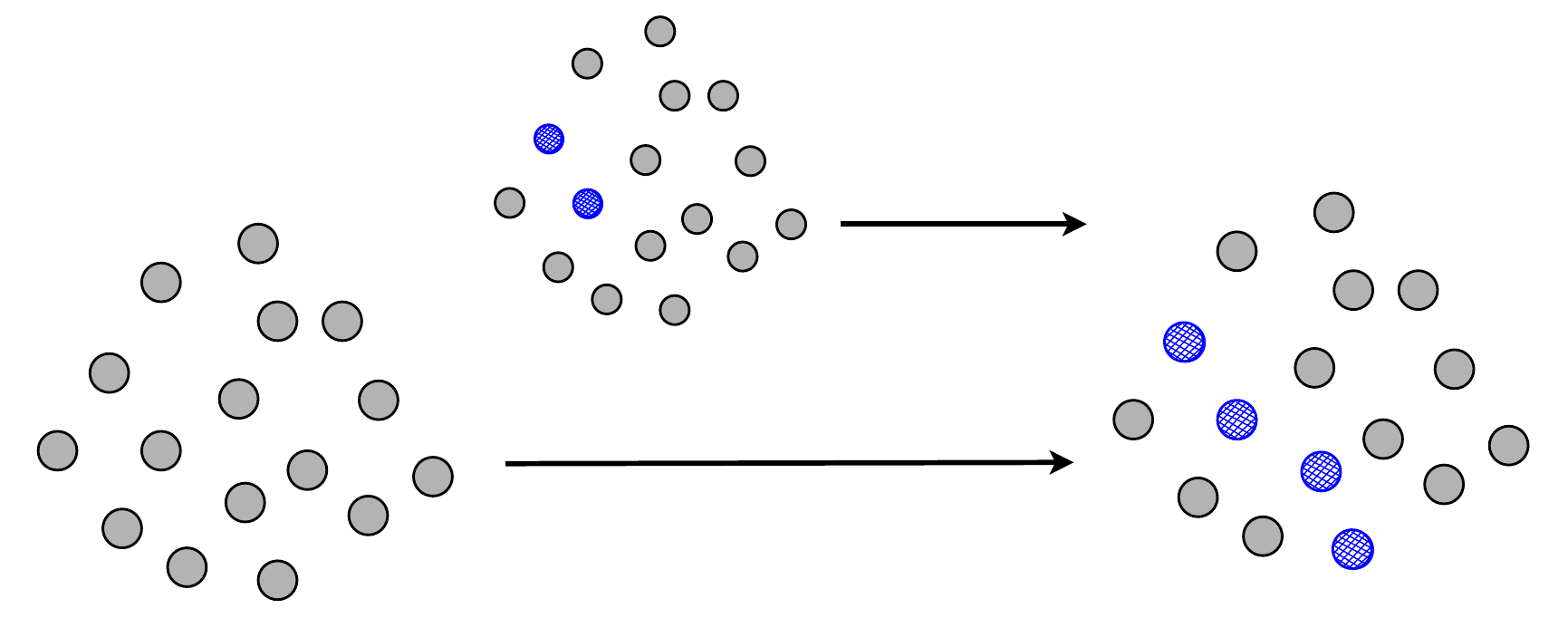}
      \caption{EncDec's input is the set of hit points from a single event, with a couple of them identified as \enquote{track seeds}. The output contains the rest of the hits associated to the track, following the given seed.}
      \label{fig:simple_flow_model_1}
    \end{subfigure}
    \qquad
    \begin{subfigure}{0.9\textwidth}
          \centering
      \includegraphics[width=0.8\linewidth]{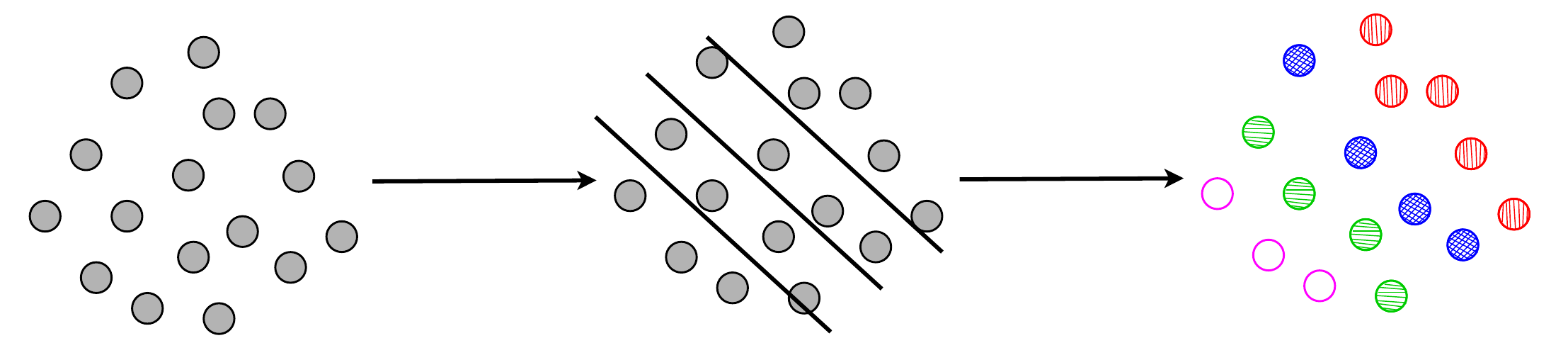}
      \caption{EncCla has learned knowledge of the classes to assign hits to. The input is the set of hit points from a single event, while the output is the collection of class IDs for each hit.}
      \label{fig:simple_flow_model_2}
    \end{subfigure}
    \qquad
    \begin{subfigure}{0.9\textwidth}
          \centering
      \includegraphics[width=0.8\linewidth]{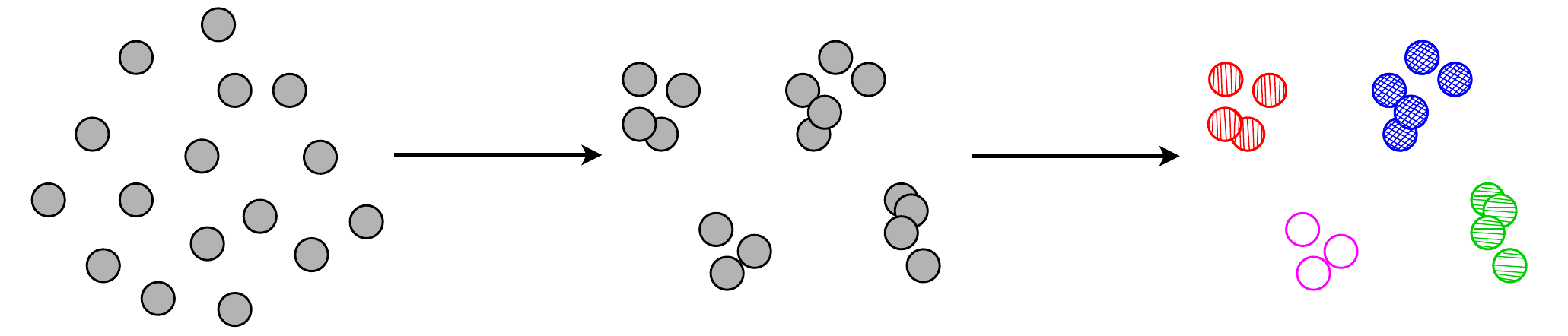}
      \caption{EncReg's input is the set of hit points from a single event, while the output is the regressed track parameters per hit. HDBSCAN collects the clusters of hits based on proximity in the track parameter space.}
      \label{fig:simple_flow_model_3}
    \end{subfigure}
    \caption{Intuitive visualisations of the inner-workings of three Transformer model designs, EncDec, EncCla and EncReg, respectively. We showcase in a simplistic manner the hit processing applied by each pipeline. Hits are represented by dots, with gray dots representing no track associations, and dots with the same colour belonging to the same track.}
    \label{fig:flow_diagrams_simple}
\end{figure}

The input sequence to the Transformers is a full event, with the tokens corresponding to hit coordinates $(x, y, z)$, \emph{which we do not discretise}. As a result of the unordered nature of the recorded hits, we do not use positional encoding for the Transformers. For models involving an \emph{encoder-only} design, i.e., EncCla and EncReg, padding is used to allow variable length inputs. Note that different events could, for instance, have variable track counts, thus resulting in variable numbers of hits. The following sections will elaborate each model design and approach in detail. A simple depiction is presented in \Cref{fig:flow_diagrams_simple}, covering Transformer-based model approaches, i.e., EncDec (\Cref{fig:simple_flow_model_1}), EncCla (\Cref{fig:simple_flow_model_2}) and EncReg (\Cref{fig:simple_flow_model_3}).

\subsubsection{Model 1 - EncDec}
\paragraph*{Overall idea}
The overall idea of this method is similar to the approach used in autoregressive large language models, such as GPT~\cite{Radford:2018:ILUG}, where the model predicts the next word in a sequence based on the preceding words. In our context, instead of predicting the next word, we predict the next hit from an initial set of hits. The model operates in an autoregressive manner, using previous predictions as inputs for subsequent predictions. This approach enables the model to sequentially build a complete track by iteratively predicting each hit based on the hits predicted so far. In contrast to GPT, we use an encoder-decoder design, which is well-suited for handling the sequential dependencies inherent in track reconstruction. This approach enables the model to sequentially build a complete track by iteratively predicting each hit based on the hits predicted so far.

\paragraph*{Details}
This model closely resembles the original Transformer architecture proposed by Vaswani et al.~\cite{Vaswani:2017:Attention}. As such, it has an encoder and a decoder, which both make use of a self attention mechanism. The encoder encodes the full set of hits in a given event, and the decoder autoregressively predicts hits belonging to a particular track within the same event. Of particular interest are the differences with respect to the original Transformer architecture. Firstly, this model uses fixed-query attention~\cite{Lee:2019:SetTransformer} in the first encoder stack in order to ensure full positional invariance of the set of input hits. Furthermore, similarly to the encoder-only models in this paper, this model also omits positional encoding in the encoder, as the positions of hits are explicitly defined by the coordinates of the hits, which are fed to the model directly. The decoder does use positional encoding on the other hand, as for the constructed track the order of hits is relevant. The output format is also unique to this model, as rather than predicting a single token with SoftMaxed probabilities, the decoder outputs a length three vector with $(x, y, z)$ coordinates of the next hit in the track.

\paragraph*{Advantages}
This model was tailored to the task of predicting the location of the next hit in a track given a set of prior input hits belonging to that track. As such, for this specific task, it achieves reasonable performance. It achieves 83\% accuracy in predicting where a next hit is going to be within a 5\% margin of error on the TrackML 10-50 tracks data set. With this in mind, it could potentially aid other tracking models, for example as a post-processing step to find missing hits for already constructed tracks, or to discard obviously wrong hits from already constructed tracks.

\paragraph*{Challenges}
This model differs from the others presented in this work due to the fact that it requires a seed (a short starting sequence of hits) from which to build the track. To do the full reconstruction from hits to tracks, it would thus require a preprocessing step to construct track seeds. Furthermore, whereas the other models in the present research reconstruct tracks in a one-step approach---at once creating all tracks in an event---this model builds tracks one by one. Building tracks one-by-one has the disadvantage that as soon as a single hit is predicted incorrectly, the model starts to diverge from the correct track path, and it is very unlikely to predict any following hits correctly. This leaves much more room for error, which is clear from the relatively worse FitAccuracy scores for this model in \Cref{tab:fit_accuracy_scores}.

\paragraph*{Training notes}
This model was hyperparameter-optimised using the weights-and-biases platform~\cite{WB:2023}. The optimal model on the 10-50 tracks TrackML data set has 9 million trainable parameters. 

\subsubsection{Model 2 - EncCla}
\paragraph*{Overall idea}
Here, we use the encoder component of the Transformer model as a classifier. It is important to note that the hits are not discretised at the beginning; instead, real values on a continuous 3D space are input into the model. Although discretisation may be beneficial in certain cases, our approach utilises the continuous nature of the data to improve model performance and accuracy. This encoder classifier design enables the model to effectively map hit data to corresponding track classifications, facilitating robust track reconstruction.

\paragraph*{Details}
This model takes a sequence of all hit coordinates from a single event as input, and outputs a sequence of class labels for each hit. The class label of a track is defined by discretizing the track parameter space into a fixed number of bins. This is done by binning each track parameter using a quantile-based approach, ensuring that each bin contains roughly equal number of hits. The class labels are created from all unique combinations of track parameter bins, each uniquely indexed.

The model has an input layer that projects hit coordinates into a higher-dimensional embedding space. It is followed by a number of encoder blocks and an output layer. To handle variable input lengths, the model uses padding up to the maximum number of hits in the current batch. Consequently, masking is used to ensure that the attention mechanism ignores the padding values. The output layer classifies each hit into a track candidate class by producing the probability distribution of all class labels for each hit and selecting the label with the highest probability as the predicted class. The hit is assigned to the track class with the highest probability, determined by the ArgMax of the SoftMax output of the classifier. The multi-class classification is trained using a cross-entropy loss function.

\paragraph*{Advantages}
The main advantage of the model is that it is a one-step model, assigning all hits to tracks in the full event in a single inference step. This makes its time complexity during inference lower than models that require a full inference step per hit prediction, such as the EncDec model.

\paragraph*{Challenges}
The model requires the a priori construction of track classes. This can only be done up to a finite granularity and can hinder clustering in high density environments.

\paragraph*{Training notes}
For the TrackML data set with 200-500 tracks, track parameters include $phi$, $theta$, $q$, and $p$. The model used 30 bins for $phi$, 30 bins for $theta$, and 3 bins for $p$. $q$ values are binned into the only two possible values: -1 and 1. All the combination of track parameter bins make up 5\,400 track classes. The largest model has 1.5 million parameters.

\subsubsection{Model 3 - EncReg}
The third model under consideration is another encoder-only Transformer design. It is also a sequence-to-sequence model, the input of which is the hit coordinates of a single event. The output is the corresponding regressed track parameters. The model has an input layer identical to the previous model. The output is then of size (batches, max number of hits, number of track parameters). It is important to note that different data sets require different models, due to the difference in regressed parameters describing the tracks and data complexity due to increasing number of tracks.

To obtain the hit classification, a clustering algorithm is run on the regressed track parameter space. We make use of the HDBSCAN clustering algorithm, which can identify clusters of varying densities~\cite{Campello:2013:DCHD}, and is relatively time efficient (in $O(n^2)$~\cite{Stewart:2022:IHCA}). It has two parameters, which we optimise for each specific data set.

Because of the large memory footprint of the attention mechanism and the constraints of the available GPU memory, the size of the events that we can work with is limited. We also identify that for larger events, training a single model is extremely costly in terms of time and computational resources. Therefore, for the largest data set utilised, we train two models: one with exact attention, and one with Flash attention (EncReg-FA), where measures are taken to improve the memory consumption of the attention computation, and speed up training and hyperparameter tuning. Moreover, for EncReg-FA, we also make use of mixed precision training, which the Flash Attention implementation relies on\footnote{This is a reliance imposed by the utilised framework, PyTorch.}. The model's weights, biases and gradients as well as the data passed to it are of type float-16 wherever possible, except during the operations which can greatly benefit from the data being of type float-32, i.e., values computed by large reductions such as batch normalisation. In addition to this being necessitated by Flash Attention, using lower precision in ML computation is another proven way of dealing with the bottleneck of GPU memory~\cite{Gupta:2015:DLLN}. 

\paragraph*{Advantages}
Similarly to the EncCla model this is a one-step model, making its time complexity during inference lower than models that require a separate inference step per hit. Its speed makes it also suitable to be used as a refiner network, which regresses track parameters per cluster, i.e., per reconstructed track, and identifies falsely associated hits, increasing the purity of the cluster.

\paragraph*{Challenges}
Perhaps the biggest challenge for the EncReg model is the discovery of track parameters that sufficiently define a track and can be learned by the model. What coordinate system they should be in, dealing with angle symmetry, different weighting of the tracks' contribution to the loss, etc., are some examples of things to consider. Another challenge is the evaluation of the model: as accuracy cannot be calculated for the regressed values, its performance is indirectly evaluated based on the formed clusters in the stage following it, or visual inspection of the regressed parameters plotted against the ground truth.

\paragraph*{Training notes}
The hyperparameters of the EncReg model are not fully optimised, but a brief search for suitable values is conducted. This Transformer has a learning rate of 0.001 and uses the Adam optimiser. Most models have 6 encoder layers, except the ones trained on the largest data used, TrackML 200-500 tracks, when there are 7 layers; each encoder layer has dropout of 0.1. The number of attention heads, embedding dimensionality and dimensionality of the fully connected layers are data set-specific. The largest model, EncReg-FA, has about 900\,000 parameters.

\subsubsection{Model 4 - U-Net}
The last model implemented has supposed an alternative methodology to Transformer-like models. Due to the nature of the pixel regression task under consideration, the utilisation of a U-Net-based model appears to be a suitable option for a fruitful methodology. Vanilla U-Net models usually consider a set of the ordinary well-known dense convolutional layers, which are the core block building the network and where the major part of the computational effort occurs. For this use case, the input data should be preprocessed into a multi-dimensional tensor of size $(n_{\mathrm{batches}},1,\mathrm{width},\mathrm{height},\mathrm{depth})$, i.e., a three-dimensional tensor encompassing only one channel which contains discrete values of 0 and 1, indicating whether there is background or hit. Due to the sparse nature of this information, a more suitable U-Net model can be build if the convolutional operations are considered to occur directly on the sparse domain, thereby ignoring the background data governing the overall tensor representing the events. Consequently, since the hit occupancy for a certain event will be very low in general, the convolutional layers have been substituted by their sparse option~\cite{Graham:2017:SSCN, Graham:2018:3DSS}, thereby giving name to the sparse U-Net model that has been considered.

The task performed by this model is a classification process, being the original physical parameter space of the tracks binned according to a quantile-based procedure, using 30 quantiles, i.e., bins, per parameter. The output of the network is then the probabilities of belonging of each pixel to each class, i.e., to each bin. As a consequence, the vanilla Cross-entropy loss has been used as a cost function. Even though mixed precision can also be considered and implemented into this approach, there is not a real necessity since by focusing only on hit data using sparse tensors it is possible to save both memory and computational effort in convolutional operations. By this process, a label for each hit is obtained without the consideration of any post-clustering process.

\paragraph*{Advantages}
The model processes the entire event as a \emph{sparse image}, meaning that it labels all hits (coordinates) at once without the need of attention layers or iterative pipelines. Despite their implementation is not straightforward due to the necessity of using sparse convolutions, attention procedures could be considered as a step forward leading to the construction of a modified U-Net architecture as seen previously in some literature~\cite{AL_Qurri:2023:IUAM}. The usefulness of sparse convolutional operations rely on both conceptual simplicity and computing speed. Studies such as~\cite{Abratenko:2021:SSSC} show state-of-the-art performance addressing particle two-dimensional space segmentation scenarios.

\paragraph*{Challenges}
Perhaps the most challenging aspect of this U-Net-based approach is the considered preprocessing. In the preprocessing step, when converting the original set of $\{x,y,z\}$ numerical values to the set of $\{i,j,k\}$ to create the sparse tensor of coordinates, a scale factor different from one can be considered. In addition, also during this step, it is possible to carry out an interpolation procedure to add more hits between the original ones. The scaling factor determines the physical spacing between the original points in the three-dimensional data. Conversely, the number of points used for interpolation between consecutive points enhances the amount of information provided to the network. Notably, these parameters were considered exclusively during the training process, leaving the test data unaltered. Different interpolation methods can be employed. In this case, the cubic interpolation method from the \emph{SciPy library} has been utilised. As a consequence, both factors will introduce hyperparameters that need exploration, since a bigger scale factor will demand a higher amount of interpolated data, and vice-versa. However, a fixed scale factor of $\times 10$ and also $10$ interpolated hits between each pair of original coordinates have led to the best results that are presented in this manuscript.

\paragraph*{Training notes}
Regarding the hyperparameters of the entire pipeline, one may expect the performance to vary with respect to different aspects. The first one refers to a certain arbitrary factor of scale that can be considered during the process of converting from the $\{x,y,z\}$ set to the $\{i,j,k\}$ one, while the second one would represent the considered amount of interpolated hits for each event. The largest model trained has a total of about 8\,000\,000 parameters. The said hyperparameters influence data processing but not the model itself. Since a balanced approach is desired, the training data was formatted as three-dimensional tensors for U-Net processing, with flexibility in scaling and interpolation. Various scaling factors ($\times 2$, $\times 5$, $\times 10$, $\times 20$) and interpolation levels (5, 10, 20, 50 points) were tested. A $\times 10$ scale with 10 interpolated points per hit provided the best trade-off between performance and training efficiency. However, it is important to note that these hyperparameters are not a priority, as they pertain to data preprocessing rather than the neural network itself.

\subsection{Model workflows}
The model design elaborated above are functional within a workflow, hence the phrase \enquote{ML-assisted solution}. Depending on the model design, there may be simple or elaborate data pre/post-processing steps involved. In most cases, these pre/post-processing steps are the main factor differentiating computational performance for workflows. Detailed computational performance results are provided in \Cref{subsec:computational_performance}.

\begin{figure}[htbp]
    \centering
    \begin{subfigure}{\textwidth}
        \centering
    	\includegraphics[width=0.8\linewidth]{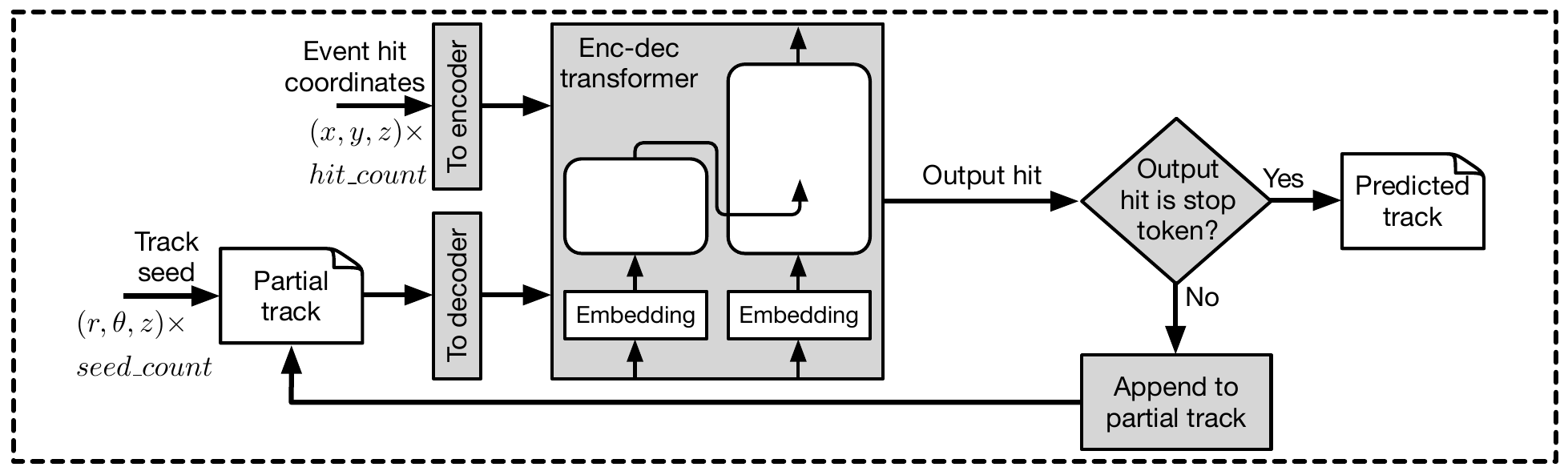}
    	\caption{Model 1 (EncDec) uses a full Transformer architecture to autoregressively append hits to a track candidate, starting with a seed. The iterations stop upon the stop token output.}
    	\label{fig:flow_model_1}
    \end{subfigure}
    \qquad
    \begin{subfigure}{\textwidth}
        \centering
    	\includegraphics[width=0.8\linewidth]{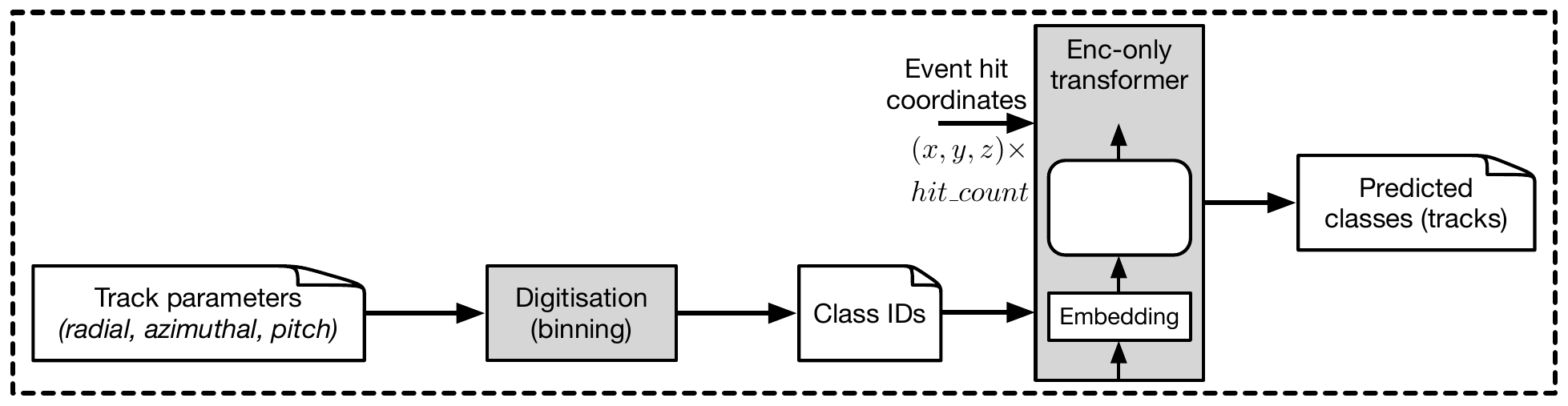}
    	\caption{Model 2 (EncCla) utilises an encoder-only Transformer design alongside a digitisation step to assign class IDs to tracks. The model assigns class IDs, i.e., bins, to each of an event's hits.}
    	\label{fig:flow_model_2}
    \end{subfigure}
    \qquad
    \begin{subfigure}{\textwidth}
        \centering
    	\includegraphics[width=0.8\linewidth]{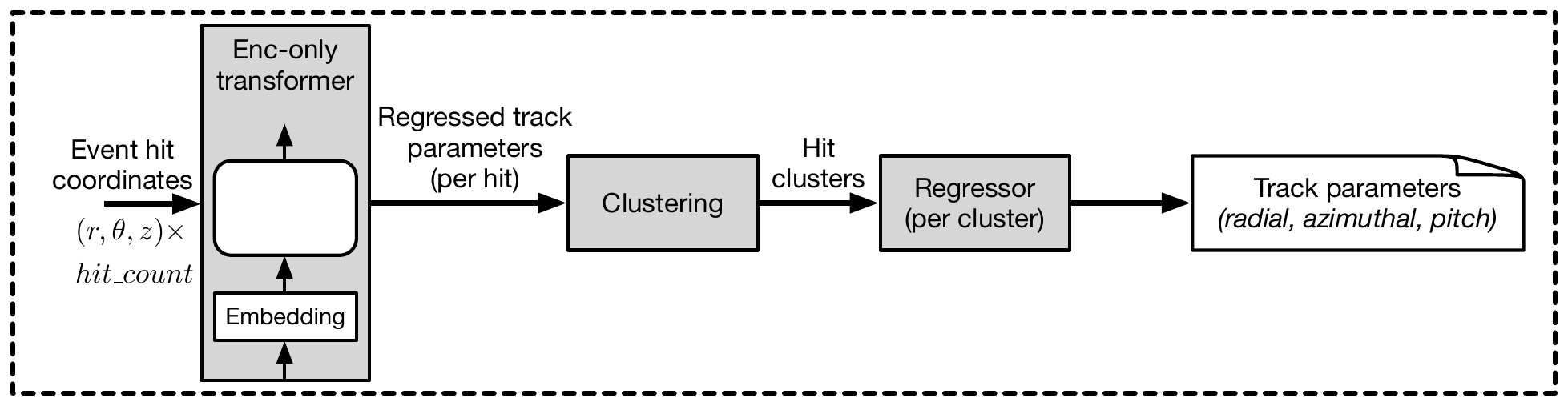}
    	\caption{Model 3 (EncReg) utilises an encoder-only Transformer design, regressing the track-defining parameters of an event's hits, followed by a clustering in the track parameter space. Per-cluster regression of track parameters for every particle is applied.}
    	\label{fig:flow_model_3}
    \end{subfigure}
    \qquad
    \begin{subfigure}{\textwidth}
        \centering
    	\includegraphics[width=0.8\linewidth]{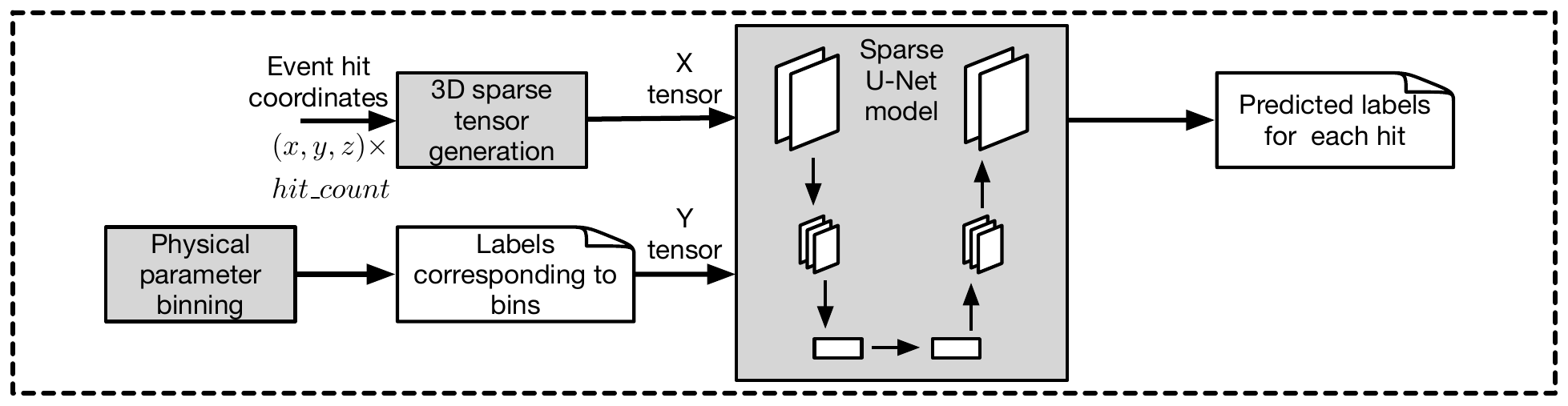}
    	\caption{Model 4 (U-Net) has its input as the set of sparse tensors representing the spatial information (hits), plus the labels corresponding to each one of the considered bins, into which the space of parameters has been partitioned. For each hit, the label of the predicted bin is given as output.}
    	\label{fig:flow_model_4}
    \end{subfigure}
    \caption{Depicting the high-level views of dedicated workflows for each ML model design.}
    \label{fig:flow_diagrams}
\end{figure}

Accordingly, \Cref{fig:flow_diagrams} depicts diagrams, covering steps and tasks within four individual workflows for our model designs, EncDec (\Cref{fig:flow_model_1}), EncCla (\Cref{fig:flow_model_2}), EncReg (\Cref{fig:flow_model_3}) and U-Net (\Cref{fig:flow_model_4}).

\section{Results and performance}
\label{sec:results}
\subsection{Prediction performance}
To measure the \emph{prediction accuracy} of our model designs, we consider a custom metric, named \emph{FitAccuracy score}. It is essentially identical to the \emph{TrackML score}~\cite{Kiehn:2019:TrackML}, with a small modification that concerns the case of REDVID-generated data sets only. The TrackML score is calculated based on the association of weights to hits in the TrackML data. However, REDVID simulations do not generate particles and thus do not consider weights, but tracks. Essentially, we replace the true particle with the true track and consider the weight value of 1 for all hits to arrive at our custom FitAccuracy value. In this fashion, we can have a single comparable scoring for all data sets and model designs. In the definition of the TrackML score and, consequently, FitAccuracy, reconstructed tracks with four or more hits are considered, and at least 50\% of a reconstructed track's hits must originate from the same truth particle for that track to be considered for the scoring. The score of a track is the sum of correctly assigned hit weights. As such, the FitAcurracy score gives an indication of the fraction of correctly reconstructed tracks, and can thus be interpreted as a (weighted) reconstruction efficiency. Available scoring for each model is provided in \Cref{tab:fit_accuracy_scores}.
\begin{table}[htbp]
    \centering
    \caption{FitAccuracy scores for different models are given per data set, focusing on linear, helical and reduced TrackML data. Note that data sets can contain different fixed, or variable (randomised) counts of linear/helical tracks per event. Note that for the EncDec, the fitaccuracy score is slightly differently defined, due to the fact that it starts out with a seed. The seed hits are not counted towards the accuracy.}
    \begin{tabular}{@{}lrrrrr@{}}
        \toprule
        					& \multicolumn{4}{c}{FitAccuracy score} \\
 		\cmidrule(lr){2-5}
        \multicolumn{1}{c}{Data set} & 
        \multicolumn{1}{c}{EncDec} & 
        \multicolumn{1}{c}{EncCla} & 
        \multicolumn{1}{c}{EncReg} & 
        \multicolumn{1}{c}{EncReg-FA} & 
        \multicolumn{1}{c}{U-Net} \\ 
        \midrule
        REDVID - 10-50 linear tracks		& 93\%			& 93\%			  & 97\%			& -			    & 68\% \\ 
        REDVID - 10-50 helical tracks		& 85\%			& 93\%		      & 92\%			& -			    & 62\% \\ 
        REDVID - 50-100 helical tracks		& 85\%			& 88\%		      & 85\%			& -			    & 57\% \\ 
        TrackML - 10-50 tracks			    & 26\%			& 94\%			  & 93\%			& -			    & - \\ 
        TrackML - 200-500 tracks		    & -			    & 78\%			  & 70\%			& 67\%			& - \\ 
        \bottomrule
	\end{tabular}
    \label{tab:fit_accuracy_scores}
\end{table}

\begin{figure}[htbp]
    \centering
    \begin{subfigure}{0.47\textwidth}
	   \includegraphics[width=\linewidth]{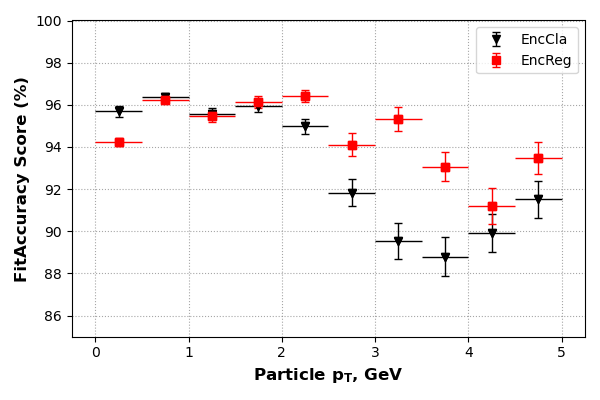}
	   \caption{FitAccuracy versus the transverse momentum $p_T$.}
	   \label{fig:fitaccuracy_pt}
    \end{subfigure}
    \qquad
    \begin{subfigure}{0.47\textwidth}
      \includegraphics[width=\linewidth]{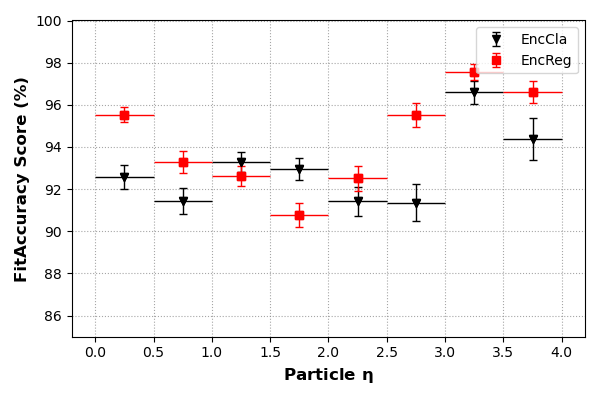}
	   \caption{FitAccuracy versus pseudorapidity $\eta$.}
	   \label{fig:fitaccurary_eta}
    \end{subfigure}
    \qquad
    \begin{subfigure}{0.47\textwidth}
      \includegraphics[width=\linewidth]{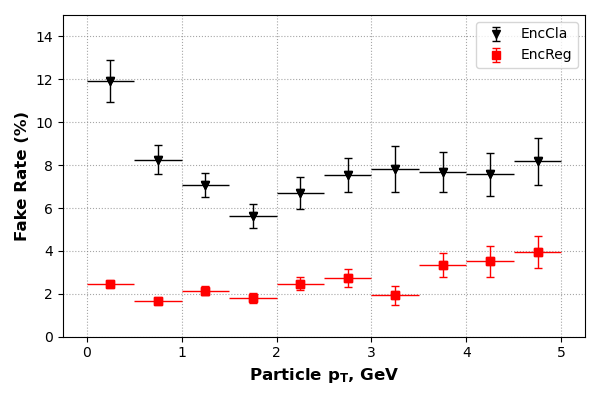}
	   \caption{Fake rate versus transverse momentum $p_T$.}
	   \label{fig:fake_rate_pt}
    \end{subfigure}
    \qquad
    \begin{subfigure}{0.47\textwidth}
      \includegraphics[width=\linewidth]{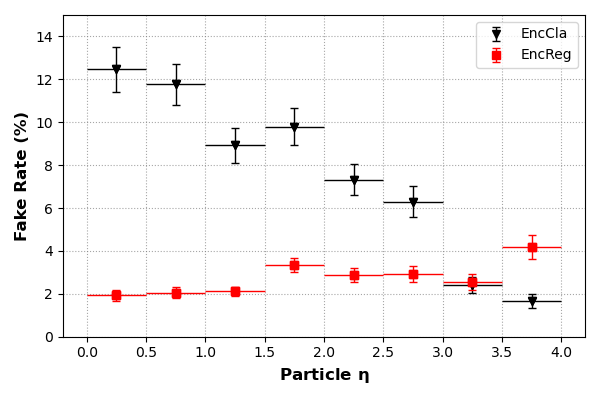}
	   \caption{Fake rate versus pseudorapidity $\eta$.}
	   \label{fig:fake_rate_eta}
    \end{subfigure}
    \caption{FitAccuracy score, as defined in Section 4, and the fake rate, which represents the fraction of predicted tracks that do not correspond to true tracks, as a function of physics observables for the EncCla and EncReg models on the TrackML 10-50 tracks data set, in black and red, respectively. (a) FitAccuracy as a function of the transverse momentum $p_T$, (b) FitAccuracy as a function of the pseudorapidity $\eta$, (c) fake rate as a function of the transverse momentum $p_T$, and (d) fake rate as a function of the pseudorapidity $\eta$. The uncertainties represent the 68\% confidence intervals and were calculated via a bootstrapping procedure. The decreasing trend of EncCla in (d) is due to the increase of bin size with $\eta$, which affects both the number of correctly predicted tracks and the total number of predicted tracks.}
    \label{fig:model_performance_physics_observables}
\end{figure}

The models with the best overall performance, EncReg and EncCla, were further analysed to assess their physics performance as functions of two observables in the TrackML 10-50 tracks data set: pseudorapidity $\eta$\footnote{Pseudorapidity is a widely used transformation of $\theta$, the angle relative to the beam axis, and is defined as $\eta \equiv -\ln \left[ \tan \left( \frac{\theta}{2} \right) \right]$.} and transverse momentum $p_T$. To evaluate model performance, two representative metrics were chosen: the FitAccuracy score, as defined in \Cref{sec:results}, and the fake rate, which represents the fraction of predicted tracks that do not correspond to true tracks. The results are presented in \Cref{fig:model_performance_physics_observables}. Both models exhibit similar trends in their FitAccuracy scores, with EncReg showing marginally better performance at higher $p_T$ values. However, the fake rate analysis reveals a more pronounced difference: the encoder classifier (EncCla) performs significantly worse across most slices of phase space, except at high $\eta$.

\subsection{Computational performance}
\label{subsec:computational_performance}
Beyond FitAccuracy, another important aspect to consider when comparing model designs and relevant workflows, is the computational effort required. The cost of inference is especially interesting, since tracking is bound to be deployed in an online (embedded in the data-taking pipeline), or semi-online fashion, and it is considered to be a time-critical use-case. Considering the workflows depicted in \Cref{fig:flow_diagrams}, we separate the computational effort for each model as the CPU side and the GPU side. Each data set, and model combination will have a different cost, as the processing and in most cases, the model training/inference, is dependent on the scale and complexity of data. Although following the same base design, often the model size has to match the size and the complexity of the considered data set.

\Cref{tab:model_benchmarks} lists the complete set of metric collections for workflow blocks executed on CPU and for training/inference loops executed on GPU, in CPU-time and GPU-time, respectively. The GPU-time measures are given as mean iteration costs of the inference loop. Note that for the EncDec model, only wall-clock times were measured, as these were anyway significantly larger than for the other models. We omit the first iteration in training and inference loops to avoid cold-start effects invoking excess delays. All Transformer models have been trained on an NVIDIA A100 GPU with 40 GB HMB2 and 18 CPU cores, on the Snellius supercomputer\footnote{\url{https://www.surf.nl/en/services/snellius-the-national-supercomputer}}. The training of the sparse U-Net model, using a similar hardware platform, has been done on an A100 GPU with 40 GB of memory from the Artemisa cluster\footnote{\url{https://artemisa.ific.uv.es/web/}}.
\begin{table}[htbp]
    \centering
    \caption{Lists detailed computational effort consisting of CPU-time and GPU-time collections for every data set and model combination. For both measurements, mean instance inference time is provided. For the case of EncDec as an exception, we provide the wall-clock collection, i.e., the execution duration. This model is slower than the alternatives, and this is sufficiently visible with the execution duration.}
    \begin{tabular}{@{}llrrr@{}}
        \toprule
        \multicolumn{1}{c}{Data set} & 
        \multicolumn{1}{c}{Model} & 
        \multicolumn{1}{c}{\begin{tabular}[c]{@{}c@{}}Inference (mean)\\ CPU side\end{tabular}} & 
        \multicolumn{1}{c}{\begin{tabular}[c]{@{}c@{}}Inference (mean)\\ GPU side\end{tabular}} &
        \multicolumn{1}{c}{\begin{tabular}[c]{@{}c@{}}Inference (mean)\\ Wall-clock\end{tabular}} \\ 
        \midrule
        \multirow{4}{*}{\begin{tabular}[l]{@{}l@{}}REDVID\\ 10-50 linear tracks\end{tabular}}		
        										& EncDec		& n/a			& n/a         	& 41 s \\
        										& EncCla		& 0.1 ms     	& 4.0 ms      	& - \\
        										& EncReg		& 8.3 ms		& 2.4 ms      	& - \\
        										& U-Net			& 8.5 ms		& 2.4 ms      	& - \\
        \midrule
        \multirow{4}{*}{\begin{tabular}[l]{@{}l@{}}REDVID\\ 10-50 helical tracks\end{tabular}} 	    
        										& EncDec		& n/a			& n/a         	& 19 s \\
        										& EncCla		& 0.1 ms       	& 4.1 ms      	& - \\
        										& EncReg		& 8.7 ms		& 2.3 ms      	& - \\
        										& U-Net			& 8.6 ms		& 2.4 ms      	& - \\
        \midrule
        \multirow{4}{*}{\begin{tabular}[l]{@{}l@{}}REDVID\\ 50-100 helical tracks\end{tabular}}	    
        										& EncDec		& n/a			& n/a         	& 27 s \\
        										& EncCla		& 0.1 ms       	& 4.3 ms      	& - \\
        										& EncReg		& 18.6 ms		& 4.1 ms      	& - \\
        										& U-Net			& 20.4 ms		& 5.6 ms      	& - \\
        \midrule
        \multirow{4}{*}{\begin{tabular}[l]{@{}l@{}}TrackML\\ 10-50 tracks\end{tabular}}		        
        										& EncDec		& n/a			& n/a         	& 16 s \\
        										& EncCla		& 0.1 ms       	& 4.0 ms      	& - \\
        										& EncReg		& 5.8 ms		& 2.2 ms      	& - \\
        										& U-Net			& n/a			& n/a         	& - \\
        \midrule
        \multirow{4}{*}{\begin{tabular}[l]{@{}l@{}}TrackML\\ 200-500 tracks\end{tabular}} 	        
        										& EncDec		& n/a			& n/a         	& - \\
        										& EncCla		& 0.1 ms       	& 7.0 ms      	& - \\
        										& EncReg		& 70.5 ms	   	& 31.9 ms     	& - \\
        										& EncReg-FA		& 72.2 ms	   	& 3.6 ms      	& - \\
        										& U-Net			& n/a			& n/a         	& - \\
        \bottomrule
	\end{tabular}
    \label{tab:model_benchmarks}
\end{table}

\begin{figure}[htbp]
    \centering
    \includegraphics[width=0.5\textwidth]{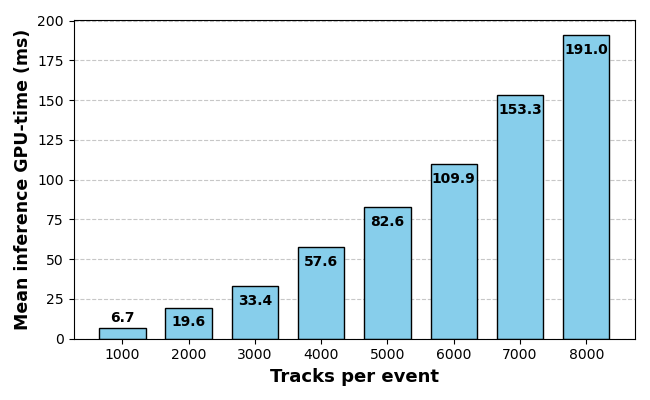}
    \caption{Mean inference GPU-time for the EncReg-FA model trained on the 200–500 tracks per event data set, plotted as a function of the number of tracks per event.}
    \label{fig:computational_scaling}
\end{figure}

\Cref{tab:model_benchmarks} can be used to infer a rough estimate for the scaling of the mean inference GPU-time as a function of the input size. However, a more detailed analysis was conducted for the EncReg-FA model, trained on the 200–500 tracks data set. This model is particularly well-suited for scaling to larger events due to its integration of FlashAttention. The results, presented in \Cref{fig:computational_scaling}, demonstrate that while the inference GPU-time does not scale linearly with input size, its growth remains sub-quadratic. Notably, even for events with 8\,000 tracks, the mean inference time remains below 200~ms, which is promising for applications to high-luminosity LHC data, where similar track multiplicities are anticipated. It is worth noting, however, that for events with higher track counts, larger models might offer improved physics performance, albeit at the cost of increased execution time. Consequently, this study should be viewed as a rough demonstration of scaling under the specific constraint of fixed model size. We do recognise that input scale is only one dimension of complexity and \Cref{fig:computational_scaling} intentionally does not take into account model prediction qualities, i.e., physics accuracy. The main point conveyed is that the indicated increase in \emph{inference} computational burden is superior to traditional algorithms.

\section{Discussion - Applying the method}
\label{sec:discussion}
\begin{figure}[htbp]
    \centering
    \begin{subfigure}{0.47\textwidth}
	   \includegraphics[width=\linewidth]{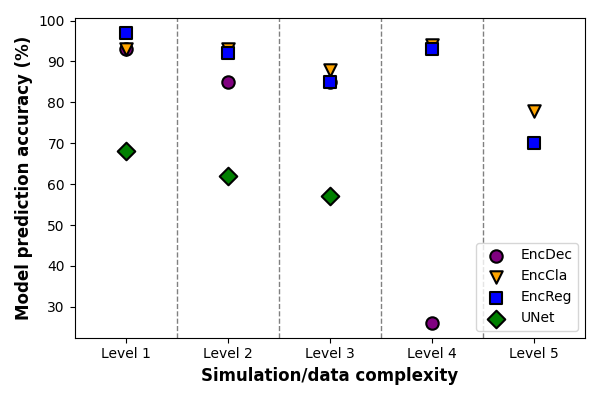}
	   \caption{Prediction accuracy (higher is better)}
	   \label{fig:model_accuracy}
    \end{subfigure}
    \qquad
    \begin{subfigure}{0.47\textwidth}
      \includegraphics[width=\linewidth]{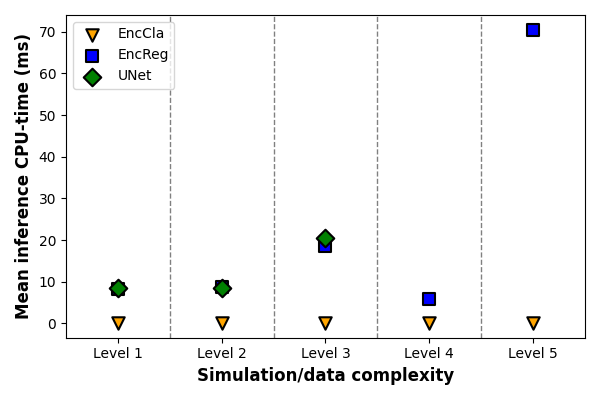}
	   \caption{CPU-time (lower is better)}
	   \label{fig:model_performance_cpu}
    \end{subfigure}
    \qquad
    \begin{subfigure}{0.47\textwidth}
      \includegraphics[width=\linewidth]{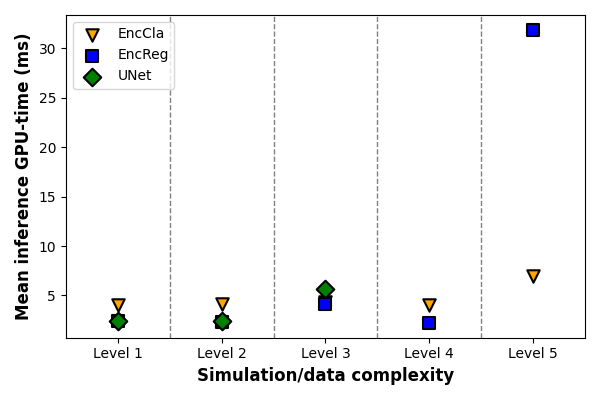}
	   \caption{GPU-time (lower is better)}
	   \label{fig:model_performance_gpu}
    \end{subfigure}
    \qquad
    \begin{subfigure}{0.47\textwidth}
      \includegraphics[width=\linewidth]{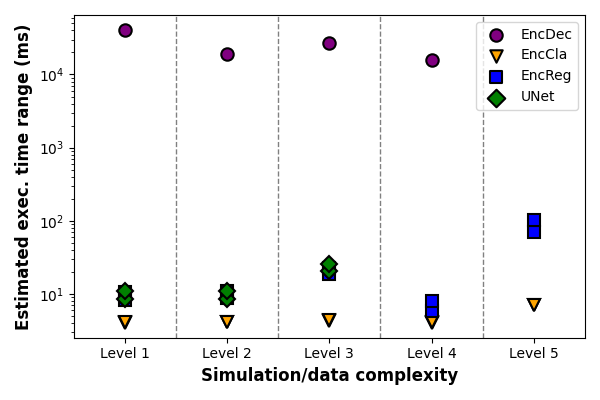}
	   \caption{Execution time (lower is better)}
	   \label{fig:model_performance_execution}
    \end{subfigure}
    \caption{The five levels of simulation/data complexity in these plots, from simple to complex, correspond to the five data sets we have considered, in the same order as provided in \Cref{sec:data_sets}. The combination of prediction accuracy and computational cost figures will be considered for early on model elimination. (a) Prediction accuracy results for four model designs at different complexity levels, (b) The CPU side of the computational effort for different model designs at different complexity levels, given as mean inference CPU-time, (c) The GPU side of the computational effort for different model designs at different complexity levels, given as mean inference GPU-time, (d) Providing an estimated range for overall inference execution time per event, based on collected CPU-time, GPU-time and wall-clock measurements. Note that an inference instance for EncDec will predict the next hit in a track, making a full event inference costly. Mean wall-clock measurement is provided for EncDec instead of a range based on CPU- and GPU-times.}
    \label{fig:model_performance_metrics}
\end{figure}

Next, we examine the performance of our different designs. Based on the two collected metric types, i.e., the model prediction accuracy results shown in \Cref{fig:model_accuracy} and the computational cost results for CPU and GPU, shown in \Cref{fig:model_performance_cpu,fig:model_performance_gpu}, respectively, we can look at how the models perform on different complex tasks. The total execution times provided in \Cref{fig:model_performance_execution} are given as a range for EncCla, EncReg and U-Net models, while EncDec is represented by wall-clock time collection. This is due to implementation differences. We consider the maximum of CPU-time and GPU-time as the lower bound and the sum of the two (CPU-time and GPU-time) as the upper bound for this range, which is an acceptable estimation. Note that under parallel operation of a given computing platform, the CPU and the GPU activity is not necessarily sequential and the lengthiest of the two would be the least execution duration, hence the consideration of a lower bound.

\paragraph*{Encoder-Decoder (EncDec) model}
The first observation is that the Encoder-Decoder model takes the most time due to its autoregressive operation. This model does an inference step per hit, which significantly increases computational time with respect to the other models, which classify all event hits in a single inference step. Additionally, its performance declines with higher complexity tasks, specifically those involving more than 100 tracks or the reduced TrackML data. To mitigate excessive training times, this model was no longer used for the TrackML 200-500 tracks data.

\paragraph*{Encoder-Classifier (EncCla) model}
The EncCla model is the fastest solution, primarily operating on the GPU and producing track classes in a single inference step. The inference time is influenced by the size of the model, the attention matrices and the number of hits per event. Furthermore, the EncCla was found to be the best-performing model, delivering high accuracy without the need for any pre or post-processing steps. This model proved to be effective across all levels of task complexity.

\paragraph*{Encoder-Regressor (EncReg) model}
The EncReg model extends the Encoder-Classifier model by adding the time required to cluster hits into tracks, making it slightly more time-consuming. Additionally, the accuracy of the EncReg model appears to be slightly lower than that of the EncCla model. Despite these differences, the EncReg model is suitable for all simulation complexities.

\paragraph*{Sparse U-Net model}
The U-Net model can also determine the track classes for all hits in one step. The inference time increases slightly with larger input matrices and model sizes. However, it was not possible to achieve good accuracy for simulations with higher complexity levels (above level 3). This method was proposed as an alternative to Transformer-based models with attention mechanisms, but its performance was found to depend on parameters such as input image scale and the number of interpolation points in the training set. The REDVID data sets did not require sparse convolutional networks, but this approach ensured scalability for TrackML data. In addition, implementation restrictions of attention mechanisms within the \emph{spconv}~\cite{Spconv:2022:SSCL} package hindered direct comparisons with Transformers. This limitation restricts the U-Net model's applicability to less complex scenarios.

By evaluating these results, we can better understand the trade-offs between computational cost and performance accuracy for each model, enabling us to choose the most suitable architecture for different particle tracking scenarios.

While negotiating the full scale and complexity of a realistic detector data set requires further research, our methodology provides an effective approach. To recapitulate, instead of picking one ML model design and iteratively improving it, we have opted for multiple designs. The layered approach to complexity allows for speedy evaluation and elimination of bad designs, making the design journey rather cost-effective. The way forward from this point on is to move forward in the complexity spectrum, represented by data sets of higher complexity and higher scale. In this particular case, models EncCla and EncReg, or improved variations of these shall be considered. Further fine-tuning along the way, in the form of post-processing models or computational steps could be considered. In short, we have the opportunity to try many design candidates, precisely because we can do it in a cost-effective way.

\section{Conclusion and future work}
\label{sec:conclusion}
\paragraph*{Model architectures}
Our main concluding remark relates to the effectiveness of the Transformer architecture for the task of particle tracking. Our experiments show potential for single-step processing of event hit data, especially with the encoder-classifier architecture. This architecture is inspired by a single-step translation of hits into track candidates. The low computational effort for the CPU side and the GPU side of event inference and good accuracy, shown in \Cref{fig:model_performance_metrics}, are strong indications for this. The total inference time for the assignment of track classes to approximately, 10\,000 hits per event (level 5) was estimated at 3-6 milliseconds (with or without the use of flash attention). It is not possible to directly compare our inference time with other approaches, as the data sets and computer architectures are not identical. However, we would like to point out again that the Kalman filter pipeline in~\cite{Atlas:2019:FTRH} required 12 seconds of CPU-time, and an optimised Kalman filter algorithm required 1.8 seconds of CPU-time per event. A benchmark for GNNs is~\cite{Ju:2021:PGDL} and reports 2.2 seconds wall-clock time (including data transfer to GPUs, pre and post-processing steps) for the full TrackML events using an NVIDIA A100 GPU. This emphasises that the one-step Transformer-encoder-classifier architecture is worth a closer look and needs to be further investigated to improve its performance through post-processing and preprocessing steps.

\paragraph*{Increasing complexity} Another conclusion is the effectiveness of our method where we consider a complexity spectrum for the problem ranging from a simple to a complex reconstruction problem. Both the prediction accuracy and the computational effort results from \Cref{fig:model_performance_metrics} provide metrics for the early elimination of underperforming machine learning designs. It goes without saying that the design and validation effort (both human and machine) for lower complexity levels is incomparably lower than when tackling the problem in its full complexity. It should be noted also that the most complex data set considered in this work is still an order of magnitude removed from the true HL-LHC data in hit multiplicity. The scaling behaviour with increasingly complex data sets, as shown in \Cref{fig:model_performance_metrics} and \Cref{tab:model_benchmarks}, looks promising in particular for the EncCla model and to a lesser extent, for the EncReg model. However, a full evaluation of the included models up to the complexity of HL-LHC data is considered follow-up work, as significant adaptations would be required including the integration of pre- and post-processing steps into the algorithm, which may substantially optimise performance. Furthermore, computational developments, such as parallelised training and inference of the Transformer models may also enable more efficient scaling.

\paragraph*{Future directions - Post-processing models}
Our single-step models should be supplemented in future research by pre and post-processing steps. Pre-processing involves the formation of clusters from hits, i.e., one could envisage a combination of the encoder-regressor model to form clusters, which are then fed into the classifier model. There are many avenues for ML-based post-processing algorithms. Among these, one could again consider a combined application of the presented methods. For example, as the EncDec model is particularly well suited to the prediction of individual hits, it could be used as a post-processing step, where it could determine for each track candidate whether it is missing hits and add those hits to that track. Other additions could be models to detect incorrect hit assignments (per track candidate), to predict track parameters (with track candidate hit set as input) and to utilise physics-informed loss functions for an improved parameterisation. Note that our current solutions are limited to formation of hit clusters matching tracks and do not cover the actual parameterisation of a track function. It would scale with the number of track candidates (this would still be better than models that scale with the number of hit candidates) and would certainly increase accuracy.

\paragraph*{Future directions - Faster models}
Other directions of future research could be oriented towards reducing the computational complexity of the Transformers through the use of recent developments in this field, for example using top-K attention, as suggested by Gupta et al.~\cite{Gupta:2021:METT} and studying the large number of efficient Transformer architectures, e.g.,~\cite{Dao:2023:FABP, Child:2019:GLSS, Beltagy:2020:LLDT, Kitaev:2020:RTET, Choromanski:2022:RAWP}. Regarding sparse U-Nets, while optimised for speed, future work could focus on further reducing computation time. This may involve studying the trade-off between performance and model size to find an optimal balance.

\begin{figure}[htbp]
    \centering
    \includegraphics[width=0.6\textwidth]{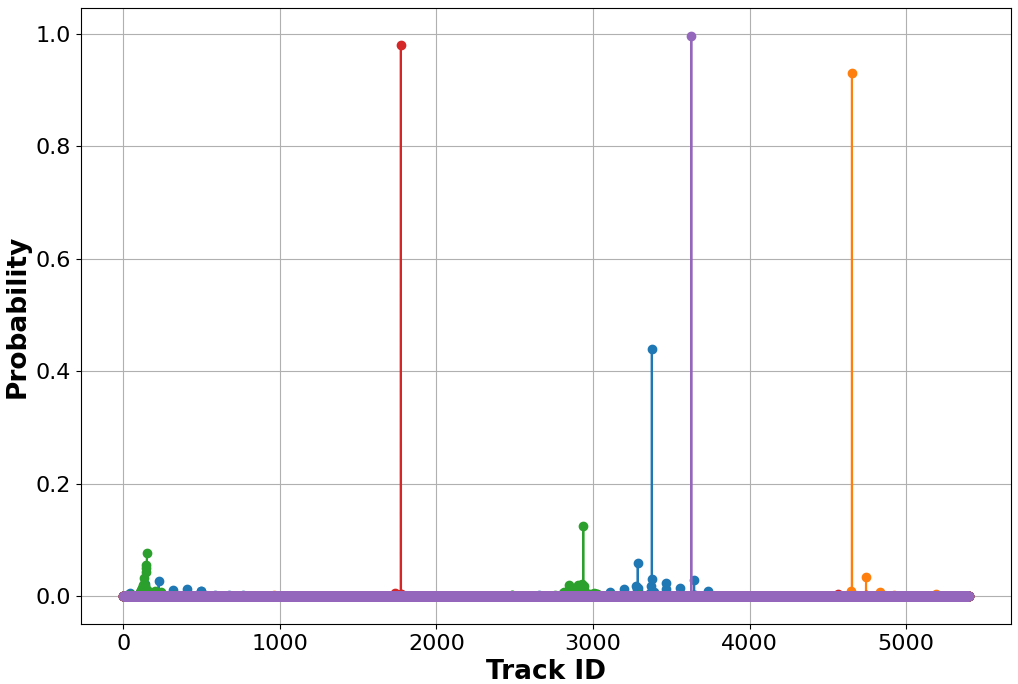}
    \caption{SoftMax outputs of the EncCla model for the first five hits for one event for all track classes, showing some hits with high confidence predictions for a track ID and others with low confidence predictions.}
    \label{fig:hit_softmax_plot}
\end{figure}

\paragraph*{Future directions - The use of the full posterior}
The EncCla provides also a unique feature that could be made use of in future developments, which is the fact that it does not just output for every hit exactly what track it belongs to, but it actually predicts a vector of probabilities per hits. As such, for each hit, it associates a probability to this hit belonging to all of the tracks. An example plot for five hits of an arbitrary event is given in \Cref{fig:hit_softmax_plot}. Such output will allow for powerful post-prediction analysis as an extra validatory step, resulting in corrections to the predicted hit associations if need be. In the context of language models, this full vector of probabilities has found extensive use, and similar strategies could be applied in this context.

\begin{acks}
This paper is a result of the collaboration between Radboud University, Nikhef, the University of Twente, SURF, the University of Valencia and the University of Amsterdam. The project has been supported by grants and computing resources from NWO (grant no.\ 62004546 and grant no.\ OCENW.GROOT.2019.032), the Dutch national e-infrastructure with the support of the SURF (grant no.\ EINF-7404), Valencian Community (grant no.\ CIAICO/2021/184), Valencian Foundation (ValgrAI), Quantum Spain project. The work of R. RdA was supported by PID2020-113644GB-I00 from the Spanish Ministerio de Ciencia e Innovación and by the PROMETEO/2022/69 from the Spanish GVA. The author(s) gratefully acknowledges the computer resources at Artemisa, funded by the European Union ERDF and Comunitat Valenciana as well as the technical support provided by the Instituto de Fisica Corpuscular, IFIC (CSIC-UV). The authors would furthermore like to thank Viacheslav Pshenov for his valuable contributions to this work, especially for model performance evaluations.
\end{acks}


\balance

\printbibliography

@article{Kiehn:2019:TrackML,
	author = {{Kiehn, Moritz} and {Amrouche, Sabrina} and {Calafiura, Paolo} and {Estrade, Victor} and {Farrell, Steven} and {Germain, Cécile} and {Gligorov, Vava} and {Golling, Tobias} and {Gray, Heather} and {Guyon, Isabelle} and {Hushchyn, Mikhail} and {Innocente, Vincenzo} and {Moyse, Edward} and {Rousseau, David} and {Salzburger, Andreas} and {Ustyuzhanin, Andrey} and {Vlimant, Jean-Roch} and {Yilnaz, Yetkin}},
	title = {The TrackML high-energy physics tracking challenge on Kaggle},
	journal = {EPJ Web Conf.},
	year = {2019},
	OPTvolume = {214},
	OPTpages = {06037},
    doi = {10.1051/epjconf/201921406037},
	OPTurl = {https://doi.org/10.1051/epjconf/201921406037}
}

@inproceedings{Vaswani:2017:Attention,
    author = {Vaswani, Ashish and Shazeer, Noam and Parmar, Niki and Uszkoreit, Jakob and Jones, Llion and Gomez, Aidan N. and Kaiser, \L{}ukasz and Polosukhin, Illia},
    title = {Attention is All You Need},
    year = {2017},
    OPTisbn = {9781510860964},
    OPTpublisher = {Curran Associates Inc.},
    OPTaddress = {Red Hook, NY, USA},
    booktitle = {Proceedings of the 31st International Conference on Neural Information Processing Systems},
    OPTpages = {6000–6010},
    OPTnumpages = {11},
    OPTlocation = {Long Beach, California, USA},
    OPTseries = {NIPS'17}
}

@inproceedings{Ronneberger:2015:UNet,
    author = {Ronneberger, Olaf and Fischer, Philipp and Brox, Thomas},
    OPTeditor = {Navab, Nassir and Hornegger, Joachim and Wells, William M. and Frangi, Alejandro F.},
    title = {U-Net: Convolutional Networks for Biomedical Image Segmentation},
    booktitle = {Medical Image Computing and Computer-Assisted Intervention -- MICCAI 2015},
    year = {2015},
    OPTpublisher = {Springer International Publishing},
    OPTaddress = {Cham},
    OPTpages = {234--241},
    OPTisbn = {978-3-319-24574-4},
    doi = {10.1007/978-3-319-24574-4_28}
}

@online{WB:2023,
    title = {Weights \& Biases: The AI Developer Platform},
    year = {2023},
    url={https://wandb.ai/}
}

@misc{Gupta:2021:METT,
    title = {Memory-efficient Transformers via Top-$k$ Attention}, 
    author = {Ankit Gupta and Guy Dar and Shaya Goodman and David Ciprut and Jonathan Berant},
    year = {2021},
    OPTeprint = {2106.06899},
    OPTarchivePrefix = {arXiv},
    OPTprimaryClass = {cs.CL},
    doi = {10.48550/arXiv.2106.06899}
}

@inproceedings{Lee:2019:SetTransformer,
    title = {Set Transformer: A Framework for Attention-based Permutation-Invariant Neural Networks},
    author = {Lee, Juho and Lee, Yoonho and Kim, Jungtaek and Kosiorek, Adam and Choi, Seungjin and Teh, Yee Whye},
    booktitle = {Proceedings of the 36th International Conference on Machine Learning},
    OPTpages = {3744--3753},
    year = {2019},
    OPTeditor = {Chaudhuri, Kamalika and Salakhutdinov, Ruslan},
    OPTvolume = {97},
    OPTseries = {Proceedings of Machine Learning Research},
    OPTmonth = {09--15 Jun},
    OPTpublisher = {PMLR},
    OPTurl = {https://proceedings.mlr.press/v97/lee19d.html}
}

@inproceedings{Gupta:2015:DLLN,
    title = {Deep Learning with Limited Numerical Precision},
    author = {Gupta, Suyog and Agrawal, Ankur and Gopalakrishnan, Kailash and Narayanan, Pritish},
    booktitle = {Proceedings of the 32nd International Conference on Machine Learning},
    OPTpages = {1737--1746},
    year = {2015},
    OPTeditor = {Bach, Francis and Blei, David},
    OPTvolume = {37},
    OPTseries = {Proceedings of Machine Learning Research},
    OPTaddress = {Lille, France},
    OPTmonth = {07--09 Jul},
    OPTpublisher = {PMLR},
    OPTurl = {https://proceedings.mlr.press/v37/gupta15.html}
}

@inproceedings{Campello:2013:DCHD,
	author = {Campello, Ricardo J. G. B. and Moulavi, Davoud and Sander, Joerg},
	OPTeditor = {Pei, Jian and Tseng, Vincent S. and Cao, Longbing and Motoda, Hiroshi and Xu, Guandong},
	title = {Density-Based Clustering Based on Hierarchical Density Estimates},
	booktitle = {Advances in Knowledge Discovery and Data Mining},
	year = {2013},
	OPTpublisher = {Springer Berlin Heidelberg},
	OPTaddress = {Berlin, Heidelberg},
	OPTpages = {160--172},
	OPTisbn = {978-3-642-37456-2},
	doi = {10.1007/978-3-642-37456-2_14}
}

@article{Stewart:2022:IHCA,
    author = {Stewart, Geoffrey and Al-Khassaweneh, Mahmood},
    title = {An Implementation of the HDBSCAN* Clustering Algorithm},
    journal = {Applied Sciences},
    OPTvolume = {12},
    year = {2022},
    OPTnumber = {5},
    OPTarticle-number = {2405},
    OPTurl = {https://www.mdpi.com/2076-3417/12/5/2405},
    OPTissn = {2076-3417},
    doi = {10.3390/app12052405}
}

@inproceedings{Dao:2022:FAFM,
    author = {Dao, Tri and Fu, Dan and Ermon, Stefano and Rudra, Atri and R\'{e}, Christopher},
    title = {FlashAttention: Fast and Memory-Efficient Exact Attention with IO-Awareness},
    booktitle = {Advances in Neural Information Processing Systems},
    OPTeditor = {S. Koyejo and S. Mohamed and A. Agarwal and D. Belgrave and K. Cho and A. Oh},
    OPTpages = {16344--16359},
    OPTpublisher = {Curran Associates, Inc.},
    OPTurl = {https://proceedings.neurips.cc/paper_files/paper/2022/file/67d57c32e20fd0a7a302cb81d36e40d5-Paper-Conference.pdf},
    OPTvolume = {35},
    year = {2022}
}

@article{Woodruff:2018:APTI,
    author = {Katherine Woodruff and on behalf of the MicroBooNE collaboration},
    title = {Automated proton track identification in MicroBooNE using gradient boosted decision trees},
    journal = {Journal of Physics: Conference Series},
    year = {2018},
    OPTmonth = {sep},
    OPTpublisher = {IOP Publishing},
    OPTvolume = {1085},
    OPTnumber = {4},
    OPTpages = {042019},
    OPTurl = {https://dx.doi.org/10.1088/1742-6596/1085/4/042019},
    doi = {10.1088/1742-6596/1085/4/042019}
}

@inproceedings{Odyurt:2024:RSHE,
	author = {Odyurt, Uraz and Swatman, Stephen Nicholas and Varbanescu, Ana-Lucia and Caron, Sascha},
	OPTeditor = {Franco, Leonardo and de Mulatier, Cl{\'e}lia and Paszynski, Maciej and Krzhizhanovskaya, Valeria V. and Dongarra, Jack J. and Sloot, Peter M. A.},
	title = {Reduced Simulations for High-Energy Physics, a Middle Ground for Data-Driven Physics Research},
	booktitle = {Computational Science -- ICCS 2024},
	year = {2024},
	OPTpublisher = {Springer Nature Switzerland},
	OPTaddress = {Cham},
	OPTpages = {84--99},
	OPTisbn = {978-3-031-63751-3},
	doi = {10.1007/978-3-031-63751-3_6}
}

@article{Collaboration:2008:ALICE,
	author = {The ALICE Collaboration and  K Aamodt and  A Abrahantes Quintana and  R Achenbach and  S Acounis and  D Adamová and  C Adler and  M Aggarwal and  F Agnese and  G Aglieri Rinella and  Z Ahammed and  A Ahmad and  N Ahmad and  S Ahmad and  A Akindinov and  P Akishin and  D Aleksandrov and  B Alessandro and  R Alfaro and  G Alfarone and  A Alici and  J Alme and  T Alt and  S Altinpinar and  W Amend and  C Andrei and  Y Andres and  A Andronic and  G Anelli and  M Anfreville and  V Angelov and  A Anzo and  C Anson and  T Anticić and  V Antonenko and  D Antonczyk and  F Antinori and  S Antinori and  P Antonioli and  L Aphecetche and  H Appelshäuser and  V Aprodu and  M Arba and  S Arcelli and  A Argentieri and  N Armesto and  R Arnaldi and  A Arefiev and  I Arsene and  A Asryan and  A Augustinus and  T C Awes and  J Äysto and  M Danish Azmi and  S Bablock and  A Badalà and  S K Badyal and  J Baechler and  S Bagnasco and  R Bailhache and  R Bala and  A Baldisseri and  A Baldit and  J Bán and  R Barbera and  P-L Barberis and  J M Barbet and  G Barnäfoldi and  V Barret and  J Bartke and  D Bartos and  M Basile and  V Basmanov and  N Bastid and  G Batigne and  B Batyunya and  J Baudot and  C Baumann and  I Bearden and  B Becker and  J Belikov and  R Bellwied and  E Belmont-Moreno and  A Belogianni and  S Belyaev and  A Benato and  J L Beney and  L Benhabib and  F Benotto and  S Beolé and  I Berceanu and  A Bercuci and  E Berdermann and  Y Berdnikov and  C Bernard and  R Berny and  J D Berst and  H Bertelsen and  L Betev and  A Bhasin and  P Baskar and  A Bhati and  N Bianchi and  J Bielčik and  J Bielčiková and  L Bimbot and  G Blanchard and  F Blanco and  F Blanco and  D Blau and  C Blume and  S Blyth and  M Boccioli and  A Bogdanov and  H Bøggild and  M Bogolyubsky and  L Boldizsár and  M Bombara and  C Bombonati and  M Bondila and  D Bonnet and  V Bonvicini and  H Borel and  F Borotto and  V Borshchov and  Y Bortoli and  O Borysov and  S Bose and  L Bosisio and  M Botje and  S Böttger and  G Bourdaud and  O Bourrion and  S Bouvier and  A Braem and  M Braun and  P Braun-Munzinger and  L Bravina and  M Bregant and  G Bruckner and  R Brun and  E Bruna and  O Brunasso and  G E Bruno and  D Bucher and  V Budilov and  D Budnikov and  H Buesching and  P Buncic and  M Burns and  S Burachas and  O Busch and  J Bushop and  X Cai and  H Caines and  F Calaon and  M Caldogno and  I Cali and  P Camerini and  R Campagnolo and  M Campbell and  X Cao and  G P Capitani and  G Cara Romeo and  M Cardenas-Montes and  H Carduner and  F Carena and  W Carena and  P Cariola and  F Carminati and  J Casado and  A Casanova Diaz and  M Caselle and  J Castillo Castellanos and  J Castor and  V Catanescu and  E Cattaruzza and  D Cavazza and  P Cerello and  S Ceresa and  V Černý and  V Chambert and  S Chapeland and  A Charpy and  D Charrier and  M Chartoire and  J L Charvet and  S Chattopadhyay and  S Chattopadhyay and  V Chepurnov and  S Chernenko and  M Cherney and  C Cheshkov and  B Cheynis and  P Chochula and  E Chiavassa and  V Chibante Barroso and  J Choi and  P Christakoglou and  P Christiansen and  C Christensen and  O A Chykalov and  C Cicalo and  L Cifarelli-Strolin and  M Ciobanu and  F Cindolo and  C Cirstoiu and  O Clausse and  J Cleymans and  O Cobanoglu and  J-P Coffin and  S Coli and  A Colla and  C Colledani and  C Combaret and  M Combet and  M Comets and  G Conesa Balbastre and  Z Conesa del Valle and  G Contin and  J Contreras and  T Cormier and  F Corsi and  P Cortese and  F Costa and  E Crescio and  P Crochet and  E Cuautle and  J Cussonneau and  M Dahlinger and  A Dainese and  H H Dalsgaard and  L Daniel and  I Das and  T Das and  A Dash and  R Da Silva and  M Davenport and  H Daues and  A De Caro and  G de Cataldo and  J De Cuveland and  A De Falco and  M de Gaspari and  P de Girolamo and  J de Groot and  D De Gruttola and  A De Haas and  N De Marco and  S De Pasquale and  P De Remigis and  D de Vaux and  G Decock and  H Delagrange and  M Del Franco and  G Dellacasa and  C Dell'Olio and  D Dell'Olio and  A Deloff and  V Demanov and  E Dénes and  G D'Erasmo and  D Derkach and  A Devaux and  D Di Bari and  A Di Bartolomeo and  C Di Giglio and  S Di Liberto and  A Di Mauro and  P Di Nezza and  M Dialinas and  L Diaz and  R Díaz Valdes and  T Dietel and  R Dima and  H Ding and  C Dinca and  R Divià and  V Dobretsov and  A Dobrin and  B Doenigus and  T Dobrowolski and  I Domínguez and  M Dorn and  S Drouet and  A E Dubey and  L Ducroux and  F Dumitrache and  E Dumonteil and  P Dupieux and  V Duta and  A Dutta Majumdar and  M Dutta Majumdar and  Th Dyhre and  L Efimov and  A Efremov and  D Elia and  D Emschermann and  C Engster and  A Enokizono and  B Espagnon and  M Estienne and  A Evangelista and  D Evans and  S Evrard and  C W Fabjan and  D Fabris and  J Faivre and  D Falchieri and  A Fantoni and  R Farano and  R Fearick and  O Fedorov and  V Fekete and  D Felea and  G Feofilov and  A Férnandez Téllez and  A Ferretti and  F Fichera and  S Filchagin and  E Filoni and  C Finck and  R Fini and  E M Fiore and  D Flierl and  M Floris and  Z Fodor and  Y Foka and  S Fokin and  P Force and  F Formenti and  E Fragiacomo and  M Fragkiadakis and  D Fraissard and  A Franco and  M Franco and  U Frankenfeld and  U Fratino and  S Fresneau and  A Frolov and  U Fuchs and  J Fujita and  C Furget and  M Furini and  M Fusco Girard and  J-J Gaardhøje and  A Gabrielli and  S Gadrat and  M Gagliardi and  A Gago and  L Gaido and  A Gallas Torreira and  M Gallio and  E Gandolfi and  P Ganoti and  M Ganti and  J Garabatos and  A Garcia Lopez and  L Garizzo and  L Gaudichet and  R Gemme and  M Germain and  A Gheata and  M Gheata and  B Ghidini and  P Ghosh and  G Giolu and  G Giraudo and  P Giubellino and  R Glasow and  P Glässel and  E G Ferreiro and  C Gonzalez Gutierrez and  L H Gonzales-Trueba and  S Gorbunov and  Y Gorbunov and  H Gos and  J Gosset and  S Gotovac and  H Gottschlag and  D Gottschalk and  V Grabski and  T Grassi and  H Gray and  O Grebenyuk and  K Grebieszkow and  C Gregory and  C Grigoras and  N Grion and  V Grigoriev and  A Grigoryan and  C Grigoryan and  S Grigoryan and  Y Grishuk and  P Gros and  J Grosse-Oetringhaus and  J-Y Grossiord and  R Grosso and  B Grynyov and  C Guarnaccia and  F Guber and  F Guerin and  R Guernane and  M Guerzoni and  A Guichard and  M Guida and  G Guilloux and  H Gulkanyan and  K Gulbrandsen and  T Gunji and  A Gupta and  V Gupta and  H-A Gustafsson and  H Gutbrod and  C Hadjidakis and  M Haiduc and  G Hamar and  H Hamagaki and  J Hamblen and  J C Hansen and  P Hardy and  D Hatzifotiadou and  J W Harris and  M Hartig and  A Harutyunyan and  A Hayrapetyan and  D Hasch and  D Hasegan and  J Hehner and  N Heine and  M Heinz and  H Helstrup and  A Herghelegiu and  S Herlant and  G Herrera Corral and  N Herrmann and  K Hetland and  P Hille and  H Hinke and  B Hippolyte and  M Hoch and  H Hoebbel and  H Hoedlmoser and  T Horaguchi and  M Horner and  P Hristov and  I Hřivnáčová and  S Hu and  C Hu Guo and  T Humanic and  A Hurtado and  D S Hwang and  J C Ianigro and  M Idzik and  S Igolkin and  R Ilkaev and  I Ilkiv and  M Imhoff and  P G Innocenti and  E Ionescu and  M Ippolitov and  M Irfan and  C Insa and  M Inuzuka and  C Ivan and  A Ivanov and  M Ivanov and  V Ivanov and  P Jacobs and  A Jacholkowski and  L Jančurová and  R Janik and  M Jasper and  C Jena and  L Jirden and  D P Johnson and  G T Jones and  C Jorgensen and  F Jouve and  P Jovanović and  A Junique and  A Jusko and  H Jung and  W Jung and  K Kadija and  A Kamal and  R Kamermans and  S Kapusta and  A Kaidalov and  V Kakoyan and  S Kalcher and  E Kang and  J Kapitan and  V Kaplin and  K Karadzhev and  O Karavichev and  T Karavicheva and  E Karpechev and  K Karpio and  A Kazantsev and  U Kebschull and  R Keidel and  M Mohsin Khan and  A Khanzadeev and  Y Kharlov and  D Kikola and  B Kileng and  D Kim and  D S Kim and  D W Kim and  H N Kim and  J S Kim and  S Kim and  J B Kinson and  S K Kiprich and  I Kisel and  S Kiselev and  A Kisiel and  T Kiss and  V Kiworra and  J Klay and  C Klein Bösing and  M Kliemant and  A Klimov and  A Klovning and  A Kluge and  R Kluit and  S Kniege and  R Kolevatov and  T Kollegger and  A Kolojvari and  V Kondratiev and  E Kornas and  E Koshurnikov and  I Kotov and  R Kour and  M Kowalski and  S Kox and  K Kozlov and  I Králik and  F Kramer and  I Kraus and  A Kravčáková and  T Krawutschke and  M Krivda and  E Kryshen and  Y Kucheriaev and  A Kugler and  C Kuhn and  P Kuijer and  L Kumar and  N Kumar and  P Kumpumaeki and  A Kurepin and  A N Kurepin and  S Kushpil and  V Kushpil and  M Kutovsky and  H Kvaerno and  M Kweon and  J-C Labbé and  F Lackner and  P Ladron de Guevara and  V Lafage and  P La Rocca and  M Lamont and  C Lara and  D T Larsen and  G Laurenti and  C Lazzeroni and  Y Le Bornec and  N Le Bris and  C Le Gailliard and  V Lebedev and  J Lecoq and  K S Lee and  S C Lee and  F Lefévre and  I Legrand and  T Lehmann and  L Leistam and  P Lenoir and  V Lenti and  H Leon and  I Leon Monzon and  P Lévai and  Q Li and  X Li and  F Librizzi and  R Lietava and  N Lindegaard and  V Lindenstruth and  C Lippmann and  M Lisa and  O M Listratenko and  F Littel and  Y Liu and  J Lo and  V Lobanov and  V Loginov and  M López Noriega and  R López-Ramírez and  E López Torres and  P M Lorenzo and  G Løvhøiden and  S Lu and  W Ludolphs and  M Lunardon and  L Luquin and  S Lusso and  J-R Lutz and  M Luvisetto and  V Lyapin and  A Maevskaya and  C Magureanu and  A Mahajan and  S Majahan and  T Mahmoud and  A Mairani and  D Mahapatra and  A Makarov and  I Makhlyueva and  M Malek and  T Malkiewicz and  D Mal'Kevich and  P Malzacher and  A Mamonov and  C Manea and  L K Mangotra and  D Maniero and  V Manko and  F Manso and  V Manzari and  Y Mao and  A Marcel and  S Marchini and  J Mareš and  G V Margagliotti and  A Margotti and  A Marin and  J-C Marin and  D Marras and  P Martinengo and  M I Martínez and  A Martinez-Davalos and  G Martínez Garcia and  S Martini and  A Marzari Chiesa and  C Marzocca and  S Masciocchi and  M Masera and  M Masetti and  N I Maslov and  A Masoni and  F Massera and  M Mast and  A Mastroserio and  Z L Matthews and  B Mayer and  G Mazza and  M D Mazzaro and  A Mazzoni and  F Meddi and  E Meleshko and  A Menchaca-Rocha and  S Meneghini and  M Meoni and  J Mercado Perez and  P Mereu and  O Meunier and  Y Miake and  A Michalon and  R Michinelli and  N Miftakhov and  M Mignone and  K Mikhailov and  J Milosevic and  Y Minaev and  F Minafra and  A Mischke and  D Miśkowiec and  V Mitsyn and  C Mitu and  B Mohanty and  D Moisa and  L Molnar and  M Mondal and  N Mondal and  L Montaño Zetina and  M Monteno and  M Morando and  M Morel and  S Moretto and  Th Morhardt and  A Morsch and  T Moukhanova and  M Mucchi and  V Muccifora and  E Mudnic and  H Müller and  W Müller and  J Munoz and  D Mura and  L Musa and  J F Muraz and  A Musso and  R Nania and  B Nandi and  E Nappi and  F Navach and  S Navin and  T Nayak and  S Nazarenko and  G Nazarov and  L Nellen and  F Nendaz and  A Nianine and  M Nicassio and  B S Nielsen and  S Nikolaev and  V Nikolic and  S Nikulin and  V Nikulin and  B Nilsen and  M Nitti and  F Noferini and  P Nomokonov and  G Nooren and  F Noto and  D Nouais and  A Nyiri and  J Nystrand and  G Odyniec and  H Oeschler and  M Oinonen and  M Oldenburg and  I Oleks and  E K Olsen and  V Onuchin and  C Oppedisano and  F Orsini and  A Ortiz-Velázquez and  C Oskamp and  A Oskarsson and  F Osmic and  L Österman and  I Otterlund and  G Ovrebekk and  K Oyama and  M Pachr and  P Pagano and  G Paić and  C Pajares and  S Pal and  S Pal and  G Pálla and  A Palmeri and  G Pancaldi and  R Panse and  A Pantaleo and  G S Pappalardo and  B Pastirčák and  C Pastore and  O Patarakin and  V Paticchio and  G Patimo and  A Pavlinov and  T Pawlak and  T Peitzmann and  Y Pénichot and  A Pepato and  H Pereira and  D Peresunko and  C Perez and  J Perez Griffo and  D Perini and  D Perrino and  W Peryt and  A Pesci and  V Peskov and  Y Pestov and  A J Peters and  V Petráček and  A Petridis and  M Petris and  V Petrov and  V Petrov and  M Petrovici and  J Peyré and  S Piano and  A Piccotti and  P Pichot and  C Piemonte and  M Pikna and  R Pilastrini and  P Pillot and  O Pinazza and  B Pini and  L Pinsky and  V Pinto Morais and  V Pismennaya and  F Piuz and  R Platt and  M Ploskon and  S Plumeri and  J Pluta and  T Pocheptsov and  P Podesta and  F Poggio and  M Poghosyan and  T Poghosyan and  K Polák and  B Polichtchouk and  P Polozov and  V Polyakov and  B Pommeresch and  F Pompei and  A Pop and  S Popescu and  F Posa and  V Pospíšil and  B Potukuchi and  J Pouthas and  S Prasad and  R Preghenella and  F Prino and  L Prodan and  G Prono and  M A Protsenko and  C A Pruneau and  A Przybyla and  I Pshenichnov and  G Puddu and  P Pujahari and  A Pulvirenti and  A Punin and  V Punin and  J Putschke and  J Quartieri and  E Quercigh and  I Rachevskaya and  A Rachevski and  A Rademakers and  S Radomski and  A Radu and  J Rak and  L Ramello and  R Raniwala and  S Raniwala and  O B Rasmussen and  J Rasson and  V Razin and  K Read and  J Real and  K Redlich and  C Reichling and  C Renard and  G Renault and  R Renfordt and  A R Reolon and  A Reshetin and  J-P Revol and  K Reygers and  H Ricaud and  L Riccati and  R A Ricci and  M Richter and  P Riedler and  L M Rigalleau and  F Riggi and  W Riegler and  E Rindel and  J Riso and  A Rivetti and  M Rizzi and  V Rizzi and  M Rodriguez Cahuantzi and  K Røed and  D Röhrich and  S Román-López and  M Romanato and  R Romita and  F Ronchetti and  P Rosinsky and  P Rosnet and  S Rossegger and  A Rossi and  V Rostchin and  F Rotondo and  F Roukoutakis and  S Rousseau and  C Roy and  D Roy and  P Roy and  L Royer and  G Rubin and  A Rubio and  R Rui and  I Rusanov and  G Russo and  V Ruuskanen and  E Ryabinkin and  A Rybicki and  S Sadovsky and  K Šafařík and  R Sahoo and  J Saini and  P Saiz and  S Salur and  S Sambyal and  V Samsonov and  L Šándor and  A Sandoval and  H Sann and  J-C Santiard and  R Santo and  R Santoro and  G Sargsyan and  P Saturnini and  E Scapparone and  F Scarlassara and  B Schackert and  C Schiaua and  R Schicker and  T Schioler and  J D Schippers and  C Schmidt and  H Schmidt and  R Schneider and  K Schossmaier and  J Schukraft and  Y Schutz and  K Schwarz and  K Schweda and  E Schyns and  G Scioli and  E Scomparin and  H Snow and  S Sedykh and  G Segato and  S Sellitto and  F Semeria and  S Senyukov and  H Seppänen and  S Serci and  L Serkin and  S Serra and  T Sesselmann and  A Sevcenco and  I Sgura and  G Shabratova and  R Shahoyan and  E Sharkov and  S Sharma and  K Shigaki and  K Shileev and  P Shukla and  A Shurygin and  M Shurygina and  Y Sibiriak and  E Siddi and  T Siemiarczuk and  M H Sigward and  A Silenzi and  D Silvermyr and  R Silvestri and  E Simili and  V Simion and  R Simon and  L Simonetti and  R Singaraju and  V Singhal and  B Sinha and  T Sinha and  M Siska and  B Sitár and  M Sitta and  B Skaali and  P Skowronski and  M Slodkowski and  N Smirnov and  L Smykov and  R Snellings and  W Snoeys and  C Soegaard and  J Soerensen and  O Sokolov and  A Soldatov and  A Soloviev and  H Soltveit and  R Soltz and  W Sommer and  C Soos and  F Soramel and  S Sorensen and  D Soyk and  M Spyropoulou-Stassinaki and  J Stachel and  F Staley and  I Stan and  A Stavinskiy and  J Steckert and  G Stefanini and  G Stefanek and  T Steinbeck and  H Stelzer and  E Stenlund and  D Stocco and  M Stockmeier and  G Stoicea and  P Stolpovsky and  P Strmeň and  J S Stutzmann and  G Su and  T Sugitate and  M Šumbera and  C Suire and  T Susa and  K Sushil Kumar and  D Swoboda and  J Symons and  I Szarka and  A Szostak and  M Szuba and  P Szymanski and  M Tadel and  C Tagridis and  L Tan and  D Tapia Takaki and  H Taureg and  A Tauro and  M Tavlet and  G Tejeda Munoz and  J Thäder and  R Tieulent and  P Timmer and  T Tolyhy and  N Topilskaya and  C Torcato de Matos and  H Torii and  L Toscano and  F Tosello and  A Tournaire and  T Traczyk and  G Tröger and  W Tromeur and  D Truesdale and  W Trzaska and  G Tsiledakis and  E Tsilis and  A Tsvetkov and  M Turcato and  R Turrisi and  M Tuveri and  T Tveter and  H Tydesjo and  L Tykarski and  K Tywoniuk and  E Ugolini and  K Ullaland and  J Urbán and  G M Urciuoli and  G L Usai and  M Usseglio and  A Vacchi and  M Vala and  F Valiev and  P Vande Vyvre and  A Van Den Brink and  N Van Eijndhoven and  N Van Der Kolk and  M van Leeuwen and  L Vannucci and  S Vanzetto and  J-P Vanuxem and  M A Vargas and  R Varma and  A Vascotto and  A Vasiliev and  M Vassiliou and  P Vasta and  V Vechernin and  M Venaruzzo and  E Vercellin and  S Vergara and  W Verhoeven and  F Veronese and  I Vetlitskiy and  R Vernet and  V Victorov and  L Vidak and  G Viesti and  O Vikhlyantsev and  Z Vilakazi and  O Villalobos Baillie and  A Vinogradov and  L Vinogradov and  Y Vinogradov and  T Virgili and  Y Viyogi and  A Vodopianov and  G Volpe and  D Vranic and  J Vrláková and  B Vulpescu and  C Wabnitz and  V Wagner and  L Wallet and  R Wan and  Y Wang and  Y Wang and  R Wheadon and  R Weis and  Q Wen and  J Wessels and  J Westergaard and  J Wiechula and  A Wiesenaecker and  J Wikne and  A Wilk and  G Wilk and  C Williams and  N Willis and  B Windelband and  R Witt and  H Woehri and  K Wyllie and  C Xu and  C Yang and  H Yang and  F Yermia and  Z Yin and  Z Yin and  B Yun Ky and  I Yushmanov and  B Yuting and  E Zabrodin and  S Zagato and  B Zagreev and  P Zaharia and  A Zalite and  G Zampa and  C Zampolli and  Y Zanevskiy and  A Zarochentsev and  O Zaudtke and  P Závada and  H Zbroszczyk and  A Zepeda and  V Zeter and  I Zgura and  M Zhalov and  D Zhou and  S Zhou and  G Zhu and  A Zichichi and  A Zinchenko and  G Zinovjev and  Y Zoccarato and  A Zubarev and  A Zucchini and  M Zuffa},
	title = {{The ALICE experiment at the CERN LHC}},
	journal = {Journal of Instrumentation},
	year = {2008},
	OPTmonth = Aug,
	OPTpublisher = {},
	OPTvolume = {3},
	OPTnumber = {08},
	OPTpages = {S08002},
	doi = {10.1088/1748-0221/3/08/S08002},
	OPTurl = {https://dx.doi.org/10.1088/1748-0221/3/08/S08002}
}

@article{Collaboration:2008:CMS,
	author = {The CMS Collaboration and  S Chatrchyan and  G Hmayakyan and  V Khachatryan and  A M Sirunyan and  W Adam and  T Bauer and  T Bergauer and  H Bergauer and  M Dragicevic and  J Erö and  M Friedl and  R Frühwirth and  V M Ghete and  P Glaser and  C Hartl and  N Hoermann and  J Hrubec and  S Hänsel and  M Jeitler and  K Kastner and  M Krammer and  I Magrans de Abril and  M Markytan and  I Mikulec and  B Neuherz and  T Nöbauer and  M Oberegger and  M Padrta and  M Pernicka and  P Porth and  H Rohringer and  S Schmid and  T Schreiner and  R Stark and  H Steininger and  J Strauss and  A Taurok and  D Uhl and  W Waltenberger and  G Walzel and  E Widl and  C-E Wulz and  V Petrov and  V Prosolovich and  V Chekhovsky and  O Dvornikov and  I Emeliantchik and  A Litomin and  V Makarenko and  I Marfin and  V Mossolov and  N Shumeiko and  A Solin and  R Stefanovitch and  J Suarez Gonzalez and  A Tikhonov and  A Fedorov and  M Korzhik and  O Missevitch and  R Zuyeuski and  W Beaumont and  M Cardaci and  E De Langhe and  E A De Wolf and  E Delmeire and  S Ochesanu and  M Tasevsky and  P Van Mechelen and  J D'Hondt and  S De Weirdt and  O Devroede and  R Goorens and  S Hannaert and  J Heyninck and  J Maes and  M U Mozer and  S Tavernier and  W Van Doninck and  L Van Lancker and  P Van Mulders and  I Villella and  C Wastiels and  C Yu and  O Bouhali and  O Charaf and  B Clerbaux and  P De Harenne and  G De Lentdecker and  J P Dewulf and  S Elgammal and  R Gindroz and  G H Hammad and  T Mahmoud and  L Neukermans and  M Pins and  R Pins and  S Rugovac and  J Stefanescu and  V Sundararajan and  C Vander Velde and  P Vanlaer and  J Wickens and  M Tytgat and  S Assouak and  J L Bonnet and  G Bruno and  J Caudron and  B De Callatay and  J De Favereau De Jeneret and  S De Visscher and  P Demin and  D Favart and  C Felix and  B Florins and  E Forton and  A Giammanco and  G Grégoire and  M Jonckman and  D Kcira and  T Keutgen and  V Lemaitre and  D Michotte and  O Militaru and  S Ovyn and  T Pierzchala and  K Piotrzkowski and  V Roberfroid and  X Rouby and  N Schul and  O Van der Aa and  N Beliy and  E Daubie and  P Herquet and  G Alves and  M E Pol and  M H G Souza and  M Vaz and  D De Jesus Damiao and  V Oguri and  A Santoro and  A Sznajder and  E De Moraes Gregores and  R L Iope and  S F Novaes and  T Tomei and  T Anguelov and  G Antchev and  I Atanasov and  J Damgov and  N Darmenov and  L Dimitrov and  V Genchev and  P Iaydjiev and  A Marinov and  S Piperov and  S Stoykova and  G Sultanov and  R Trayanov and  I Vankov and  C Cheshkov and  A Dimitrov and  M Dyulendarova and  I Glushkov and  V Kozhuharov and  L Litov and  M Makariev and  E Marinova and  S Markov and  M Mateev and  I Nasteva and  B Pavlov and  P Petev and  P Petkov and  V Spassov and  Z Toteva and  V Velev and  V Verguilov and  J G Bian and  G M Chen and  H S Chen and  M Chen and  C H Jiang and  B Liu and  X Y Shen and  H S Sun and  J Tao and  J Wang and  M Yang and  Z Zhang and  W R Zhao and  H L Zhuang and  Y Ban and  J Cai and  Y C Ge and  S Liu and  H T Liu and  L Liu and  S J Qian and  Q Wang and  Z H Xue and  Z C Yang and  Y L Ye and  J Ying and  P J Li and  J Liao and  Z L Xue and  D S Yan and  H Yuan and  C A Carrillo Montoya and  J C Sanabria and  N Godinovic and  I Puljak and  I Soric and  Z Antunovic and  M Dzelalija and  K Marasovic and  V Brigljevic and  K Kadija and  S Morovic and  R Fereos and  C Nicolaou and  A Papadakis and  F Ptochos and  P A Razis and  D Tsiakkouri and  Z Zinonos and  A Hektor and  M Kadastik and  K Kannike and  E Lippmaa and  M Müntel and  M Raidal and  L Rebane and  P A Aarnio and  E Anttila and  K Banzuzi and  P Bulteau and  S Czellar and  N Eiden and  C Eklund and  P Engstrom and  A Heikkinen and  A Honkanen and  J Härkönen and  V Karimäki and  H M Katajisto and  R Kinnunen and  J Klem and  J Kortesmaa and  M Kotamäki and  A Kuronen and  T Lampén and  K Lassila-Perini and  V Lefébure and  S Lehti and  T Lindén and  P R Luukka and  S Michal and  F Moura Brigido and  T Mäenpää and  T Nyman and  J Nystén and  E Pietarinen and  K Skog and  K Tammi and  E Tuominen and  J Tuominiemi and  D Ungaro and  T P Vanhala and  L Wendland and  C Williams and  M Iskanius and  A Korpela and  G Polese and  T Tuuva and  G Bassompierre and  A Bazan and  P Y David and  J Ditta and  G Drobychev and  N Fouque and  J P Guillaud and  V Hermel and  A Karneyeu and  T Le Flour and  S Lieunard and  M Maire and  P Mendiburu and  P Nedelec and  J P Peigneux and  M Schneegans and  D Sillou and  J P Vialle and  M Anfreville and  J P Bard and  P Besson and  E Bougamont and  M Boyer and  P Bredy and  R Chipaux and  M Dejardin and  D Denegri and  J Descamps and  B Fabbro and  J L Faure and  S Ganjour and  F X Gentit and  A Givernaud and  P Gras and  G Hamel de Monchenault and  P Jarry and  C Jeanney and  F Kircher and  M C Lemaire and  Y Lemoigne and  B Levesy and  E Locci and  J P Lottin and  I Mandjavidze and  M Mur and  J P Pansart and  A Payn and  J Rander and  J M Reymond and  J Rolquin and  F Rondeaux and  A Rosowsky and  J Y A Rousse and  Z H Sun and  J Tartas and  A Van Lysebetten and  P Venault and  P Verrecchia and  M Anduze and  J Badier and  S Baffioni and  M Bercher and  C Bernet and  U Berthon and  J Bourotte and  A Busata and  P Busson and  M Cerutti and  D Chamont and  C Charlot and  C Collard and  A Debraine and  D Decotigny and  L Dobrzynski and  O Ferreira and  Y Geerebaert and  J Gilly and  C Gregory and  L Guevara Riveros and  M Haguenauer and  A Karar and  B Koblitz and  D Lecouturier and  A Mathieu and  G Milleret and  P Miné and  P Paganini and  P Poilleux and  N Pukhaeva and  N Regnault and  T Romanteau and  I Semeniouk and  Y Sirois and  C Thiebaux and  J C Vanel and  A Zabi and  J L Agram and  A Albert and  L Anckenmann and  J Andrea and  F Anstotz and  A M Bergdolt and  J D Berst and  R Blaes and  D Bloch and  J M Brom and  J Cailleret and  F Charles and  E Christophel and  G Claus and  J Coffin and  C Colledani and  J Croix and  E Dangelser and  N Dick and  F Didierjean and  F Drouhin and  W Dulinski and  J P Ernenwein and  R Fang and  J C Fontaine and  G Gaudiot and  W Geist and  D Gelé and  T Goeltzenlichter and  U Goerlach and  P Graehling and  L Gross and  C Guo Hu and  J M Helleboid and  T Henkes and  M Hoffer and  C Hoffmann and  J Hosselet and  L Houchu and  Y Hu and  D Huss and  C Illinger and  F Jeanneau and  P Juillot and  T Kachelhoffer and  M R Kapp and  H Kettunen and  L Lakehal Ayat and  A C Le Bihan and  A Lounis and  C Maazouzi and  V Mack and  P Majewski and  D Mangeol and  J Michel and  S Moreau and  C Olivetto and  A Pallarès and  Y Patois and  P Pralavorio and  C Racca and  Y Riahi and  I Ripp-Baudot and  P Schmitt and  J P Schunck and  G Schuster and  B Schwaller and  M H Sigward and  J L Sohler and  J Speck and  R Strub and  T Todorov and  R Turchetta and  P Van Hove and  D Vintache and  A Zghiche and  M Ageron and  J E Augustin and  C Baty and  G Baulieu and  M Bedjidian and  J Blaha and  A Bonnevaux and  G Boudoul and  P Brunet and  E Chabanat and  E C Chabert and  R Chierici and  V Chorowicz and  C Combaret and  D Contardo and  R Della Negra and  P Depasse and  O Drapier and  M Dupanloup and  T Dupasquier and  H El Mamouni and  N Estre and  J Fay and  S Gascon and  N Giraud and  C Girerd and  G Guillot and  R Haroutunian and  B Ille and  M Lethuillier and  N Lumb and  C Martin and  H Mathez and  G Maurelli and  S Muanza and  P Pangaud and  S Perries and  O Ravat and  E Schibler and  F Schirra and  G Smadja and  S Tissot and  B Trocme and  S Vanzetto and  J P Walder and  Y Bagaturia and  D Mjavia and  A Mzhavia and  Z Tsamalaidze and  V Roinishvili and  R Adolphi and  G Anagnostou and  R Brauer and  W Braunschweig and  H Esser and  L Feld and  W Karpinski and  A Khomich and  K Klein and  C Kukulies and  K Lübelsmeyer and  J Olzem and  A Ostaptchouk and  D Pandoulas and  G Pierschel and  F Raupach and  S Schael and  A Schultz von Dratzig and  G Schwering and  R Siedling and  M Thomas and  M Weber and  B Wittmer and  M Wlochal and  F Adamczyk and  A Adolf and  G Altenhöfer and  S Bechstein and  S Bethke and  P Biallass and  O Biebel and  M Bontenackels and  K Bosseler and  A Böhm and  M Erdmann and  H Faissner and  B Fehr and  H Fesefeldt and  G Fetchenhauer and  J Frangenheim and  J H Frohn and  J Grooten and  T Hebbeker and  S Hermann and  E Hermens and  G Hilgers and  K Hoepfner and  C Hof and  E Jacobi and  S Kappler and  M Kirsch and  P Kreuzer and  R Kupper and  H R Lampe and  D Lanske and  R Mameghani and  A Meyer and  S Meyer and  T Moers and  E Müller and  R Pahlke and  B Philipps and  D Rein and  H Reithler and  W Reuter and  P Rütten and  S Schulz and  H Schwarthoff and  W Sobek and  M Sowa and  T Stapelberg and  H Szczesny and  H Teykal and  D Teyssier and  H Tomme and  W Tomme and  M Tonutti and  O Tsigenov and  J Tutas and  J Vandenhirtz and  H Wagner and  M Wegner and  C Zeidler and  F Beissel and  M Davids and  M Duda and  G Flügge and  M Giffels and  T Hermanns and  D Heydhausen and  S Kalinin and  S Kasselmann and  G Kaussen and  T Kress and  A Linn and  A Nowack and  L Perchalla and  M Poettgens and  O Pooth and  P Sauerland and  A Stahl and  D Tornier and  M H Zoeller and  U Behrens and  K Borras and  A Flossdorf and  D Hatton and  B Hegner and  M Kasemann and  R Mankel and  A Meyer and  J Mnich and  C Rosemann and  C Youngman and  W D Zeuner and  F Bechtel and  P Buhmann and  E Butz and  G Flucke and  R H Hamdorf and  U Holm and  R Klanner and  U Pein and  N Schirm and  P Schleper and  G Steinbrück and  R Van Staa and  R Wolf and  B Atz and  T Barvich and  P Blüm and  F Boegelspacher and  H Bol and  Z Y Chen and  S Chowdhury and  W De Boer and  P Dehm and  G Dirkes and  M Fahrer and  U Felzmann and  M Frey and  A Furgeri and  E Gregoriev and  F Hartmann and  F Hauler and  S Heier and  K Kärcher and  B Ledermann and  S Mueller and  Th Müller and  D Neuberger and  C Piasecki and  G Quast and  K Rabbertz and  A Sabellek and  A Scheurer and  F P Schilling and  H J Simonis and  A Skiba and  P Steck and  A Theel and  W H Thümmel and  A Trunov and  A Vest and  T Weiler and  C Weiser and  S Weseler and  V Zhukov and  M Barone and  G Daskalakis and  N Dimitriou and  G Fanourakis and  C Filippidis and  T Geralis and  C Kalfas and  K Karafasoulis and  A Koimas and  A Kyriakis and  S Kyriazopoulou and  D Loukas and  A Markou and  C Markou and  N Mastroyiannopoulos and  C Mavrommatis and  J Mousa and  I Papadakis and  E Petrakou and  I Siotis and  K Theofilatos and  S Tzamarias and  A Vayaki and  G Vermisoglou and  A Zachariadou and  L Gouskos and  G Karapostoli and  P Katsas and  A Panagiotou and  C Papadimitropoulos and  X Aslanoglou and  I Evangelou and  P Kokkas and  N Manthos and  I Papadopoulos and  F A Triantis and  G Bencze and  L Boldizsar and  G Debreczeni and  C Hajdu and  P Hidas and  D Horvath and  P Kovesarki and  A Laszlo and  G Odor and  G Patay and  F Sikler and  G Veres and  G Vesztergombi and  P Zalan and  A Fenyvesi and  J Imrek and  J Molnar and  D Novak and  J Palinkas and  G Szekely and  N Beni and  A Kapusi and  G Marian and  B Radics and  P Raics and  Z Szabo and  Z Szillasi and  Z L Trocsanyi and  G Zilizi and  H S Bawa and  S B Beri and  V Bhandari and  V Bhatnagar and  M Kaur and  J M Kohli and  A Kumar and  B Singh and  J B Singh and  S Arora and  S Bhattacharya and  S Chatterji and  S Chauhan and  B C Choudhary and  P Gupta and  M Jha and  K Ranjan and  R K Shivpuri and  A K Srivastava and  R K Choudhury and  D Dutta and  M Ghodgaonkar and  S Kailas and  S K Kataria and  A K Mohanty and  L M Pant and  P Shukla and  A Topkar and  T Aziz and  Sunanda Banerjee and  S Bose and  S Chendvankar and  P V Deshpande and  M Guchait and  A Gurtu and  M Maity and  G Majumder and  K Mazumdar and  A Nayak and  M R Patil and  S Sharma and  K Sudhakar and  B S Acharya and  Sudeshna Banerjee and  S Bheesette and  S Dugad and  S D Kalmani and  V R Lakkireddi and  N K Mondal and  N Panyam and  P Verma and  H Arfaei and  M Hashemi and  M Mohammadi Najafabadi and  A Moshaii and  S Paktinat Mehdiabadi and  M Felcini and  M Grunewald and  K Abadjiev and  M Abbrescia and  L Barbone and  P Cariola and  F Chiumarulo and  A Clemente and  A Colaleo and  D Creanza and  N De Filippis and  M De Palma and  G De Robertis and  G Donvito and  R Ferorelli and  L Fiore and  M Franco and  D Giordano and  R Guida and  G Iaselli and  N Lacalamita and  F Loddo and  G Maggi and  M Maggi and  N Manna and  B Marangelli and  M S Mennea and  S My and  S Natali and  S Nuzzo and  G Papagni and  C Pinto and  A Pompili and  G Pugliese and  A Ranieri and  F Romano and  G Roselli and  G Sala and  G Selvaggi and  L Silvestris and  P Tempesta and  R Trentadue and  S Tupputi and  G Zito and  G Abbiendi and  W Bacchi and  C Battilana and  A C Benvenuti and  M Boldini and  D Bonacorsi and  S Braibant-Giacomelli and  V D Cafaro and  P Capiluppi and  A Castro and  F R Cavallo and  C Ciocca and  G Codispoti and  M Cuffiani and  I D'Antone and  G M Dallavalle and  F Fabbri and  A Fanfani and  S Finelli and  P Giacomelli and  V Giordano and  M Giunta and  C Grandi and  M Guerzoni and  L Guiducci and  S Marcellini and  G Masetti and  A Montanari and  F L Navarria and  F Odorici and  A Paolucci and  G Pellegrini and  A Perrotta and  A M Rossi and  T Rovelli and  G P Siroli and  G Torromeo and  R Travaglini and  G P Veronese and  S Albergo and  M Chiorboli and  S Costa and  M Galanti and  G Gatto Rotondo and  N Giudice and  N Guardone and  F Noto and  R Potenza and  M A Saizu and  G Salemi and  C Sutera and  A Tricomi and  C Tuve and  L Bellucci and  M Brianzi and  G Broccolo and  E Catacchini and  V Ciulli and  C Civinini and  R D'Alessandro and  E Focardi and  S Frosali and  C Genta and  G Landi and  P Lenzi and  A Macchiolo and  F Maletta and  F Manolescu and  C Marchettini and  L Masetti and  S Mersi and  M Meschini and  C Minelli and  S Paoletti and  G Parrini and  E Scarlini and  G Sguazzoni and  L Benussi and  M Bertani and  S Bianco and  M Caponero and  D Colonna and  L Daniello and  F Fabbri and  F Felli and  M Giardoni and  A La Monaca and  B Ortenzi and  M Pallotta and  A Paolozzi and  C Paris and  L Passamonti and  D Pierluigi and  B Ponzio and  C Pucci and  A Russo and  G Saviano and  P Fabbricatore and  S Farinon and  M Greco and  R Musenich and  S Badoer and  L Berti and  M Biasotto and  S Fantinel and  E Frizziero and  U Gastaldi and  M Gulmini and  F Lelli and  G Maron and  S Squizzato and  N Toniolo and  S Traldi and  S Banfi and  R Bertoni and  M Bonesini and  L Carbone and  G B Cerati and  F Chignoli and  P D'Angelo and  A De Min and  P Dini and  F M Farina and  F Ferri and  P Govoni and  S Magni and  M Malberti and  S Malvezzi and  R Mazza and  D Menasce and  V Miccio and  L Moroni and  P Negri and  M Paganoni and  D Pedrini and  A Pullia and  S Ragazzi and  N Redaelli and  M Rovere and  L Sala and  S Sala and  R Salerno and  T Tabarelli de Fatis and  V Tancini and  S Taroni and  A Boiano and  F Cassese and  C Cassese and  A Cimmino and  B D'Aquino and  L Lista and  D Lomidze and  P Noli and  P Paolucci and  G Passeggio and  D Piccolo and  L Roscilli and  C Sciacca and  A Vanzanella and  P Azzi and  N Bacchetta and  L Barcellan and  M Bellato and  M Benettoni and  D Bisello and  E Borsato and  A Candelori and  R Carlin and  L Castellani and  P Checchia and  L Ciano and  A Colombo and  E Conti and  M Da Rold and  F Dal Corso and  M De Giorgi and  M De Mattia and  T Dorigo and  U Dosselli and  C Fanin and  G Galet and  F Gasparini and  U Gasparini and  A Giraldo and  P Giubilato and  F Gonella and  A Gresele and  A Griggio and  P Guaita and  A Kaminskiy and  S Karaevskii and  V Khomenkov and  D Kostylev and  S Lacaprara and  I Lazzizzera and  I Lippi and  M Loreti and  M Margoni and  R Martinelli and  S Mattiazzo and  M Mazzucato and  A T Meneguzzo and  L Modenese and  F Montecassiano and  A Neviani and  M Nigro and  A Paccagnella and  D Pantano and  A Parenti and  M Passaseo and  R Pedrotta and  M Pegoraro and  G Rampazzo and  S Reznikov and  P Ronchese and  A Sancho Daponte and  P Sartori and  I Stavitskiy and  M Tessaro and  E Torassa and  A Triossi and  S Vanini and  S Ventura and  L Ventura and  M Verlato and  M Zago and  F Zatti and  P Zotto and  G Zumerle and  P Baesso and  G Belli and  U Berzano and  S Bricola and  A Grelli and  G Musitelli and  R Nardò and  M M Necchi and  D Pagano and  S P Ratti and  C Riccardi and  P Torre and  A Vicini and  P Vitulo and  C Viviani and  D Aisa and  S Aisa and  F Ambroglini and  M M Angarano and  E Babucci and  D Benedetti and  M Biasini and  G M Bilei and  S Bizzaglia and  M T Brunetti and  B Caponeri and  B Checcucci and  R Covarelli and  N Dinu and  L Fanò and  L Farnesini and  M Giorgi and  P Lariccia and  G Mantovani and  F Moscatelli and  D Passeri and  A Piluso and  P Placidi and  V Postolache and  R Santinelli and  A Santocchia and  L Servoli and  D Spiga and  P Azzurri and  G Bagliesi and  G Balestri and  A Basti and  R Bellazzini and  L Benucci and  J Bernardini and  L Berretta and  S Bianucci and  T Boccali and  A Bocci and  L Borrello and  F Bosi and  F Bracci and  A Brez and  F Calzolari and  R Castaldi and  U Cazzola and  M Ceccanti and  R Cecchi and  C Cerri and  A S Cucoanes and  R Dell'Orso and  D Dobur and  S Dutta and  F Fiori and  L Foà and  A Gaggelli and  S Gennai and  A Giassi and  S Giusti and  D Kartashov and  A Kraan and  L Latronico and  F Ligabue and  S Linari and  T Lomtadze and  G A Lungu and  G Magazzu and  P Mammini and  F Mariani and  G Martinelli and  M Massa and  A Messineo and  A Moggi and  F Palla and  F Palmonari and  G Petragnani and  G Petrucciani and  A Profeti and  F Raffaelli and  D Rizzi and  G Sanguinetti and  S Sarkar and  G Segneri and  D Sentenac and  A T Serban and  A Slav and  P Spagnolo and  G Spandre and  R Tenchini and  S Tolaini and  G Tonelli and  A Venturi and  P G Verdini and  M Vos and  L Zaccarelli and  S Baccaro and  L Barone and  A Bartoloni and  B Borgia and  G Capradossi and  F Cavallari and  A Cecilia and  D D'Angelo and  I Dafinei and  D Del Re and  E Di Marco and  M Diemoz and  G Ferrara and  C Gargiulo and  S Guerra and  M Iannone and  E Longo and  M Montecchi and  M Nuccetelli and  G Organtini and  A Palma and  R Paramatti and  F Pellegrino and  S Rahatlou and  C Rovelli and  F Safai Tehrani and  A Zullo and  G Alampi and  N Amapane and  R Arcidiacono and  S Argiro and  M Arneodo and  R Bellan and  F Benotto and  C Biino and  S Bolognesi and  M A Borgia and  C Botta and  A Brasolin and  N Cartiglia and  R Castello and  G Cerminara and  R Cirio and  M Cordero and  M Costa and  D Dattola and  F Daudo and  G Dellacasa and  N Demaria and  G Dughera and  F Dumitrache and  R Farano and  G Ferrero and  E Filoni and  G Kostyleva and  H E Larsen and  C Mariotti and  M Marone and  S Maselli and  E Menichetti and  P Mereu and  E Migliore and  G Mila and  V Monaco and  M Musich and  M Nervo and  M M Obertino and  R Panero and  A Parussa and  N Pastrone and  C Peroni and  G Petrillo and  A Romero and  M Ruspa and  R Sacchi and  M Scalise and  A Solano and  A Staiano and  P P Trapani and  D Trocino and  V Vaniev and  A Vilela Pereira and  A Zampieri and  S Belforte and  F Cossutti and  G Della Ricca and  B Gobbo and  C Kavka and  A Penzo and  Y E Kim and  S K Nam and  D H Kim and  G N Kim and  J C Kim and  D J Kong and  S R Ro and  D C Son and  S Y Park and  Y J Kim and  J Y Kim and  I T Lim and  M Y Pac and  S J Lee and  S Y Jung and  J T Rhee and  S H Ahn and  B S Hong and  Y K Jeng and  M H Kang and  H C Kim and  J H Kim and  T J Kim and  K S Lee and  J K Lim and  D H Moon and  I C Park and  S K Park and  M S Ryu and  K-S Sim and  K J Son and  S J Hong and  Y I Choi and  H Castilla Valdez and  A Sanchez Hernandez and  S Carrillo Moreno and  A Morelos Pineda and  A Aerts and  P Van der Stok and  H Weffers and  P Allfrey and  R N C Gray and  M Hashimoto and  D Krofcheck and  A J Bell and  N Bernardino Rodrigues and  P H Butler and  S Churchwell and  R Knegjens and  S Whitehead and  J C Williams and  Z Aftab and  U Ahmad and  I Ahmed and  W Ahmed and  M I Asghar and  S Asghar and  G Dad and  M Hafeez and  H R Hoorani and  I Hussain and  N Hussain and  M Iftikhar and  M S Khan and  K Mehmood and  A Osman and  H Shahzad and  A R Zafar and  A Ali and  A Bashir and  A M Jan and  A Kamal and  F Khan and  M Saeed and  S Tanwir and  M A Zafar and  J Blocki and  A Cyz and  E Gladysz-Dziadus and  S Mikocki and  M Rybczynski and  J Turnau and  Z Wlodarczyk and  P Zychowski and  K Bunkowski and  M Cwiok and  H Czyrkowski and  R Dabrowski and  W Dominik and  K Doroba and  A Kalinowski and  K Kierzkowski and  M Konecki and  J Krolikowski and  I M Kudla and  M Pietrusinski and  K Pozniak and  W Zabolotny and  P Zych and  R Gokieli and  L Goscilo and  M Górski and  K Nawrocki and  P Traczyk and  G Wrochna and  P Zalewski and  K T Pozniak and  R Romaniuk and  W M Zabolotny and  R Alemany-Fernandez and  C Almeida and  N Almeida and  A S Araujo Vila Verde and  T Barata Monteiro and  M Bluj and  S Da Mota Silva and  A David Tinoco Mendes and  M Freitas Ferreira and  M Gallinaro and  M Husejko and  A Jain and  M Kazana and  P Musella and  R Nobrega and  J Rasteiro Da Silva and  P Q Ribeiro and  M Santos and  P Silva and  S Silva and  I Teixeira and  J P Teixeira and  J Varela and  G Varner and  N Vaz Cardoso and  I Altsybeev and  K Babich and  A Belkov and  I Belotelov and  P Bunin and  S Chesnevskaya and  V Elsha and  Y Ershov and  I Filozova and  M Finger and  M Finger Jr and  A Golunov and  I Golutvin and  N Gorbounov and  I Gramenitski and  V Kalagin and  A Kamenev and  V Karjavin and  S Khabarov and  V Khabarov and  Y Kiryushin and  V Konoplyanikov and  V Korenkov and  G Kozlov and  A Kurenkov and  A Lanev and  V Lysiakov and  A Malakhov and  I Melnitchenko and  V V Mitsyn and  K Moisenz and  P Moisenz and  S Movchan and  E Nikonov and  D Oleynik and  V Palichik and  V Perelygin and  A Petrosyan and  E Rogalev and  V Samsonov and  M Savina and  R Semenov and  S Sergeev and  S Shmatov and  S Shulha and  V Smirnov and  D Smolin and  A Tcheremoukhine and  O Teryaev and  E Tikhonenko and  A Urkinbaev and  S Vasil'ev and  A Vishnevskiy and  A Volodko and  N Zamiatin and  A Zarubin and  P Zarubin and  E Zubarev and  N Bondar and  Y Gavrikov and  V Golovtsov and  Y Ivanov and  V Kim and  V Kozlov and  V Lebedev and  G Makarenkov and  F Moroz and  P Neustroev and  G Obrant and  E Orishchin and  A Petrunin and  Y Shcheglov and  A Shchetkovskiy and  V Sknar and  V Skorobogatov and  I Smirnov and  V Sulimov and  V Tarakanov and  L Uvarov and  S Vavilov and  G Velichko and  S Volkov and  A Vorobyev and  D Chmelev and  D Druzhkin and  A Ivanov and  V Kudinov and  O Logatchev and  S Onishchenko and  A Orlov and  V Sakharov and  V Smetannikov and  A Tikhomirov and  S Zavodthikov and  Yu Andreev and  A Anisimov and  V Duk and  S Gninenko and  N Golubev and  D Gorbunov and  M Kirsanov and  N Krasnikov and  V Matveev and  A Pashenkov and  A Pastsyak and  V E Postoev and  A Sadovski and  A Skassyrskaia and  Alexander Solovey and  Anatoly Solovey and  D Soloviev and  A Toropin and  S Troitsky and  A Alekhin and  A Baldov and  V Epshteyn and  V Gavrilov and  N Ilina and  V Kaftanov and  V Karpishin and  I Kiselevich and  V Kolosov and  M Kossov and  A Krokhotin and  S Kuleshov and  A Oulianov and  A Pozdnyakov and  G Safronov and  S Semenov and  N Stepanov and  V Stolin and  E Vlasov and  V Zaytsev and  E Boos and  M Dubinin and  L Dudko and  A Ershov and  G Eyyubova and  A Gribushin and  V Ilyin and  V Klyukhin and  O Kodolova and  N A Kruglov and  A Kryukov and  I Lokhtin and  L Malinina and  V Mikhaylin and  S Petrushanko and  L Sarycheva and  V Savrin and  L Shamardin and  A Sherstnev and  A Snigirev and  K Teplov and  I Vardanyan and  A M Fomenko and  N Konovalova and  V Kozlov and  A I Lebedev and  N Lvova and  S V Rusakov and  A Terkulov and  V Abramov and  S Akimenko and  A Artamonov and  A Ashimova and  I Azhgirey and  S Bitioukov and  O Chikilev and  K Datsko and  A Filine and  A Godizov and  P Goncharov and  V Grishin and  A Inyakin and  V Kachanov and  A Kalinin and  A Khmelnikov and  D Konstantinov and  A Korablev and  V Krychkine and  A Krinitsyn and  A Levine and  I Lobov and  V Lukanin and  Y Mel'nik and  V Molchanov and  V Petrov and  V Petukhov and  V Pikalov and  A Ryazanov and  R Ryutin and  V Shelikhov and  V Skvortsov and  S Slabospitsky and  A Sobol and  A Sytine and  V Talov and  L Tourtchanovitch and  S Troshin and  N Tyurin and  A Uzunian and  A Volkov and  S Zelepoukine and  V Lukyanov and  G Mamaeva and  Z Prilutskaya and  I Rumyantsev and  S Sokha and  S Tataurschikov and  I Vasilyev and  P Adzic and  I Anicin and  M Djordjevic and  D Jovanovic and  D Maletic and  J Puzovic and  N Smiljkovic and  E Aguayo Navarrete and  M Aguilar-Benitez and  J Ahijado Munoz and  J M Alarcon Vega and  J Alberdi and  J Alcaraz Maestre and  M Aldaya Martin and  P Arce and  J M Barcala and  J Berdugo and  C L Blanco Ramos and  C Burgos Lazaro and  J Caballero Bejar and  E Calvo and  M Cerrada and  M Chamizo Llatas and  J J Chercoles Catalán and  N Colino and  M Daniel and  B De La Cruz and  A Delgado Peris and  C Fernandez Bedoya and  A Ferrando and  M C Fouz and  D Francia Ferrero and  J Garcia Romero and  P Garcia-Abia and  O Gonzalez Lopez and  J M Hernandez and  M I Josa and  J Marin and  G Merino and  A Molinero and  J J Navarrete and  J C Oller and  J Puerta Pelayo and  J C Puras Sanchez and  J Ramirez and  L Romero and  C Villanueva Munoz and  C Willmott and  C Yuste and  C Albajar and  J F de Trocóniz and  I Jimenez and  R Macias and  R F Teixeira and  J Cuevas and  J Fernández Menéndez and  I Gonzalez Caballero and  J Lopez-Garcia and  H Naves Sordo and  J M Vizan Garcia and  I J Cabrillo and  A Calderon and  D Cano Fernandez and  I Diaz Merino and  J Duarte Campderros and  M Fernandez and  J Fernandez Menendez and  C Figueroa and  L A Garcia Moral and  G Gomez and  F Gomez Casademunt and  J Gonzalez Sanchez and  R Gonzalez Suarez and  C Jorda and  P Lobelle Pardo and  A Lopez Garcia and  A Lopez Virto and  J Marco and  R Marco and  C Martinez Rivero and  P Martinez Ruiz del Arbol and  F Matorras and  P Orviz Fernandez and  A Patino Revuelta and  T Rodrigo and  D Rodriguez Gonzalez and  A Ruiz Jimeno and  L Scodellaro and  M Sobron Sanudo and  I Vila and  R Vilar Cortabitarte and  M Barbero and  D Goldin and  B Henrich and  L Tauscher and  S Vlachos and  M Wadhwa and  D Abbaneo and  S M Abbas and  I Ahmed and  S Akhtar and  M I Akhtar and  E Albert and  M Alidra and  S Ashby and  P Aspell and  E Auffray and  P Baillon and  A Ball and  S L Bally and  N Bangert and  R Barillère and  D Barney and  S Beauceron and  F Beaudette and  G Benelli and  R Benetta and  J L Benichou and  W Bialas and  A Bjorkebo and  D Blechschmidt and  C Bloch and  P Bloch and  S Bonacini and  J Bos and  M Bosteels and  V Boyer and  A Branson and  H Breuker and  R Bruneliere and  O Buchmuller and  D Campi and  T Camporesi and  A Caner and  E Cano and  E Carrone and  A Cattai and  J P Chatelain and  M Chauvey and  T Christiansen and  M Ciganek and  S Cittolin and  J Cogan and  A Conde Garcia and  H Cornet and  E Corrin and  M Corvo and  S Cucciarelli and  B Curé and  D D'Enterria and  A De Roeck and  T de Visser and  C Delaere and  M Delattre and  C Deldicque and  D Delikaris and  D Deyrail and  S Di Vincenzo and  A Domeniconi and  S Dos Santos and  G Duthion and  L M Edera and  A Elliott-Peisert and  M Eppard and  F Fanzago and  M Favre and  H Foeth and  R Folch and  N Frank and  S Fratianni and  M A Freire and  A Frey and  A Fucci and  W Funk and  A Gaddi and  F Gagliardi and  M Gastal and  M Gateau and  J C Gayde and  H Gerwig and  A Ghezzi and  D Gigi and  K Gill and  A S Giolo-Nicollerat and  J P Girod and  F Glege and  W Glessing and  R Gomez-Reino Garrido and  R Goudard and  R Grabit and  J P Grillet and  P Gutierrez Llamas and  E Gutierrez Mlot and  J Gutleber and  R Hall-wilton and  R Hammarstrom and  M Hansen and  J Harvey and  A Hervé and  J Hill and  H F Hoffmann and  A Holzner and  A Honma and  D Hufnagel and  M Huhtinen and  S D Ilie and  V Innocente and  W Jank and  P Janot and  P Jarron and  M Jeanrenaud and  P Jouvel and  R Kerkach and  K Kloukinas and  L J Kottelat and  J C Labbé and  D Lacroix and  X Lagrue and  C Lasseur and  E Laure and  J F Laurens and  P Lazeyras and  J M Le Goff and  M Lebeau and  P Lecoq and  F Lemeilleur and  M Lenzi and  N Leonardo and  C Leonidopoulos and  M Letheren and  M Liendl and  F Limia-Conde and  L Linssen and  C Ljuslin and  B Lofstedt and  R Loos and  J A Lopez Perez and  C Lourenco and  A Lyonnet and  A Machard and  R Mackenzie and  N Magini and  G Maire and  L Malgeri and  R Malina and  M Mannelli and  A Marchioro and  J Martin and  F Meijers and  P Meridiani and  E Meschi and  T Meyer and  A Meynet Cordonnier and  J F Michaud and  L Mirabito and  R Moser and  F Mossiere and  J Muffat-Joly and  M Mulders and  J Mulon and  E Murer and  P Mättig and  A Oh and  A Onnela and  M Oriunno and  L Orsini and  J A Osborne and  C Paillard and  I Pal and  G Papotti and  G Passardi and  A Patino-Revuelta and  V Patras and  B Perea Solano and  E Perez and  G Perinic and  J F Pernot and  P Petagna and  P Petiot and  P Petit and  A Petrilli and  A Pfeiffer and  C Piccut and  M Pimiä and  R Pintus and  M Pioppi and  A Placci and  L Pollet and  H Postema and  M J Price and  R Principe and  A Racz and  E Radermacher and  R Ranieri and  G Raymond and  P Rebecchi and  J Rehn and  S Reynaud and  H Rezvani Naraghi and  D Ricci and  M Ridel and  M Risoldi and  P Rodrigues Simoes Moreira and  A Rohlev and  G Roiron and  G Rolandi and  P Rumerio and  O Runolfsson and  V Ryjov and  H Sakulin and  D Samyn and  L C Santos Amaral and  H Sauce and  E Sbrissa and  P Scharff-Hansen and  P Schieferdecker and  W D Schlatter and  B Schmitt and  H G Schmuecker and  M Schröder and  C Schwick and  C Schäfer and  I Segoni and  P Sempere Roldán and  S Sgobba and  A Sharma and  P Siegrist and  C Sigaud and  N Sinanis and  T Sobrier and  P Sphicas and  M Spiropulu and  G Stefanini and  A Strandlie and  F Szoncsó and  B G Taylor and  O Teller and  A Thea and  E Tournefier and  D Treille and  P Tropea and  J Troska and  E Tsesmelis and  A Tsirou and  J Valls and  I Van Vulpen and  M Vander Donckt and  F Vasey and  M Vazquez Acosta and  L Veillet and  P Vichoudis and  G Waurick and  J P Wellisch and  P Wertelaers and  M Wilhelmsson and  I M Willers and  M Winkler and  M Zanetti and  W Bertl and  K Deiters and  P Dick and  W Erdmann and  D Feichtinger and  K Gabathuler and  Z Hochman and  R Horisberger and  Q Ingram and  H C Kaestli and  D Kotlinski and  S König and  P Poerschke and  D Renker and  T Rohe and  T Sakhelashvili and  A Starodumov and  V Aleksandrov and  F Behner and  I Beniozef and  B Betev and  B Blau and  A M Brett and  L Caminada and  Z Chen and  N Chivarov and  D Da Silva Di Calafiori and  S Dambach and  G Davatz and  V Delachenal and  R Della Marina and  H Dimov and  G Dissertori and  M Dittmar and  L Djambazov and  M Dröge and  C Eggel and  J Ehlers and  R Eichler and  M Elmiger and  G Faber and  K Freudenreich and  J F Fuchs and  G M Georgiev and  C Grab and  C Haller and  J Herrmann and  M Hilgers and  W Hintz and  Hans Hofer and  Heinz Hofer and  U Horisberger and  I Horvath and  A Hristov and  C Humbertclaude and  B Iliev and  W Kastli and  A Kruse and  J Kuipers and  U Langenegger and  P Lecomte and  E Lejeune and  G Leshev and  C Lesmond and  B List and  P D Luckey and  W Lustermann and  J D Maillefaud and  C Marchica and  A Maurisset and  B Meier and  P Milenovic and  M Milesi and  F Moortgat and  I Nanov and  A Nardulli and  F Nessi-Tedaldi and  B Panev and  L Pape and  F Pauss and  E Petrov and  G Petrov and  M M Peynekov and  D Pitzl and  T Punz and  P Riboni and  J Riedlberger and  A Rizzi and  F J Ronga and  P A Roykov and  U Röser and  D Schinzel and  A Schöning and  A Sourkov and  K Stanishev and  S Stoenchev and  F Stöckli and  H Suter and  P Trüb and  S Udriot and  D G Uzunova and  I Veltchev and  G Viertel and  H P von Gunten and  S Waldmeier-Wicki and  R Weber and  M Weber and  J Weng and  M Wensveen and  F Wittgenstein and  K Zagoursky and  E Alagoz and  C Amsler and  V Chiochia and  C Hoermann and  C Regenfus and  P Robmann and  T Rommerskirchen and  A Schmidt and  S Steiner and  D Tsirigkas and  L Wilke and  S Blyth and  Y H Chang and  E A Chen and  A Go and  C C Hung and  C M Kuo and  S W Li and  W Lin and  P Chang and  Y Chao and  K F Chen and  Z Gao and  G W S Hou and  Y B Hsiung and  Y J Lei and  S W Lin and  R S Lu and  J G Shiu and  Y M Tzeng and  K Ueno and  Y Velikzhanin and  C C Wang and  M-Z Wang and  S Aydin and  A Azman and  M N Bakirci and  S Basegmez and  S Cerci and  I Dumanoglu and  S Erturk and  E Eskut and  A Kayis Topaksu and  H Kisoglu and  P Kurt and  K Ozdemir and  N Ozdes Koca and  H Ozkurt and  S Ozturk and  A Polatöz and  K Sogut and  H Topakli and  M Vergili and  G Önengüt and  H Gamsizkan and  S Sekmen and  M Serin-Zeyrek and  R Sever and  M Zeyrek and  M Deliomeroglu and  E Gülmez and  E Isiksal and  M Kaya and  O Kaya and  S Ozkorucuklu and  N Sonmez and  B Grinev and  V Lyubynskiy and  V Senchyshyn and  L Levchuk and  S Lukyanenko and  D Soroka and  P Sorokin and  S Zub and  A Anjum and  N Baker and  T Hauer and  R McClatchey and  M Odeh and  D Rogulin and  A Solomonides and  J J Brooke and  R Croft and  D Cussans and  D Evans and  R Frazier and  N Grant and  M Hansen and  R D Head and  G P Heath and  H F Heath and  C Hill and  B Huckvale and  J Jackson and  C Lynch and  C K Mackay and  S Metson and  S J Nash and  D M Newbold and  A D Presland and  M G Probert and  E C Reid and  V J Smith and  R J Tapper and  R Walton and  E Bateman and  K W Bell and  R M Brown and  B Camanzi and  I T Church and  D J A Cockerill and  J E Cole and  J F Connolly and  J A Coughlan and  P S Flower and  P Ford and  V B Francis and  M J French and  S B Galagedera and  W Gannon and  A P R Gay and  N I Geddes and  R J S Greenhalgh and  R N J Halsall and  W J Haynes and  J A Hill and  F R Jacob and  P W Jeffreys and  L L Jones and  B W Kennedy and  A L Lintern and  A B Lodge and  A J Maddox and  Q R Morrissey and  P Murray and  G N Patrick and  C A X Pattison and  M R Pearson and  S P H Quinton and  G J Rogers and  J G Salisbury and  A A Shah and  C H Shepherd-Themistocleous and  B J Smith and  M Sproston and  R Stephenson and  S Taghavi and  I R Tomalin and  M J Torbet and  J H Williams and  W J Womersley and  S D Worm and  F Xing and  M Apollonio and  F Arteche and  R Bainbridge and  G Barber and  P Barrillon and  J Batten and  R Beuselinck and  P M Brambilla Hall and  D Britton and  W Cameron and  D E Clark and  I W Clark and  D Colling and  N Cripps and  G Davies and  M Della Negra and  G Dewhirst and  S Dris and  C Foudas and  J Fulcher and  D Futyan and  D J Graham and  S Greder and  S Greenwood and  G Hall and  J F Hassard and  J Hays and  G Iles and  V Kasey and  M Khaleeq and  J Leaver and  P Lewis and  B C MacEvoy and  O Maroney and  E M McLeod and  D G Miller and  J Nash and  A Nikitenko and  E Noah Messomo and  M Noy and  A Papageorgiou and  M Pesaresi and  K Petridis and  D R Price and  X Qu and  D M Raymond and  A Rose and  S Rutherford and  M J Ryan and  F Sciacca and  C Seez and  P Sharp and  G Sidiropoulos and  M Stettler and  M Stoye and  J Striebig and  M Takahashi and  H Tallini and  A Tapper and  C Timlin and  L Toudup and  T Virdee and  S Wakefield and  P Walsham and  D Wardrope and  M Wingham and  Y Zhang and  O Zorba and  C Da Via and  I Goitom and  P R Hobson and  D C Imrie and  I Reid and  C Selby and  O Sharif and  L Teodorescu and  S J Watts and  I Yaselli and  E Hazen and  A Heering and  A Heister and  C Lawlor and  D Lazic and  E Machado and  J Rohlf and  L Sulak and  F Varela Rodriguez and  S X Wu and  A Avetisyan and  T Bose and  L Christofek and  D Cutts and  S Esen and  R Hooper and  G Landsberg and  M Narain and  D Nguyen and  T Speer and  K V Tsang and  R Breedon and  M Case and  M Chertok and  J Conway and  P T Cox and  J Dolen and  R Erbacher and  Y Fisyak and  E Friis and  G Grim and  B Holbrook and  W Ko and  A Kopecky and  R Lander and  F C Lin and  A Lister and  S Maruyama and  D Pellett and  J Rowe and  M Searle and  J Smith and  A Soha and  M Squires and  M Tripathi and  R Vasquez Sierra and  C Veelken and  V Andreev and  K Arisaka and  Y Bonushkin and  S Chandramouly and  D Cline and  R Cousins and  S Erhan and  J Hauser and  M Ignatenko and  C Jarvis and  B Lisowski and  C Matthey and  B Mohr and  J Mumford and  S Otwinowski and  Y Pischalnikov and  G Rakness and  P Schlein and  Y Shi and  B Tannenbaum and  J Tucker and  V Valuev and  R Wallny and  H G Wang and  X Yang and  Y Zheng and  J Andreeva and  J Babb and  S Campana and  D Chrisman and  R Clare and  J Ellison and  D Fortin and  J W Gary and  W Gorn and  G Hanson and  G Y Jeng and  S C Kao and  J G Layter and  F Liu and  H Liu and  A Luthra and  G Pasztor and  H Rick and  A Satpathy and  B C Shen and  R Stringer and  V Sytnik and  P Tran and  S Villa and  R Wilken and  S Wimpenny and  D Zer-Zion and  J G Branson and  J A Coarasa Perez and  E Dusinberre and  R Kelley and  M Lebourgeois and  J Letts and  E Lipeles and  B Mangano and  T Martin and  M Mojaver and  J Muelmenstaedt and  M Norman and  H P Paar and  A Petrucci and  H Pi and  M Pieri and  A Rana and  M Sani and  V Sharma and  S Simon and  A White and  F Würthwein and  A Yagil and  A Affolder and  A Allen and  C Campagnari and  M D'Alfonso and  A Dierlamm and  J Garberson and  D Hale and  J Incandela and  P Kalavase and  S A Koay and  D Kovalskyi and  V Krutelyov and  S Kyre and  J Lamb and  S Lowette and  M Nikolic and  V Pavlunin and  F Rebassoo and  J Ribnik and  J Richman and  R Rossin and  Y S Shah and  D Stuart and  S Swain and  J R Vlimant and  D White and  M Witherell and  A Bornheim and  J Bunn and  J Chen and  G Denis and  P Galvez and  M Gataullin and  I Legrand and  V Litvine and  Y Ma and  R Mao and  D Nae and  I Narsky and  H B Newman and  T Orimoto and  C Rogan and  S Shevchenko and  C Steenberg and  X Su and  M Thomas and  V Timciuc and  F van Lingen and  J Veverka and  B R Voicu and  A Weinstein and  R Wilkinson and  Y Xia and  Y Yang and  L Y Zhang and  K Zhu and  R Y Zhu and  T Ferguson and  D W Jang and  S Y Jun and  M Paulini and  J Russ and  N Terentyev and  H Vogel and  I Vorobiev and  M Bunce and  J P Cumalat and  M E Dinardo and  B R Drell and  W T Ford and  K Givens and  B Heyburn and  D Johnson and  U Nauenberg and  K Stenson and  S R Wagner and  L Agostino and  J Alexander and  F Blekman and  D Cassel and  S Das and  J E Duboscq and  L K Gibbons and  B Heltsley and  C D Jones and  V Kuznetsov and  J R Patterson and  D Riley and  A Ryd and  S Stroiney and  W Sun and  J Thom and  J Vaughan and  P Wittich and  C P Beetz and  G Cirino and  V Podrasky and  C Sanzeni and  D Winn and  S Abdullin and  M A Afaq and  M Albrow and  J Amundson and  G Apollinari and  M Atac and  W Badgett and  J A Bakken and  B Baldin and  K Banicz and  L A T Bauerdick and  A Baumbaugh and  J Berryhill and  P C Bhat and  M Binkley and  I Bloch and  F Borcherding and  A Boubekeur and  M Bowden and  K Burkett and  J N Butler and  H W K Cheung and  G Chevenier and  F Chlebana and  I Churin and  S Cihangir and  W Dagenhart and  M Demarteau and  D Dykstra and  D P Eartly and  J E Elias and  V D Elvira and  D Evans and  I Fisk and  J Freeman and  I Gaines and  P Gartung and  F J M Geurts and  L Giacchetti and  D A Glenzinski and  E Gottschalk and  T Grassi and  D Green and  C Grimm and  Y Guo and  O Gutsche and  A Hahn and  J Hanlon and  R M Harris and  T Hesselroth and  S Holm and  B Holzman and  E James and  H Jensen and  M Johnson and  U Joshi and  B Klima and  S Kossiakov and  K Kousouris and  J Kowalkowski and  T Kramer and  S Kwan and  C M Lei and  M Leininger and  S Los and  L Lueking and  G Lukhanin and  S Lusin and  K Maeshima and  J M Marraffino and  D Mason and  P McBride and  T Miao and  S Moccia and  N Mokhov and  S Mrenna and  S J Murray and  C Newman-Holmes and  C Noeding and  V O'Dell and  M Paterno and  D Petravick and  R Pordes and  O Prokofyev and  N Ratnikova and  A Ronzhin and  V Sekhri and  E Sexton-Kennedy and  I Sfiligoi and  T M Shaw and  E Skup and  R P Smith and  W J Spalding and  L Spiegel and  M Stavrianakou and  G Stiehr and  A L Stone and  I Suzuki and  P Tan and  W Tanenbaum and  L E Temple and  S Tkaczyk and  L Uplegger and  E W Vaandering and  R Vidal and  R Wands and  H Wenzel and  J Whitmore and  E Wicklund and  W M Wu and  Y Wu and  J Yarba and  V Yarba and  F Yumiceva and  J C Yun and  T Zimmerman and  D Acosta and  P Avery and  V Barashko and  P Bartalini and  D Bourilkov and  R Cavanaugh and  S Dolinsky and  A Drozdetskiy and  R D Field and  Y Fu and  I K Furic and  L Gorn and  D Holmes and  B J Kim and  S Klimenko and  J Konigsberg and  A Korytov and  K Kotov and  P Levchenko and  A Madorsky and  K Matchev and  G Mitselmakher and  Y Pakhotin and  C Prescott and  L Ramond and  P Ramond and  M Schmitt and  B Scurlock and  J Stasko and  H Stoeck and  D Wang and  J Yelton and  V Gaultney and  L Kramer and  L M Lebolo and  S Linn and  P Markowitz and  G Martinez and  J L Rodriguez and  T Adams and  A Askew and  O Atramentov and  M Bertoldi and  W G D Dharmaratna and  Y Gershtein and  S V Gleyzer and  S Hagopian and  V Hagopian and  C J Jenkins and  K F Johnson and  H Prosper and  D Simek and  J Thomaston and  M Baarmand and  L Baksay and  S Guragain and  M Hohlmann and  H Mermerkaya and  R Ralich and  I Vodopiyanov and  M R Adams and  I M Anghel and  L Apanasevich and  O Barannikova and  V E Bazterra and  R R Betts and  C Dragoiu and  E J Garcia-Solis and  C E Gerber and  D J Hofman and  R Hollis and  A Iordanova and  S Khalatian and  C Mironov and  E Shabalina and  A Smoron and  N Varelas and  U Akgun and  E A Albayrak and  A S Ayan and  R Briggs and  K Cankocak and  W Clarida and  A Cooper and  P Debbins and  F Duru and  M Fountain and  E McCliment and  J P Merlo and  A Mestvirishvili and  M J Miller and  A Moeller and  C R Newsom and  E Norbeck and  J Olson and  Y Onel and  L Perera and  I Schmidt and  S Wang and  T Yetkin and  E W Anderson and  H Chakir and  J M Hauptman and  J Lamsa and  B A Barnett and  B Blumenfeld and  C Y Chien and  G Giurgiu and  A Gritsan and  D W Kim and  C K Lae and  P Maksimovic and  M Swartz and  N Tran and  P Baringer and  A Bean and  J Chen and  D Coppage and  O Grachov and  M Murray and  V Radicci and  J S Wood and  V Zhukova and  D Bandurin and  T Bolton and  K Kaadze and  W E Kahl and  Y Maravin and  D Onoprienko and  R Sidwell and  Z Wan and  B Dahmes and  J Gronberg and  J Hollar and  D Lange and  D Wright and  C R Wuest and  D Baden and  R Bard and  S C Eno and  D Ferencek and  N J Hadley and  R G Kellogg and  M Kirn and  S Kunori and  E Lockner and  F Ratnikov and  F Santanastasio and  A Skuja and  T Toole and  L Wang and  M Wetstein and  B Alver and  M Ballintijn and  G Bauer and  W Busza and  G Gomez Ceballos and  K A Hahn and  P Harris and  M Klute and  I Kravchenko and  W Li and  C Loizides and  T Ma and  S Nahn and  C Paus and  S Pavlon and  J Piedra Gomez and  C Roland and  G Roland and  M Rudolph and  G Stephans and  K Sumorok and  S Vaurynovich and  E A Wenger and  B Wyslouch and  D Bailleux and  S Cooper and  P Cushman and  A De Benedetti and  A Dolgopolov and  P R Dudero and  R Egeland and  G Franzoni and  W J Gilbert and  D Gong and  J Grahl and  J Haupt and  K Klapoetke and  I Kronkvist and  Y Kubota and  J Mans and  R Rusack and  S Sengupta and  B Sherwood and  A Singovsky and  P Vikas and  J Zhang and  M Booke and  L M Cremaldi and  R Godang and  R Kroeger and  M Reep and  J Reidy and  D A Sanders and  P Sonnek and  D Summers and  S Watkins and  K Bloom and  B Bockelman and  D R Claes and  A Dominguez and  M Eads and  M Furukawa and  J Keller and  T Kelly and  C Lundstedt and  S Malik and  G R Snow and  D Swanson and  K M Ecklund and  I Iashvili and  A Kharchilava and  A Kumar and  M Strang and  G Alverson and  E Barberis and  O Boeriu and  G Eulisse and  T McCauley and  Y Musienko and  S Muzaffar and  I Osborne and  S Reucroft and  J Swain and  L Taylor and  L Tuura and  B Gobbi and  M Kubantsev and  A Kubik and  R A Ofierzynski and  M Schmitt and  E Spencer and  S Stoynev and  M Szleper and  M Velasco and  S Won and  K Andert and  B Baumbaugh and  B A Beiersdorf and  L Castle and  J Chorny and  A Goussiou and  M Hildreth and  C Jessop and  D J Karmgard and  T Kolberg and  J Marchant and  N Marinelli and  M McKenna and  R Ruchti and  M Vigneault and  M Wayne and  D Wiand and  B Bylsma and  L S Durkin and  J Gilmore and  J Gu and  P Killewald and  T Y Ling and  C J Rush and  V Sehgal and  G Williams and  N Adam and  S Chidzik and  P Denes and  P Elmer and  A Garmash and  D Gerbaudo and  V Halyo and  J Jones and  D Marlow and  J Olsen and  P Piroué and  D Stickland and  C Tully and  J S Werner and  T Wildish and  S Wynhoff and  Z Xie and  X T Huang and  A Lopez and  H Mendez and  J E Ramirez Vargas and  A Zatserklyaniy and  A Apresyan and  K Arndt and  V E Barnes and  G Bolla and  D Bortoletto and  A Bujak and  A Everett and  M Fahling and  A F Garfinkel and  L Gutay and  N Ippolito and  Y Kozhevnikov and  A T Laasanen and  C Liu and  V Maroussov and  S Medved and  P Merkel and  D H Miller and  J Miyamoto and  N Neumeister and  A Pompos and  A Roy and  A Sedov and  I Shipsey and  V Cuplov and  N Parashar and  P Bargassa and  S J Lee and  J H Liu and  D Maronde and  M Matveev and  T Nussbaum and  B P Padley and  J Roberts and  A Tumanov and  A Bodek and  H Budd and  J Cammin and  Y S Chung and  P De Barbaro and  R Demina and  G Ginther and  Y Gotra and  S Korjenevski and  D C Miner and  W Sakumoto and  P Slattery and  M Zielinski and  A Bhatti and  L Demortier and  K Goulianos and  K Hatakeyama and  C Mesropian and  E Bartz and  S H Chuang and  J Doroshenko and  E Halkiadakis and  P F Jacques and  D Khits and  A Lath and  A Macpherson and  R Plano and  K Rose and  S Schnetzer and  S Somalwar and  R Stone and  T L Watts and  G Cerizza and  M Hollingsworth and  J Lazoflores and  G Ragghianti and  S Spanier and  A York and  A Aurisano and  A Golyash and  T Kamon and  C N Nguyen and  J Pivarski and  A Safonov and  D Toback and  M Weinberger and  N Akchurin and  L Berntzon and  K W Carrell and  K Gumus and  C Jeong and  H Kim and  S W Lee and  B G Mc Gonagill and  Y Roh and  A Sill and  M Spezziga and  R Thomas and  I Volobouev and  E Washington and  R Wigmans and  E Yazgan and  T Bapty and  D Engh and  C Florez and  W Johns and  T Keskinpala and  E Luiggi Lopez and  S Neema and  S Nordstrom and  S Pathak and  P Sheldon and  D Andelin and  M W Arenton and  M Balazs and  M Buehler and  S Conetti and  B Cox and  R Hirosky and  M Humphrey and  R Imlay and  A Ledovskoy and  D Phillips II and  H Powell and  M Ronquest and  R Yohay and  M Anderson and  Y W Baek and  J N Bellinger and  D Bradley and  P Cannarsa and  D Carlsmith and  I Crotty and  S Dasu and  F Feyzi and  T Gorski and  L Gray and  K S Grogg and  M Grothe and  M Jaworski and  P Klabbers and  J Klukas and  A Lanaro and  C Lazaridis and  J Leonard and  R Loveless and  M Magrans de Abril and  A Mohapatra and  G Ott and  W H Smith and  M Weinberg and  D Wenman and  G S Atoian and  S Dhawan and  V Issakov and  H Neal and  A Poblaguev and  M E Zeller and  G Abdullaeva and  A Avezov and  M I Fazylov and  E M Gasanov and  A Khugaev and  Y N Koblik and  M Nishonov and  K Olimov and  A Umaraliev and  B S Yuldashev},
	title = {{The CMS experiment at the CERN LHC}},
	journal = {Journal of Instrumentation},
	year = {2008},
	OPTmonth = {aug},
	OPTpublisher = {},
	OPTvolume = {3},
	OPTnumber = {08},
	OPTpages = {S08004},
	doi = {10.1088/1748-0221/3/08/S08004},
	OPTurl = {https://dx.doi.org/10.1088/1748-0221/3/08/S08004}
}

@article{Collaboration:2008:LHCb,
	author = {The LHCb Collaboration and  A Augusto Alves Jr and  L M Andrade Filho and  A F Barbosa and  I Bediaga and  G Cernicchiaro and  G Guerrer and  H P Lima Jr and  A A Machado and  J Magnin and  F Marujo and  J M de Miranda and  A Reis and  A Santos and  A Toledo and  K Akiba and  S Amato and  B de Paula and  L de Paula and  T da Silva and  M Gandelman and  J H Lopes and  B Maréchal and  D Moraes and  E Polycarpo and  F Rodrigues and  J Ballansat and  Y Bastian and  D Boget and  I De Bonis and  V Coco and  P Y David and  D Decamp and  P Delebecque and  C Drancourt and  N Dumont-Dayot and  C Girard and  B Lieunard and  M N Minard and  B Pietrzyk and  T Rambure and  G Rospabe and  S T'Jampens and  Z Ajaltouni and  G Bohner and  R Bonnefoy and  D Borras and  C Carloganu and  H Chanal and  E Conte and  R Cornat and  M Crouau and  E Delage and  O Deschamps and  P Henrard and  P Jacquet and  C Lacan and  J Laubser and  J Lecoq and  R Lefèvre and  M Magne and  M Martemiyanov and  M-L Mercier and  S Monteil and  V Niess and  P Perret and  G Reinmuth and  A Robert and  S Suchorski and  K Arnaud and  E Aslanides and  J Babel and  C Benchouk and  J-P Cachemiche and  J Cogan and  F Derue and  B Dinkespiler and  P-Y Duval and  V Garonne and  S Favard and  R Le Gac and  F Leon and  O Leroy and  P-L Liotard and  F Marin and  M Menouni and  P Ollive and  S Poss and  A Roche and  M Sapunov and  L Tocco and  B Viaud and  A Tsaregorodtsev and  Y Amhis and  G Barrand and  S Barsuk and  C Beigbeder and  R Beneyton and  D Breton and  O Callot and  D Charlet and  B D'Almagne and  O Duarte and  F Fulda-Quenzer and  A Jacholkowska and  B Jean-Marie and  J Lefrancois and  F Machefert and  P Robbe and  M-H Schune and  V Tocut and  I Videau and  M Benayoun and  P David and  L Del Buono and  G Gilles and  M Domke and  H Futterschneider and  Ch Ilgner and  P Kapusta and  M Kolander and  R Krause and  M Lieng and  M Nedos and  K Rudloff and  S Schleich and  R Schwierz and  B Spaan and  K Wacker and  K Warda and  M Agari and  C Bauer and  D Baumeister and  N Bulian and  H P Fuchs and  W Fallot-Burghardt and  T Glebe and  W Hofmann and  K T Knöpfle and  S Löchner and  A Ludwig and  F Maciuc and  F Sanchez Nieto and  M Schmelling and  B Schwingenheuer and  E Sexauer and  N J Smale and  U Trunk and  H Voss and  J Albrecht and  S Bachmann and  J Blouw and  M Deissenroth and  H Deppe and  H B Dreis and  F Eisele and  T Haas and  S Hansmann-Menzemer and  S Hennenberger and  J Knopf and  M Moch and  A Perieanu and  S Rabenecker and  A Rausch and  C Rummel and  R Rusnyak and  M Schiller and  U Stange and  U Uwer and  M Walter and  R Ziegler and  G Avoni and  G Balbi and  F Bonifazi and  D Bortolotti and  A Carbone and  I D'Antone and  D Galli and  D Gregori and  I Lax and  U Marconi and  G Peco and  V Vagnoni and  G Valenti and  S Vecchi and  W Bonivento and  A Cardini and  S Cadeddu and  V DeLeo and  C Deplano and  S Furcas and  A Lai and  R Oldeman and  D Raspino and  B Saitta and  N Serra and  W Baldini and  S Brusa and  S Chiozzi and  A Cotta Ramusino and  F Evangelisti and  A Franconieri and  S Germani and  A Gianoli and  L Guoming and  L Landi and  R Malaguti and  C Padoan and  C Pennini and  M Savriè and  S Squerzanti and  T Zhao and  M Zhu and  A Bizzeti and  G Graziani and  M Lenti and  M Lenzi and  F Maletta and  S Pennazzi and  G Passaleva and  M Veltri and  M Alfonsi and  M Anelli and  A Balla and  A Battisti and  G Bencivenni and  P Campana and  M Carletti and  P Ciambrone and  G Corradi and  E Dané and  A DiVirgilio and  P DeSimone and  G Felici and  C Forti and  M Gatta and  G Lanfranchi and  F Murtas and  M Pistilli and  M Poli Lener and  R Rosellini and  M Santoni and  A Saputi and  A Sarti and  A Sciubba and  A Zossi and  M Ameri and  S Cuneo and  F Fontanelli and  V Gracco and  G Miní and  M Parodi and  A Petrolini and  M Sannino and  A Vinci and  M Alemi and  C Arnaboldi and  T Bellunato and  M Calvi and  F Chignoli and  A De Lucia and  G Galotta and  R Mazza and  C Matteuzzi and  M Musy and  P Negri and  D Perego and  G Pessina and  G Auriemma and  V Bocci and  A Buccheri and  G Chiodi and  S Di Marco and  F Iacoangeli and  G Martellotti and  R Nobrega and  A Pelosi and  G Penso and  D Pinci and  W Rinaldi and  A Rossi and  R Santacesaria and  C Satriano and  G Carboni and  M Iannilli and  A Massafferri Rodrigues and  R Messi and  G Paoluzzi and  G Sabatino and  E Santovetti and  A Satta and  J Amoraal and  G van Apeldoorn and  R Arink and  N van Bakel and  H Band and  Th Bauer and  A Berkien and  M van Beuzekom and  E Bos and  Ch Bron and  L Ceelie and  M Doets and  R van der Eijk and  J-P Fransen and  P de Groen and  V Gromov and  R Hierck and  J Homma and  B Hommels and  W Hoogland and  E Jans and  F Jansen and  L Jansen and  M Jaspers and  B Kaan and  B Koene and  J Koopstra and  F Kroes and  M Kraan and  J Langedijk and  M Merk and  S Mos and  B Munneke and  J Palacios and  A Papadelis and  A Pellegrino and  O van Petten and  T du Pree and  E Roeland and  W Ruckstuhl and  A Schimmel and  H Schuijlenburg and  T Sluijk and  J Spelt and  J Stolte and  H Terrier and  N Tuning and  A Van Lysebetten and  P Vankov and  J Verkooijen and  B Verlaat and  W Vink and  H de Vries and  L Wiggers and  G Ybeles Smit and  N Zaitsev and  M Zupan and  A Zwart and  J van den Brand and  H J Bulten and  M de Jong and  T Ketel and  S Klous and  J Kos and  B M'charek and  F Mul and  G Raven and  E Simioni and  J Cheng and  G Dai and  Z Deng and  Y Gao and  G Gong and  H Gong and  J He and  L Hou and  J Li and  W Qian and  B Shao and  T Xue and  Z Yang and  M Zeng and  B Muryn and  K Ciba and  A Oblakowska-Mucha and  J Blocki and  K Galuszka and  L Hajduk and  J Michalowski and  Z Natkaniec and  G Polok and  M Stodulski and  M Witek and  K Brzozowski and  A Chlopik and  P Gawor and  Z Guzik and  A Nawrot and  A Srednicki and  K Syryczynski and  M Szczekowski and  D V Anghel and  A Cimpean and  C Coca and  F Constantin and  P Cristian and  D D Dumitru and  D T Dumitru and  G Giolu and  C Kusko and  C Magureanu and  Gh Mihon and  M Orlandea and  C Pavel and  R Petrescu and  S Popescu and  T Preda and  A Rosca and  V L Rusu and  R Stoica and  S Stoica and  P D Tarta and  S Filippov and  Yu Gavrilov and  L Golyshkin and  E Gushchin and  O Karavichev and  V Klubakov and  L Kravchuk and  V Kutuzov and  S Laptev and  S Popov and  A Aref'ev and  B Bobchenko and  V Dolgoshein and  V Egorychev and  A Golutvin and  O Gushchin and  A Konoplyannikov and  I Korolko and  T Kvaratskheliya and  I Machikhiliyan and  S Malyshev and  E Mayatskaya and  M Prokudin and  D Rusinov and  V Rusinov and  P Shatalov and  L Shchutska and  E Tarkovskiy and  A Tayduganov and  K Voronchev and  O Zhiryakova and  A Bobrov and  A Bondar and  S Eidelman and  A Kozlinsky and  L Shekhtman and  K S Beloous and  R I Dzhelyadin and  Yu V Gelitsky and  Yu P Gouz and  K G Kachnov and  A S Kobelev and  V D Matveev and  V P Novikov and  V F Obraztsov and  A P Ostankov and  V I Romanovsky and  V I Rykalin and  A P Soldatov and  M M Soldatov and  E N Tchernov and  O P Yushchenko and  B Bochin and  N Bondar and  O Fedorov and  V Golovtsov and  S Guets and  A Kashchuk and  V Lazarev and  O Maev and  P Neustroev and  N Sagidova and  E Spiridenkov and  S Volkov and  An Vorobyev and  A Vorobyov and  E Aguilo and  S Bota and  M Calvo and  A Comerma and  X Cano and  A Dieguez and  A Herms and  E Lopez and  S Luengo and  J Garra and  Ll Garrido and  D Gascon and  A Gaspar de Valenzuela and  C Gonzalez and  R Graciani and  E Grauges and  A Perez Calero and  E Picatoste and  J Riera and  M Rosello and  H Ruiz and  X Vilasis and  X Xirgu and  B Adeva and  X Cid Vidal and  D MartÉnez Santos and  D Esperante Pereira and  J L Fungueiriño Pazos and  A Gallas Torreira and  C Lois Gómez and  A Pazos Alvarez and  E Pérez Trigo and  M Pló Casasús and  C Rodriguez Cobo and  P Rodríguez Pérez and  J J Saborido and  M Seco and  P Vazquez Regueiro and  P Bartalini and  A Bay and  M-O Bettler and  F Blanc and  J Borel and  B Carron and  C Currat and  G Conti and  O Dormond and  Y Ermoline and  P Fauland and  L Fernandez and  R Frei and  G Gagliardi and  N Gueissaz and  G Haefeli and  A Hicheur and  C Jacoby and  P Jalocha and  S Jimenez-Otero and  J-P Hertig and  M Knecht and  F Legger and  L Locatelli and  J-R Moser and  M Needham and  L Nicolas and  A Perrin-Giacomin and  J-P Perroud and  C Potterat and  F Ronga and  O Schneider and  T Schietinger and  D Steele and  L Studer and  M Tareb and  M T Tran and  J van Hunen and  K Vervink and  S Villa and  N Zwahlen and  R Bernet and  A Büchler and  J Gassner and  F Lehner and  T Sakhelashvili and  C Salzmann and  P Sievers and  S Steiner and  O Steinkamp and  U Straumann and  J van Tilburg and  A Vollhardt and  D Volyanskyy and  M Ziegler and  A Dovbnya and  Yu Ranyuk and  I Shapoval and  M Borisova and  V Iakovenko and  V Kyva and  O Kovalchuk and  O Okhrimenko and  V Pugatch and  Yu Pylypchenko and  M Adinolfi and  N H Brook and  R D Head and  J P Imong and  K A Lessnoff and  F C D Metlica and  A J Muir and  J H Rademacker and  A Solomin and  P M Szczypka and  C Barham and  C Buszello and  J Dickens and  V Gibson and  S Haines and  K Harrison and  C R Jones and  S Katvars and  U Kerzel and  C Lazzeroni and  Y Y Li and  G Rogers and  J Storey and  H Skottowe and  S A Wotton and  T J Adye and  C J Densham and  S Easo and  B Franek and  P Loveridge and  D Morrow and  J V Morris and  R Nandakumar and  J Nardulli and  A Papanestis and  G N Patrick and  S Ricciardi and  M L Woodward and  Z Zhang and  R J U Chamonal and  P J Clark and  P Clarke and  S Eisenhardt and  N Gilardi and  A Khan and  Y M Kim and  R Lambert and  J Lawrence and  A Main and  J McCarron and  C Mclean and  F Muheim and  A F Osorio-Oliveros and  S Playfer and  N Styles and  Y Xie and  A Bates and  L Carson and  F da Cunha Marinho and  F Doherty and  L Eklund and  M Gersabeck and  L Haddad and  A A Macgregor and  J Melone and  F McEwan and  D M Petrie and  S K Paterson and  C Parkes and  A Pickford and  B Rakotomiaramanana and  E Rodrigues and  A F Saavedra and  F J P Soler and  T Szumlak and  S Viret and  L Allebone and  O Awunor and  J Back and  G Barber and  C Barnes and  B Cameron and  D Clark and  I Clark and  P Dornan and  A Duane and  C Eames and  U Egede and  M Girone and  S Greenwood and  R Hallam and  R Hare and  A Howard and  S Jolly and  V Kasey and  M Khaleeq and  P Koppenburg and  D Miller and  R Plackett and  D Price and  W Reece and  P Savage and  T Savidge and  B Simmons and  G Vidal-Sitjes and  D Websdale and  A Affolder and  J S Anderson and  S F Biagi and  T J V Bowcock and  J L Carroll and  G Casse and  P Cooke and  S Donleavy and  L Dwyer and  K Hennessy and  T Huse and  D Hutchcroft and  D Jones and  M Lockwood and  M McCubbin and  R McNulty and  D Muskett and  A Noor and  G D Patel and  K Rinnert and  T Shears and  N A Smith and  G Southern and  I Stavitski and  P Sutcliffe and  M Tobin and  S M Traynor and  P Turner and  M Whitley and  M Wormald and  V Wright and  J H Bibby and  S Brisbane and  M Brock and  M Charles and  C Cioffi and  V V Gligorov and  T Handford and  N Harnew and  F Harris and  M J J John and  M Jones and  J Libby and  L Martin and  I A McArthur and  R Muresan and  C Newby and  B Ottewell and  A Powell and  N Rotolo and  R S Senanayake and  L Somerville and  A Soroko and  P Spradlin and  P Sullivan and  I Stokes-Rees and  S Topp-Jorgensen and  F Xing and  G Wilkinson and  M Artuso and  I Belyaev and  S Blusk and  G Lefeuvre and  N Menaa and  R Menaa-Sia and  R Mountain and  T Skwarnicki and  S Stone and  J C Wang and  L Abadie and  G Aglieri-Rinella and  E Albrecht and  J André and  G Anelli and  N Arnaud and  A Augustinus and  F Bal and  M C Barandela Pazos and  A Barczyk and  M Bargiotti and  J Batista Lopes and  O Behrendt and  S Berni and  P Binko and  V Bobillier and  A Braem and  L Brarda and  J Buytaert and  L Camilleri and  M Cambpell and  G Castellani and  F Cataneo and  M Cattaneo and  B Chadaj and  P Charpentier and  S Cherukuwada and  E Chesi and  J Christiansen and  R Chytracek and  M Clemencic and  J Closier and  P Collins and  P Colrain and  O Cooke and  B Corajod and  G Corti and  C D'Ambrosio and  B Damodaran and  C David and  S de Capua and  G Decreuse and  H Degaudenzi and  H Dijkstra and  J-P Droulez and  D Duarte Ramos and  J P Dufey and  R Dumps and  D Eckstein and  M Ferro-Luzzi and  F Fiedler and  F Filthaut and  W Flegel and  R Forty and  C Fournier and  M Frank and  C Frei and  B Gaidioz and  C Gaspar and  J-C Gayde and  P Gavillet and  A Go and  G Gracia Abril and  J-S Graulich and  P-A Giudici and  A Guirao Elias and  P Guglielmini and  T Gys and  F Hahn and  S Haider and  J Harvey and  B Hay and  J-A Hernando Morata and  J Herranz Alvarez and  E van Herwijnen and  H J Hilke and  G von Holtey and  W Hulsbergen and  R Jacobsson and  O Jamet and  C Joram and  B Jost and  N Kanaya and  J Knaster Refolio and  S Koestner and  M Koratzinos and  R Kristic and  D Lacarrère and  C Lasseur and  T Lastovicka and  M Laub and  D Liko and  C Lippmann and  R Lindner and  M Losasso and  A Maier and  K Mair and  P Maley and  P Mato Vila and  G Moine and  J Morant and  M Moritz and  J Moscicki and  M Muecke and  H Mueller and  T Nakada and  N Neufeld and  J Ocariz and  C Padilla Aranda and  U Parzefall and  M Patel and  M Pepe-Altarelli and  D Piedigrossi and  M Pivk and  W Pokorski and  S Ponce and  F Ranjard and  W Riegler and  J Renaud and  S Roiser and  A Rossi and  L Roy and  T Ruf and  D Ruffinoni and  S Saladino and  A Sambade Varela and  R Santinelli and  S Schmelling and  B Schmidt and  T Schneider and  A Schöning and  A Schopper and  J Seguinot and  W Snoeys and  A Smith and  A C Smith and  P Somogyi and  R Stoica and  W Tejessy and  F Teubert and  E Thomas and  J Toledo Alarcon and  O Ullaland and  A Valassi and  P Vannerem and  R Veness and  P Wicht and  D Wiedner and  W Witzeling and  A Wright and  K Wyllie and  T Ypsilantis},
	title = {{The LHCb Detector at the LHC}},
	journal = {Journal of Instrumentation},
	year = {2008},
	OPTmonth = {aug},
	OPTpublisher = {},
	OPTvolume = {3},
	OPTnumber = {08},
	OPTpages = {S08005},
	doi = {10.1088/1748-0221/3/08/S08005},
	OPTurl = {https://dx.doi.org/10.1088/1748-0221/3/08/S08005}
}

@techreport{Edmonds:2008:FATS,
	author = {Edmonds, K and Fleischmann, S and Lenz, T and Magass, C and Mechnich, J and Salzburger, A},
	title = {{The Fast ATLAS Track Simulation (FATRAS)}},
	OPTinstitution = {CERN},
	reportNumber = {ATL-SOFT-PUB-2008-001, ATL-COM-SOFT-2008-002},
	OPTaddress = {Geneva},
	year = {2008},
	OPTurl = {https://cds.cern.ch/record/1091969},
	OPTnote = {All figures including auxiliary figures are available at https://atlas.web.cern.ch/Atlas/GROUPS/PHYSICS/PUBNOTES/ATL-SOFT-PUB-2008-001}
}

@techreport{Richter-Was:1998:AFSP,
	author = {Richter-Was, Elzbieta and Froidevaux, D and Poggioli, Luc},
	title = {{ATLFAST 2.0 a fast simulation package for ATLAS}},
	OPTinstitution = {CERN},
	reportNumber = {ATL-PHYS-98-131},
	OPTaddress = {Geneva},
	year = {1998},
	OPTurl = {https://cds.cern.ch/record/683751},
}

\end{document}